\documentstyle[12pt,epsfig,epsf]{article}

\parskip=\itemsep               

\newcommand{\beq}{\begin{equation}}
\newcommand{\eeq}{\end{equation}}

\newcommand{\beqar}[1]{\begin{eqnarray}\label{#1}}
\newcommand{\eeqar}{\end{eqnarray}}

\newcommand{\si}{\sigma}

\newcommand{\as}{\alpha_S}

\def\eq#1{{Eq.~(\ref{#1})}}

\def\npb#1#2#3{    {\it Nucl. Phys. }{\bf B#1} (19#2) #3}
\def\plb#1#2#3{    {\it Phys. Lett. }{\bf B#1} (19#2) #3}
\def\prd#1#2#3{    {\it Phys. Rev. }{\bf D#1} (19#2) #3}

\def\zpc#1#2#3{    {\it Z. Phys. }{\bf C#1} (19#2) #3}

\newcommand{\bas}{\bar{\alpha}_S}
\relax
\begin{document}
%
%
%
\noindent
\begin{flushright}
\parbox[t]{10em}{
TAUP-2639 - 2000\\
{\bf \today} }
\end{flushright}
\vspace{1cm}
\begin{center}
{\Huge  \bf  Energy  Dependence \,\, of\,\, 
$\mathbf{\sigma^{DD}/\sigma_{tot}}$}\\[1.5ex]
{\Huge \bf in DIS  and  Shadowing  Corrections}
 \\[4ex]
 
{\large \bf  E.  ~Gotsman, $\mathbf{{}^{a),d)}}$\footnote{
Email: gotsman@post.tau.ac.il .}\,\, E.
~Levin,$\mathbf{{}^{a),b)}}$\footnote{ Email:
leving@post.tau.ac.il; levin@mail.desy.de .}\,\,
M. ~Lublinsky,$\mathbf{{}^{c)}}$\footnote{  Email:
mal@techunix.technion.ac.il}}\\
{ \large \bf  U. ~Maor
$\mathbf{{}^{a)}}$\footnote{Email:
maor@post.tau.ac.il .}\,\,and  
\,\, K. ~Tuchin $\mathbf{{}^{a)}}$\footnote{ Email:
 tuchin@post.tau.ac.il .}}\\[4.5ex]

{\it ${}^{a)}$ HEP Department}\\
{\it  School of Physics and Astronomy}\\
{\it Raymond and Beverly Sackler Faculty of Exact Science}\\
{\it Tel Aviv University, Tel Aviv, 69978, ISRAEL}\\[1.5ex]
{\it ${}^{b)}$ DESY Theory Group}\\
{\it 22603, Hamburg, GERMANY}\\[1.5ex]
{\it ${}^{c)}$  Department of Physics}\\
{\it  Technion -- Israel Institute of   Technology}\\
{\it  Haifa 32000, ISRAEL}\\[1.5ex]
{\it ${}^{d)}$ Department of Physics and Astronomy}\\
{\it Univerity of California, Irvine}\\
{\it Irvine, CA 92697-4575, USA}\\[4.5ex]

\end{center}  
~\,\,
\vspace{1cm}
 
{\samepage
{\large \bf Abstract:}
 We gereralize the Kovchegov-McLerran formula for the
ratio $\sigma^{DD}/\sigma_{tot}$ in perturbative QCD,
using Mueller-Glauber approach
for shadowing corrections and AGK cutting rules. We investigate several
phenomenological approaches with the goal of obtaining results consisent
with experimental data. We fail to reproduce the observed weak energy
dependence of the ratio, and conclude that the soft nonperturbative
contribution present at short distances must also be included.
 
}

\section{Introduction}
\setcounter{equation}{0}

One of the intriguing phenomena, observed at HERA, is the
behaviour of the energy
dependence of the ratio $\sigma^{DD}/\sigma_{tot}$ in deep inelastic
scattering (DIS ). It appears  ( see Ref.\cite{ZEUSDATA} and  Fig. 1 )
that this ratio as a function of energy
 is almost constant  for different
masses of the diffractively produced hadrons over a  wide range of photon
virtualities  $Q^2$.  
At present there is no theoretical explanation for this striking
experimental
observation. The only valid theoretical idea on the market is the
quasiclassical
gluon field approach (see Ref. \cite{BUCH} ) in which the total 
as well as the diffractive cross section do not depend on energy.
   
 Recently,  Yu. Kovchegov and L. McLerran \cite{KOLE}  suggested that the
constantcy of the ratio $\sigma^{DD}/\sigma_{tot}$ is closely related to
strong shadowing corrections ( SC )  for  diffractive production.
They derived a formula for this ratio for the case where only a quark
- antiquark pair is produced in diffractive dissociation. On the other
hand, K. Golec - Bierat and M. W\"{u}sthoff \cite{GOWU1} suggested a
phenomenological model which incorporates two main theoretical  
ideas regarding the transition between ``hard" and ``soft" processes in
QCD
\cite{TEOID}:
(i) the appearence of a new scale which depends on energy, and is
related to the average transverse monentum of a parton in the
parton cascade; and (ii) the saturation   of the parton density at high
energies. This model is successful in
describing all the available experimental data on total
and diffractive cross sections, including the energy behaviour of
$\sigma^{DD}/\sigma_{tot}$ \cite{GOWU1} \cite{GOWU2}.

\begin{figure}[htbp]
\begin{center}
  \epsfig{file=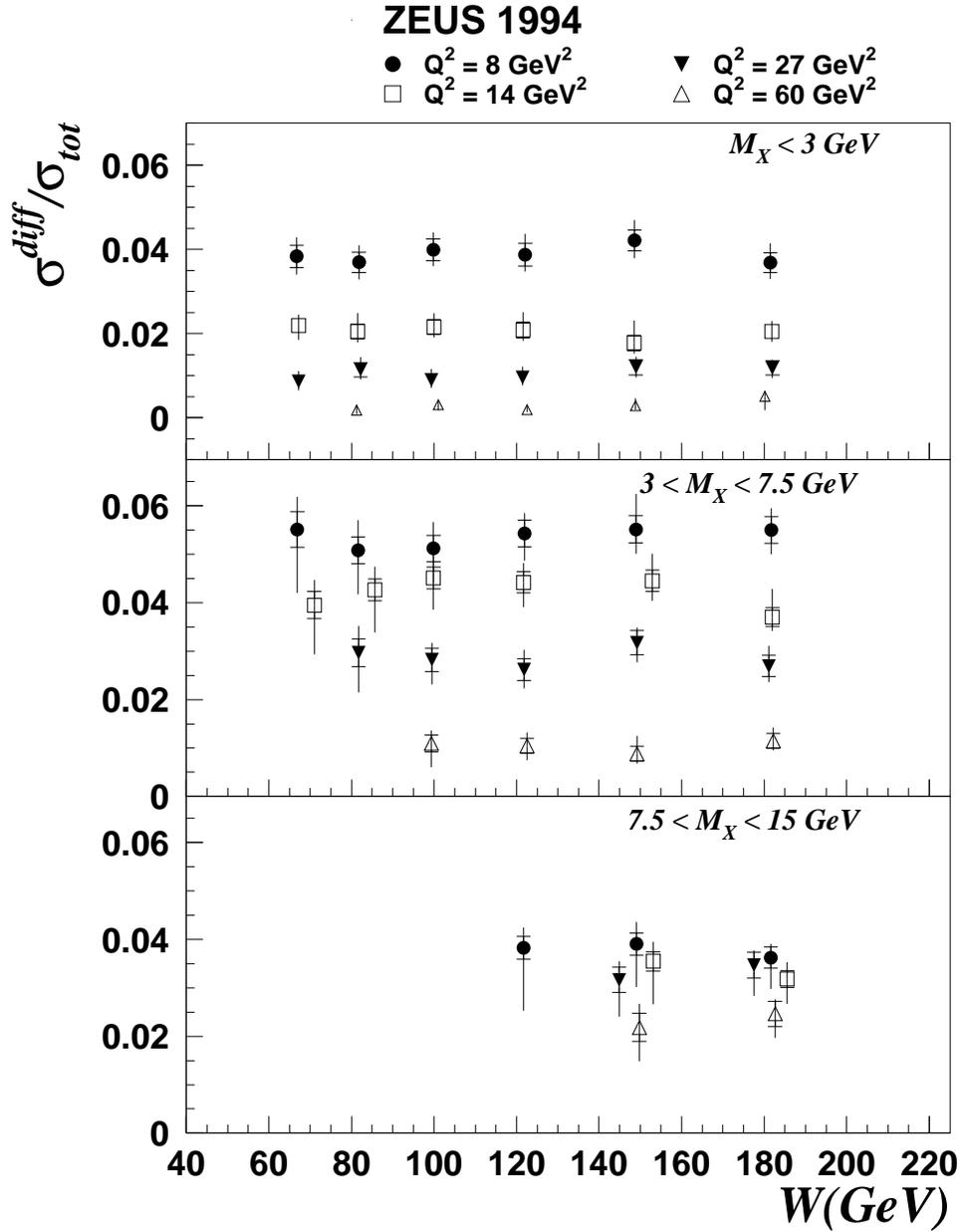,width=150mm}
  \caption[]{\it Experimantal data for the ratio
$\sigma^{DD}/\sigma_{tot}$ taken from Ref. \protect\cite{ZEUSDATA}. }
 \end{center} 
\label{fig1}
\end{figure}

  The success of the above mentioned papers
 prompted us to reexamine the energy behaviour of
$\sigma^{DD}/\sigma_{tot}$ in perturbative QCD ( pQCD ) in more detail.
Our approach is based on two main ideas used to describe the total and
diffractive
cross section for DIS:
 \begin{enumerate}
\item\,\,\,The final state of the diffractive
processes in HERA kinamatic region \cite{GLMSM},
are desribed by
the diffraction dissociation of a virtual photon ( $\gamma^*$ ) into
a  quark-antiquark  pair  and  quark-antiquark  pair plus one
extra gluon (see Fig.2 ); 

\item\,\,\,The Mueller-Glauber approach\cite{MU90}   for calculating
 SC in the total and diffractive cross sections for DIS, this was
used successfully  to describe other indications of strong SC in HERA data 
\cite{GLMSM} \cite{GLMHARD},
such as the $Q^2$ - behaviour of the $F_2$ slope and energy behaviour of
the diffractive cross section \cite{AC}. 
\end{enumerate}

\begin{figure}[htbp]
\begin{center}
  \epsfig{file=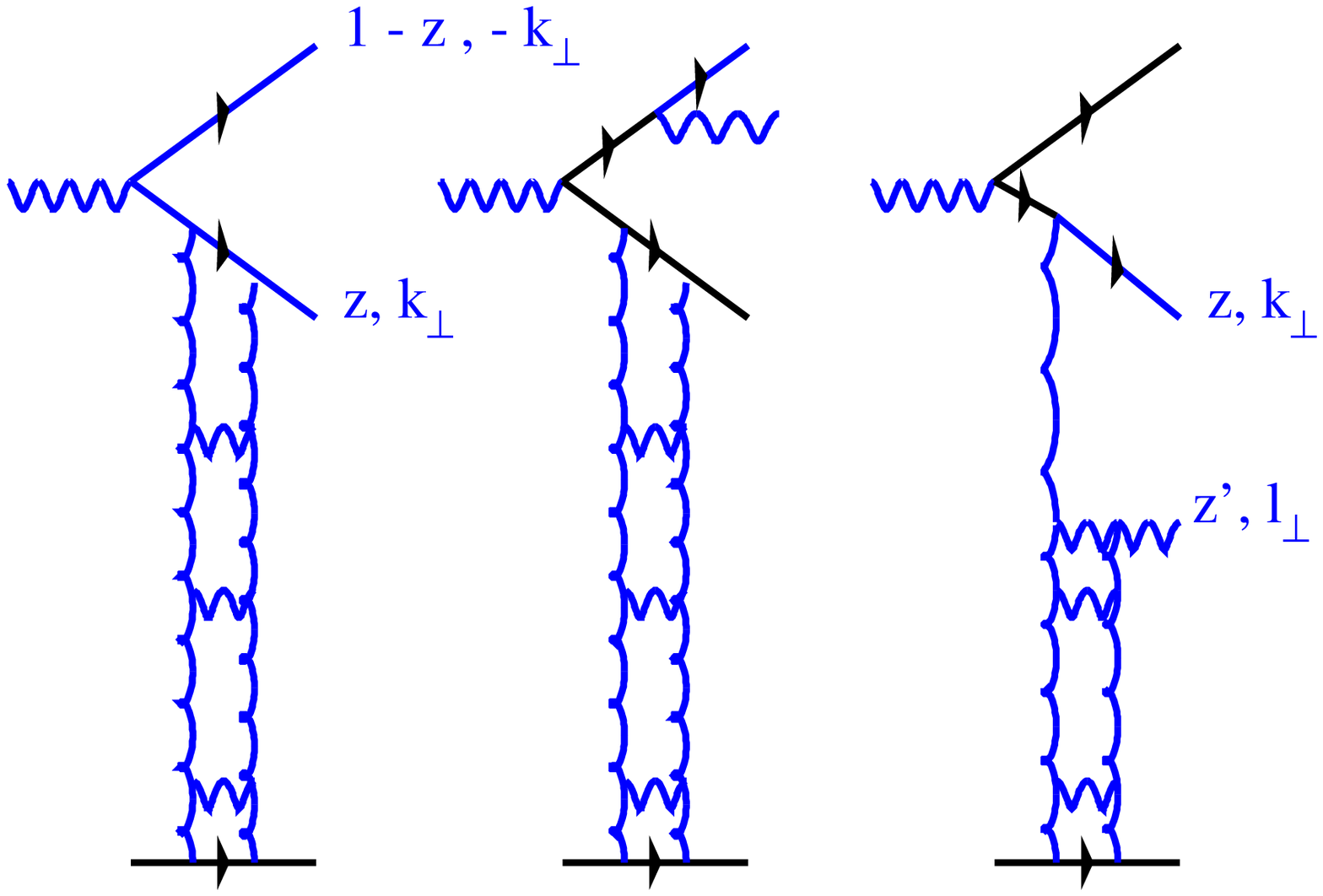,width=150mm, height=60mm}
  \caption[]{\it  Diffractive dissociation of the virtual photon into
 quark  - antiquark  pair  ( $q \bar q $ ) and  quark  - antiquark
pair plus one extra gluon ( $ q \bar q  G $)  in pQCD. }
 \end{center}
\label{fig2}
\end{figure}

The paper is organized as follows. In section 2 we give a simple
derivation of the Kovchegov and McLerran formula based on the $s$-channel
unitarity constraints. We generalize this formula for the case of
 $q \bar qG $
 final state ( see Fig.2 ) and discuss the relation between
this approach and the AGK cutting rules \cite{AGK}. Section 3 is devoted
to a numeric calculation of the ratio $\sigma^{DD}/\sigma_{tot}$ without
any
restriction on the value of produced masses. In section 4 we discuss how 
limitations on the mass range change the energy dependence of the ratio.
A summary of our results as well as a discussion on future HERA
experiments are given in section 5.   Appendix fives all formulae and all
details of our calculations.
 
\section{ Shadowing corrections in QCD}
\setcounter{equation}{0}
\subsection{Notations and definitions}
 In this paper we develop further our approach for  diffractive
production in DIS started in Ref.\cite{GLMSM}.  We  use
the same  notations and definitions as in Ref. \cite{GLMSM}.
In this paper we examine the main physical ideas of Ref. \cite{MUDIPOLE},
i.e. that the correct degrees of freedom at high enegies ( low $x$ ) are
colour
dipoles, rather than quarks and gluons which appear explicitly in the
QCD Lagrangian. The consequence of this hypothesis is that a QCD
interaction at high
energies does not change the size and energy of a colour dipole.
Hence,
the majority of our variables and observables are related to
the  distribution
and interaction of the colour dipoles in a hadron.

To facilitate  reading the paper we list the notation and definitions
which we will use  
 ( see Fig.3 ):   

\begin{figure}[htbp]
\begin{center}
  \epsfig{file=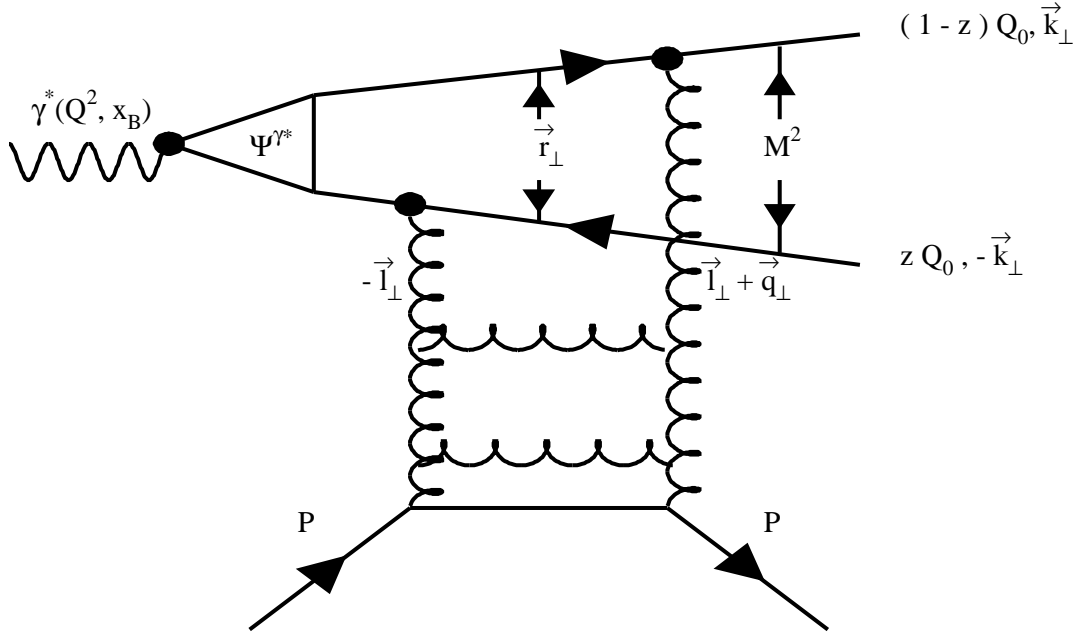,width=150mm}
  \caption[]{\it  Diffractive production of a 
quark  - antiquark  pair. }
 \end{center}
\label{fig3}
\end{figure}

\begin{enumerate}
\item\,\,\,$Q^2$ denotes the virtuality of the photon in DIS, $M$ the
produced
mass and $W$  the energy of the collision in c.m. frame;
\item\,\,\,$x_P\,\,=\,\, ( Q^2 + M^2 )/W^2 $ is the fraction of
energy 
carried by the Pomeron ( two gluon ladder in Fig.3 ). Bjorken scaling
variable is  $x_B\,\,=\,\,Q^2/W^2 $;

\item\,\,\,We use the symbol $x$ for both $x_P$ and $x_B$ .
 \item\,\,\,$\beta\,\,=\,\,Q^2/( Q^2 + M^2
)\,\,=\,\,x_B/x_P$
is the fraction of the Pomeron energy carried by the struck quark;
\item\,\,\,$k_{\perp}$ denotes the transverse momentum of the quark, and
$r_{\perp}\,\,\equiv\,\,r$  the transverse distance between the quark
and
the antiquark
i.e.  the size of the colour dipole;
\item\,\,\,$l_{\perp}$ is the transverse momentum of the gluon emitted by
the  quark ( antiquark );

\item\,\,\,$z$ is the fraction of the photon momentum in the laboratory
frame carried by the quark or antiquark;

\item\,\,\,$b_t$ is the impact parameter of the reaction and is the
variable conjugated to $q_{\perp}$, the momentum transfer from the
incoming proton to the recoiled proton. Note that $t = - q^2_{\perp}$;
\item\,\,\, Our amplitude is normalized so that
\beq \label{NORM1}
\frac{d \si}{d t }\,\,=\,\,\pi \,| f(s = W^2,t)|^2\,\,,
\eeq
with the optical theorem given by
\beq \label{OPTEO}
\si_{tot}\,\,\,=\,\,4\,\pi\,\,Im\,f (s,0 )\,\,.
\eeq
\item\,\,\, The scattering amplitude in $b_t$ space is defined by 
\beq \label{B}
a^{el}(s,b_t)\,\,=\,\,\frac{1}{2\,\pi}\,\int\,d^2\,q_{\perp}\,\,e^{ -
i\,\vec{\mathbf{q_{\perp}}}\cdot\vec{\mathbf{b_t}}}\,\,f(s,t = -
q^2_{\perp} )\,\,.
\eeq
\item\,\,\, The $s$-channel unitarity constraint  then has the form
\beq \label{UNITARITY}
2\,Im\,a^{el}(s,b_t)\,\,\,=\,\,\,|a^{el}(s,b_t)|^2\,\,\,+\,\,\,G^{in}(s,b_t
)\,\,,
\eeq
where $G^{in} $ denotes the contribution of the all inelastic processes.

Therefore, in the impact parameter representation:
\begin{eqnarray}
&
\si_{tot}\,\,\,=\,\,\,2\,\,\int\,\,d^2\,b_t\,\,
Im\,a^{el}(s,b_t)\,\,;&\label{CR1}\\
&
\si_{el}\,\,\,=\,\,\,\int\,\,d^2\,b_t\,\,
|\,a^{el}(s,b_t)\,|^2\,\,;&\label{CR2}\\
&
\si_{in}\,\,\,=\,\,\,\,\,\int\,\,d^2\,b_t\,\,
G^{in}(s,b_t)\,\,.&\label{CR3}
\end{eqnarray}

\item\,\,\,$x_BG(x_B,Q^2)$ is the gluon distribution of the nucleon;
 
\item\,\,\,$\si_{dipole} (x_P, r_{\perp} )$ is the total cross section
 for dipole - nucleon  scattering, and is given by ( see Ref.
\cite{GLMSM} and references therein ) 
\beq \label{DIPOLECR}
\si_{dipole}( x_P, r_{\perp} )\,\,\,=\,\,\,\frac{\pi^2 \as}{3 }
r^2_{\perp}\,\, x_P
G(x_P,\frac{4}{r^2_{\perp}})\,\,,
\eeq
where $ x_P G(x_P,\frac{4}{r^2_{\perp}}) $ is the number of gluons with 
$Q^2 = \frac{4}{r^2_{\perp}}$ and energy $x_P$ in the nucleon.

\item\,\,\, To characterize the strength of the colour dipole interaction
we introduce 
\beq  \label{KAPPA}
\kappa_{dipole}\,\,\,=\,\,\frac{\si_{dipole}}{\pi R^2}\,\,=\,\,\frac{\pi^2
\as}{3\,\pi\,R^2 }r^2_{\perp}\,\,
x_PG^{DGLAP}(x_P,\frac{4}{r^2_{\perp}})\,\,,
\eeq
where $R$ is the nonperturbative parameter which is related to the
correlation radius of the gluons in a hadron. We determine its value from
 high energy phenomenology and HERA experimental data ( see section 3
for discussion ). 

To illustrate the physical meaning of \eq{KAPPA} we rewrite it 
 in the form:
\beq \label{KAPPA1}
\kappa_{dipole}\,\,\,=\,\,\si_0\,\,\rho(r_{\perp},x)\,\,,
\eeq
where
\beq \label{DIDENSE}
\rho(r_{\perp},x)\,\,\,=\,\,\,\frac{xG^{DGLAP}(x,\frac{4}{r^2_{\perp}})}{\pi
R^2}\,\,
\eeq 
is the density of colour dipoles  of size $r_{\perp}$ and energy
$x$ in the transverse plane.
 In \eq{KAPPA1}, $\si_0$ denotes
  the cross section for the interaction of a dipole
of size $r_{\perp}$ with a point - like probe
$$
\si_0 \,\,=\,\,\frac{\pi^2 \as}{3 }\,\,r^2_{\perp}\,\,.
$$
Hence, $\kappa_{dipole}$ is a packing factor for dipoles of  size
$r_{\perp}$ in a proton. If $\kappa_{dipole}$ is small,  we have a diluted
gas of
colour dipoles in a proton, but at low $x$ the gluon density increases
\cite{AC} and $\kappa_{dipole}\,\,\longrightarrow  \,\,1$. In a such
kinematic region our parton cascade becomes a dense system of colour
dipoles which should be treated non-perturbatively;

\item\,\,\, The amplitude for colour dipole scattering on a
nucleon is given by
\beq \label{DIAM}
a^{el}_{dipole}(s,r_{\perp};b_t)\,\,=\,\, 
\frac{\pi^2 \as}{3 }
r^2_{\perp}\,\, x_P
G(x_P,\frac{4}{r^2_{\perp}};b_t)\,\,,
\eeq
where $ x_P G(x_P,\frac{4}{r^2_{\perp}};b_t) $ is the number of gluons at
fixed impact parameter $b_t$.

\item\,\,\,It can be shown ( see Ref.\cite{TEOID} and references therein )
that we can write
\beq \label{BFACT}
 x_P G^{DGLAP}(x_P,\frac{4}{r^2_{\perp}};b_t )\,\,\,=\,\,\, x_P
G^{DGLAP}(x_P,\frac{4}{r_{\perp}})\,\,S(b_t)\,\,,
\eeq
if $x_P G^{DGLAP}(x_P,\frac{4}{r^2_{\perp}};b_t )$ satisfies the DGLAP
evolution equations\cite{DGLAP}.

In \eq{BFACT} $S(b_t)$ is the nucleon profile function which is a pure
nonperturbative ingredient in our calculations.  We assume a
Gaussian form for
\beq \label{PROFILEGA}
S(b_t)\,\,\,=\,\,\,\frac{1}{\pi\,R^2}\,\,e^{-\frac{b^2_t}{R^2}}\,\,,
\eeq
where $R$ has been discussed above;
\item\,\,\,
For the exchange of one ladder ( ``hard" Pomeron ) as shown in Fig. 3,
 $a^{el}_{dipole}$ can be written as
\beq \label{OMEGA}
\Omega^P_{dipole}\,\,=\,\, 
a^{el}_{dipole}(x,r_{\perp};b_t)\,\,\,=\,\,\,
\kappa^{DGLAP}_{dipole}(x,r_{\perp})\,\,
e^{-\frac{b^2_t}{R^2}}\,\,
\eeq
\item\,\,\,$\Psi^{\gamma^*} (z,r_{\perp};Q^2)$ is the wave function of the
quark - antiquark pair with the transverse distance $r_{\perp}$ between
a  quark and  an antiquark and  with a fraction of energy $z$ ( colour
dipole
of the size $r_{\perp}$ ). This wave function depends on the polarization
of the virtual photon and it has been calculated previously in
\cite{MU90} \cite{WAFU}.

\begin{eqnarray}
&
\Psi^{\gamma^*}_L (z,r_{\perp};Q^2 )\,\,\,=\,\,\,Q\,z\,( 1 - z
)\,K_0(a\,r_{\perp})\,\,;& \label{WF1}\\
&
\Psi^{\gamma^*}_T (z,r_{\perp};Q^2
)\,\,\,=\,\,\,i\,a\,K_1(a\,r_{\perp})\,
\frac{\mathbf{\vec{r}_{\perp}}}{r_{\perp}}\,;&
\label{WF2}
\end{eqnarray}
where $a^2 = z(1 - z) Q^2 + m^2_q $ and subscripts $T$ and $L$ denote the
transverse and longitudinal polarizations of the photon, respectively.
\item\,\,\,In  our calculations we only require  the probabilty to find
a quark-antiquark pair with the size $r_{\perp}$ inside a virtual photon,
namely
\begin{eqnarray} 
P^{\gamma^*}(z,r_{\perp};Q^2)&=&\frac{\alpha_{em} N_c}{2
\pi^2}
\,\sum_f \,Z^2_f \sum_{\lambda_1,\lambda_2}\,\{\, | \Psi_T |^2\,\,+\,\,|
\Psi_L|^2 \,\}\,\,\label{PROBPH}\\
&=&\frac{\alpha_{em} N_c}{2 \pi^2}
\sum_f Z^2_f \,\{\,( z^2 + ( 1 - z )^2 )a^2 K^2_1( a\,r_{\perp}
)\,+\,4\,Q^2\,z^2( 1 - z )^2 K^2_0(
a\,r_{\perp})\,\}.\nonumber
\end{eqnarray}
 
 \end{enumerate}
\subsection{Shadowing corrections for penetration of $\mathbf{q \bar q}$
pair through the target.}
\subsubsection{General approach}
The physics underlying our approach has been formulated and developed in
Refs.
\cite{LERY87} \cite{MU90}. During its  passage through the target,
the distance  $r_{\perp}$ between a quark and an antiquark can vary by an
amount $\Delta r_{\perp}\,\,\propto\,\, R  k_{\perp}/E$, where $E$ is the
pair energy and $R$ is the size of the target. Since the quark's
transverse momentum $k_{\perp} \,\,\propto\,\,1/r_{\perp}$, 
 the relation 
\beq \label{CONDI}
\Delta\,r_{\perp}\,\,\propto\,\,R\,\frac{k_{\perp}}{E}\,\,\approx
\,\,R\,\frac{1}{r_{\perp}\,\,E}\,\,
\ll\,\,r_{\perp}
 \eeq
holds if
\beq \label{CONDI1}
r^2\,s\,\,\gg\,\,2 m R\,\,,
\eeq
where $s =W^2= 2 m E$ with $m$ being the mass of the hadron.

\eq{CONDI1} can be rewritten in terms of $x_P$, namely,
\beq \label{CONDIX}
x_P\,\,\ll\,\,\frac{2}{ ( 1 - \beta ) m R}\,\,.
\eeq
From \eq{CONDIX} it follows  that $r_{\perp}$ is a good degree of freedom
\cite{MU90}  for high
energy scattering.

 We can therefore write the total cross section for the interaction of the
virtual photon with the target as follows:
\begin{eqnarray}
\si_{tot}(\gamma^* + p )\,\,&=& \,\,\int \,d z
\,\int \,d^2\,r_{\perp}\,
P^{\gamma^*}(z,r_{\perp};Q^2)\,\,
\si_{dipole}(x_B,r_{\perp})\,\,\label{TOTCR1}\\
  & = & 2\,  \int\,d^2 b_t \,\,\int
dz\,\int\,d^2\,r_{\perp}\,\,P^{\gamma^*}(z,r_{\perp};Q^2)\,\,
Im\,a^{el}_{dipole}(x_B,r_{\perp}; b_t)\,\,.\label{TOTCR2}
\end{eqnarray}
The amplitude for diffractive production of a $q \bar q $ - pair is equal
to
\beq \label{AMDD}
a^{DD}(\gamma^* + p \,\rightarrow\,q + \bar q )\,\,=\,\,\int d^2
\,r_{\perp}\,\,\Psi^{\gamma^*}(z,r_{\perp};Q^2)\,
a^{el}_{dipole}(x_B,r_{\perp};b_t)
\,\,\Psi^{q \bar q}(k_{\perp},z,r_{\perp} )\,\,,
\eeq
where $\Psi^{q \bar q}(k_{\perp},z,r_{\perp} )$ is the wave function of
the quark-antiquark pair with fixed momentum $k_{\perp}$ and fraction of
energy $z$.  To calculate  the total cross section of the diffractive
production we should integrate over all $k_{\perp}$ and $z$.  Using the
completeness of the $q \bar q $  wave function one  obtains
\beq \label{DDCR}
\si^{DD}(\gamma^* + p \,\rightarrow\,q + \bar q
)\,\,\,=\,\,\int\,d^2\,b_t\,\,\int
dz\,\int\,d^2\,r_{\perp}\,
P^{\gamma^*}(z,r_{\perp};Q^2)\,| a^{el}_{dipole}(x_B,r_{\perp}; b_t)
|^2\,\,.
\eeq
Utilizing the unitarity constraint we obtain a prediction for
the ratio 
\beq \label{RAPRED}
{\Re}\,\,=\,\,\frac{\si^{DD}}{\si_{tot}}\,\,=\,\,\frac{\int\,d^2\,b_t 
\,\int\,dz \,\int\,d^2\,r_{\perp}\,\,
P^{\gamma^*}(z,r_{\perp};Q^2)\,| a^{el}_{dipole}(x_B,r_{\perp}; b_t)|^2}
{ 2\,  \int\,d^2 b_t \,\int\,dz
\,\int\,d^2\,r_{\perp}\,\,P^{\gamma^*}((z,r_{\perp};Q^2)\,\,
Im\,a^{el}_{dipole}(x_B,r_{\perp}; b_t)}\,\,.
\eeq
\eq{RAPRED} is a general prediction for the ratio $\Re$
\cite{MU98}\cite{KOLE}. It shows that the diffractive dissociation and
total cross sections are related through unitarity. However, \eq{RAPRED}
is
too general to be used for pratical estimates.

 Assuming that the dipole - proton amplitude is mainly imaginary at
high energy the unitarity constraint of \eq{UNITARITY} has a general
solution
\begin{eqnarray}
& a^{el}_{dipole}( x, r_{\perp};b_t)\,\,=\,\,i\,\left(\,1\,\,-\,\,e^{-
\frac{\Omega( x, r_{\perp};b_t)}{2}}\,\right)\,\,;&\label{U1}\\
&
G^{in}_{dipole}( x, r_{\perp};b_t)\,\,=\,\,1\,\,-\,\,e^{- \Omega( x,
r_{\perp};b_t)}\,\,;&\label{U2}
\end{eqnarray}
where $\Omega$ is arbitrary real function.

Using \eq{U1} and \eq{U2}, \eq{RAPRED} reduces to
\beq \label{RGEN}
\Re\,\,=\,\,\frac{\si^{DD}}{\si_{tot}}\,\,=\,\,\frac{\int\,d^2\,b_t \int
\,dz \int \,d^2  r_{\perp}
P^{\gamma^*}(z,r_{\perp};Q^2)\, \left(\,1\,\,-\,\,e^{-  
\frac{\Omega( x, r_{\perp};b_t)}{2}}\,\right)^2}
{ 2\,  \int\,d^2 b_t \,\int\,dz
\,\int\,d^2\,r_{\perp}\,\,P^{\gamma^*}(z,r_{\perp};Q^2)\,\,
\left(\,1\,\,-\,\,e^{-\frac{\Omega( x, r_{\perp};b_t)}{2}}\,\right)
 }\,\,.
\eeq
 The main goal of  our microscopic approach based on
QCD is to find $\Omega$ .

\subsubsection{Mueller-Glauber approach}
One way to get a more detailed  picture of the interaction, is to consider
the
dipole - proton interaction in the Eikonal model, which is closely related
to 
Mueller - Glauber approach \cite{MU90}.

The main assumption of this model is to identify the function
$\Omega$ in 
\eq{RGEN} with the exchange of the ``hard" Pomeron ( gluon ladder ) 
given by \eq{OMEGA} ( see Fig. 3 ). Since the gluon distribution
given by the DGLAP evolution equations  originates from inelastic
processes of gluon emission, we assume an oversimplified structure of the
final states in the Eikonal model, namely,
it consists of only a proton and a
quark-antiquark pair ( ``elastic" scattering ), and an inelastic state
with a
large number of emitted gluons ($N_G\,\,\propto\,\,\ln(1/x )$ . In
particular, we neglect the rich structure of the diffraction
dissociation processes and simplify them to the final state of $p + q
\bar
q $. For example, we neglect the diffractive  production of an excited
nucleon in DIS. 

We will discuss the accuracy of our approach in the next
subsection, where we expand our model to include an 
excitation of the target.  Our accuracy is restricted by 
the assumption of only a quark - antiquark pair and a nucleon in the
final
state for
diffractive processes. The rough estimate for the contribution of all
excitations of the nucleon is 
\beq \label{EXIT}
\frac{\si^{DD}( \gamma^* + p \,\rightarrow\,\,q \bar q + N^*)}
{\si^{DD}( \gamma^* + p \,\rightarrow\,\,q \bar q + N)}\,\,\,=\,\,
\sqrt{\frac{\si^{DD}( p  + p \,\rightarrow\,\,p  + N^*)}{2\,\,\si_{el}( p  +
p
\,\rightarrow\,\,p  + p)}}\,\,\approx\,\,0.2 \div 0.3\,\,.
\eeq
consequently, we have to consider the nucleon excitations or, at least,
to
discuss them in the HERA kinematic region.

Substituting $\Omega = \Omega^P$  ( see \eq{OMEGA} )
in \eq{RGEN} we obtain the Kovchegov - McLerran formula
 \cite{MU90} \cite{AGLFR}, then for large $Q^2$ we have 
\beq \label{PGSIMPLE}
\int^1_0\,dz\,\,P_T^{\gamma^*}(z,r_{\perp};Q^2)\,\,=\,\,
\frac{4\,\alpha_{em}N_c}{3
\,\pi^2\,Q^2}\,\times\,\frac{1}{r^4_t}\,\,.
\eeq

We can see from \eq{RGEN} and \eq{PGSIMPLE} that
$$
\si^{DD}\,\propto\,\,R^2Q^2/<  r^2_{\perp}  >\,\,;
$$
$$
\si_{tot}\,\propto\,\,R^2Q^2/<  r^2_{\perp}  >\,\,;
$$
 we can evaluate $<  r^2_{\perp}  >$ from the condition
\beq \label{AVR}
\Omega^P_{dipole}(b_t = 0 )  \,\,\,=\,\,\,
\kappa^{DGLAP}_{dipole}(x,< r_{\perp}> )\,\, = \,\,1\,\,.
\eeq
If we assume that $x G(x, \frac{4}{r^2_{\perp} })$ depends rather smoothly
 on $r^2_{\perp}$, and substitute $\frac{4}{r^2_{\perp}} = Q^2 $ as
an estimate, we have
\beq \label{REST}
\frac{1}{< r_{\perp}>} \,\,\propto\,\,xG(x,Q^2)\,\,.
\eeq
\eq{REST} gives the ratio $\Re$ constant as well as
$\si^{DD}\,\propto\,\,R^2Q^2\,\times\,x_BG(x_B,Q^2)$. This simple
estimate,
given in Ref. \cite{KOLE}, illustrates that the SC lead to a constant
ratio $\Re$, or, vice versa, the constant ratio $\Re$ can be a strong
argument for substantial SC. To examine this point we calculate the ratio
$\Re$ using the same assumption that $\frac{4}{r^2_{\perp}} = Q^2 $ in the
gluon distribution. Using the result of the explicit calculation in
Ref.\cite{AGL} we obtain:
\beq \label{AGLEST}
\Re\,\,=\,\,\frac{\si^{DD}}{\si_{tot}}\,\,=\,\,1\,\,-\,\,\frac{\ln
(\kappa_{dipole} )
\,\,+\,\,C\,\,+\,\,(\,1 \,+\,\kappa_{dipole}\,)\,E_1( \kappa_{dipole})
\,+\,1\,-\,e^{-
\kappa_{dipole}} }{2\,[\ln(
\frac{\kappa_{dipole}}{2} )
\,\,+\,\,C\,\,+\,\,(\,1 \,+\,\frac{\kappa_{dipole}}{2}\,)\,E_1(
\,\frac{\kappa_{dipole}}{2})
\,+\,1\,-\,e^{-
\frac{\kappa_{dipole}}{2}} \,]}\,\,.
\eeq

\begin{figure}[htbp]
\begin{tabular}{c c }  
  \epsfig{file=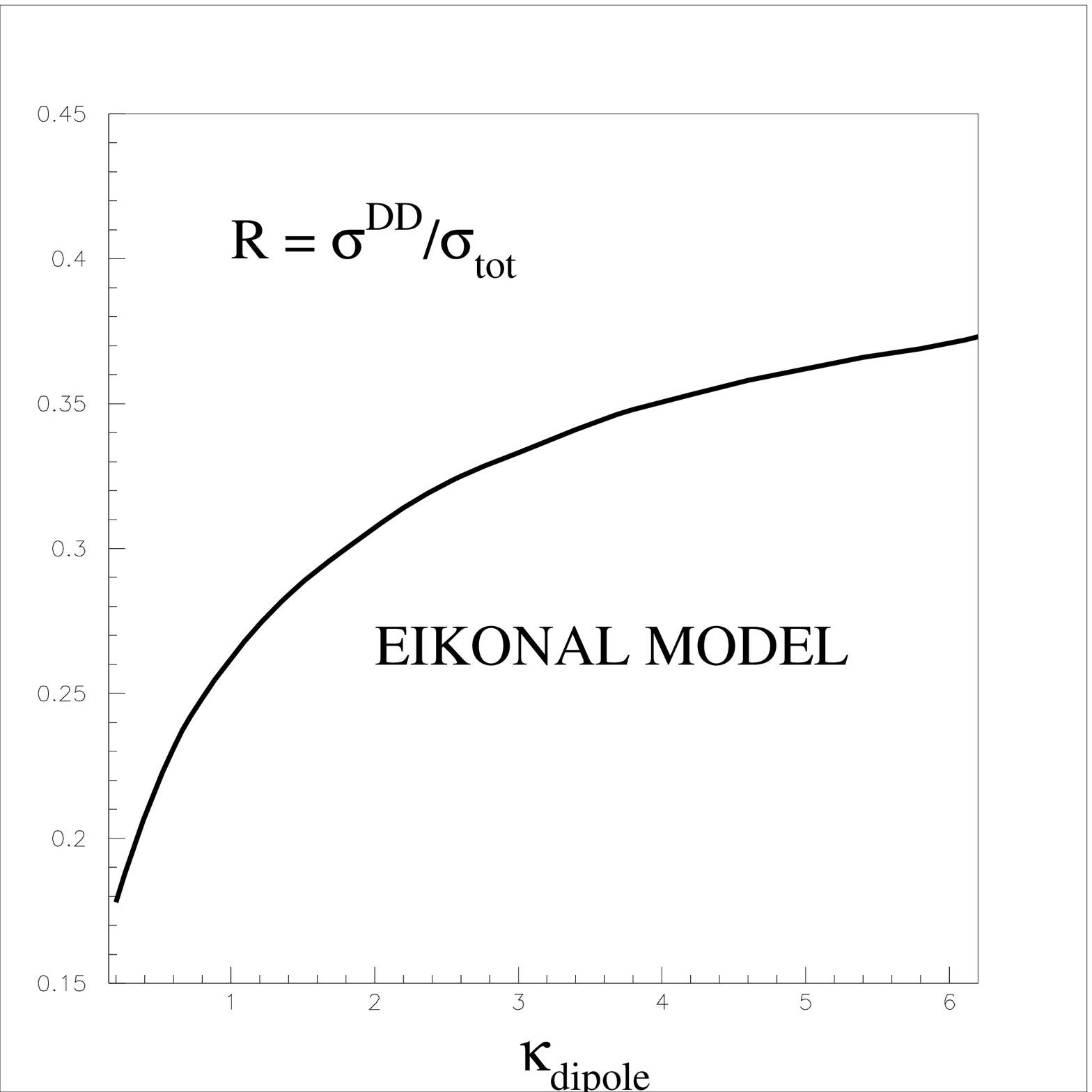,width=80mm} &
\epsfig{file=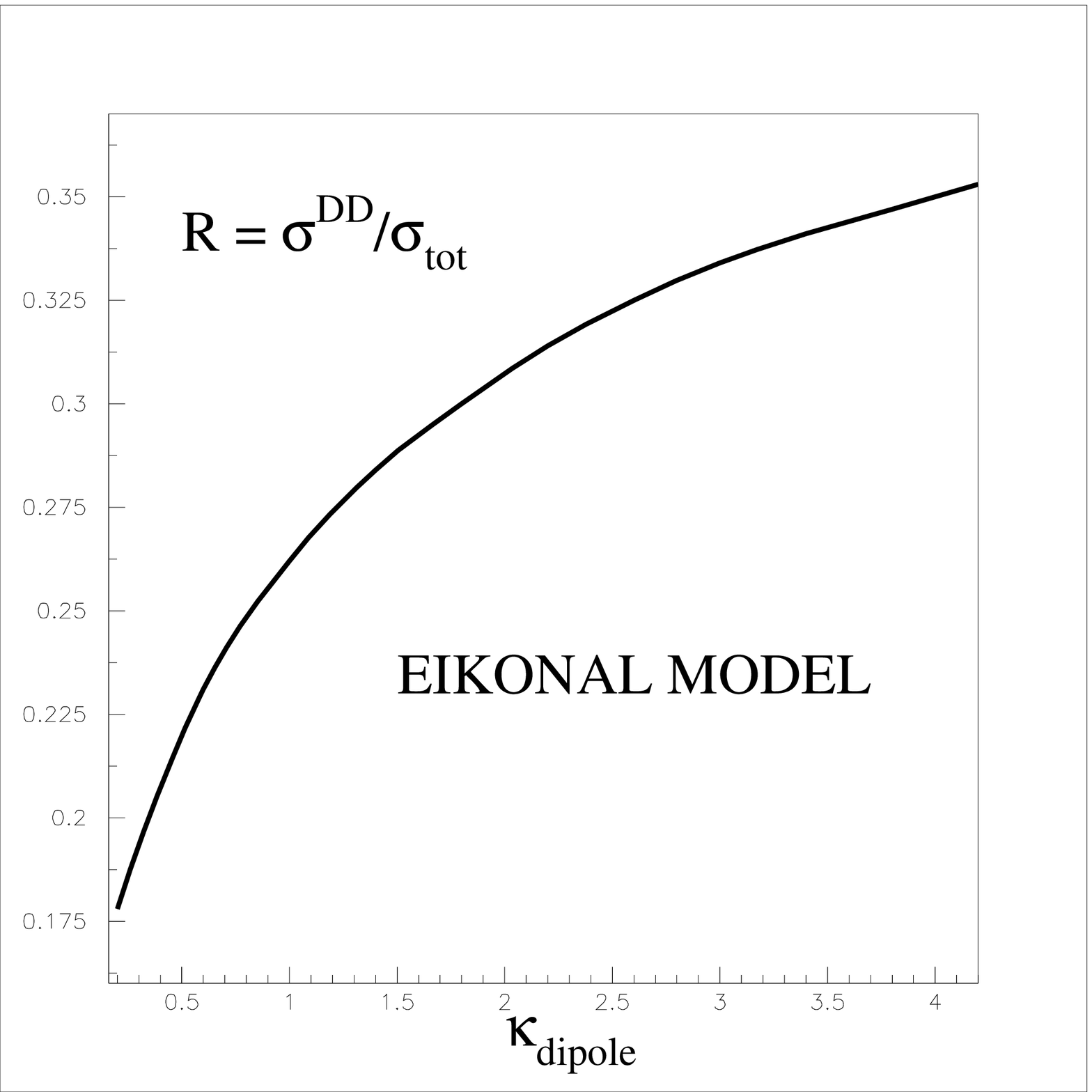,width=80mm}\\
 Fig. 4-a &
Fig. 4-b\\
\end{tabular}
  \caption[]{\it Ratio $\sigma^{DD}/\sigma_{tot}$ calculated using
\protect\eq{AGLEST} in the Eikonal model for $q \bar q$ diffractive
production. }
\label{fig4}   
\end{figure}

In Fig.4 the ratio $\Re$, given by \eq{AGLEST}, is plotted as a function of 
$\kappa_{dipole}$.  One can see that at large $\kappa_{dipole}$
this ratio has a smooth dependence on $\kappa_{dipole}$. However, the
values
of $\kappa_{dipole}$ that we are dealing with are not very large (
$ \kappa_{dipole}\,\,\approx\,\,1$,   see
Fig.5, where $\kappa_{dipole}$ is calculated using the GRV
parameterization for the gluon distribution \cite{GRV} ).   
Fig.4-a shows that  for $\kappa_{dipole} = 0.2 - 2$  we cannot expect that
\eq{AGLEST} to yield a more or less constant ratio $\Re$.

\begin{figure}[htbp]
\begin{center}
  \epsfig{file=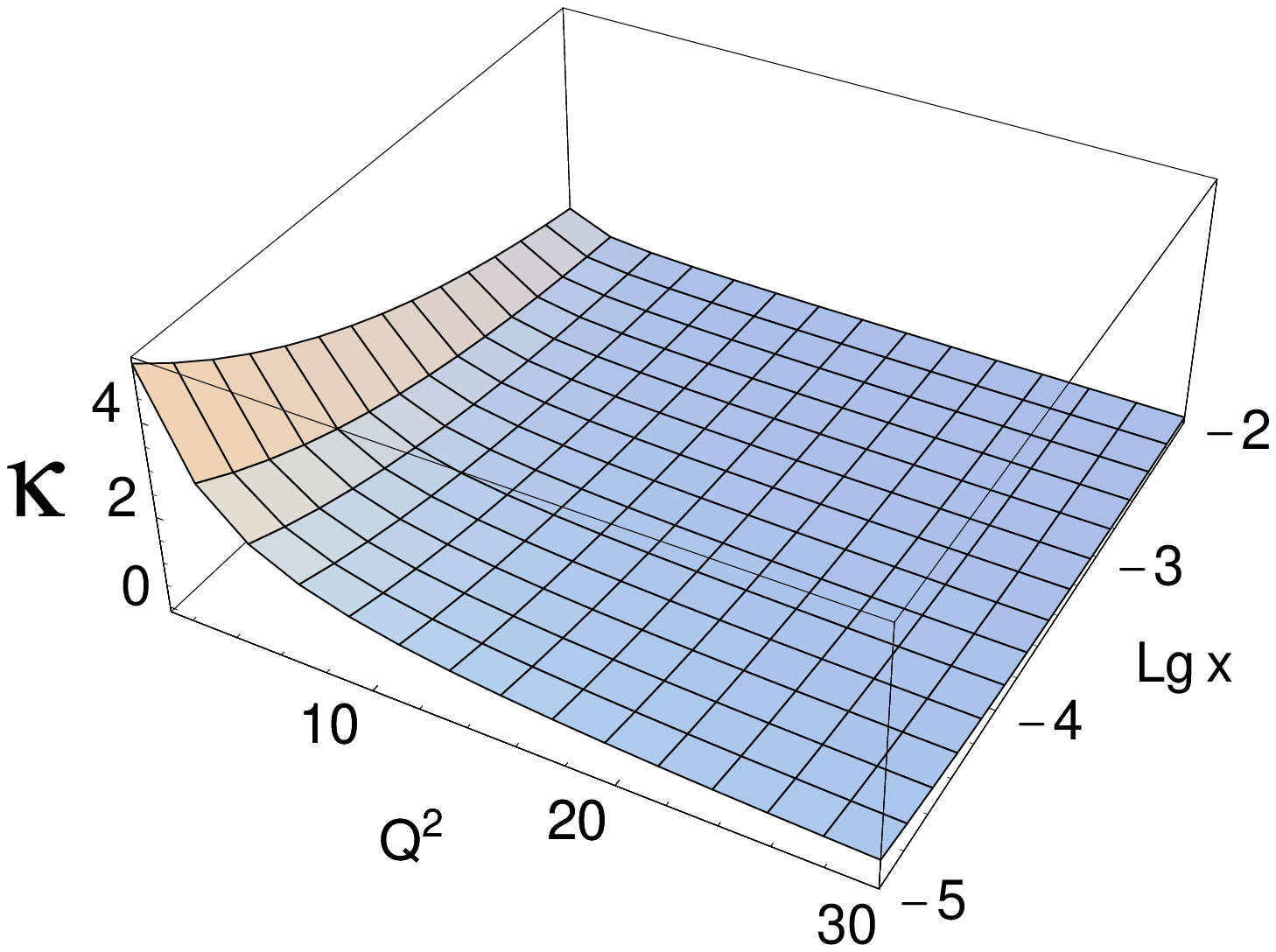,width=130mm} 
  \caption[]{\it $\kappa_{dipole}$ calculated in HERA kinematic region,
using  the GRV-94 parameterization \protect\cite{GRV} for the gluon
distrubution. $lg x  = log_{10}(x)$.}
 \end{center}
\label{fig5}
\end{figure}

However, we would like to draw  the reader's attention to the fact that
the ratio
differs from the small $\kappa_{dipole} \,\,\ll\,\,1$ limit
where $\Re\,\,\propto\,\,\kappa_{dipole}$. 

 These simple estimates indicate that the SC are essential, but
they are still not sufficiently strong to use the asymptotic formulae. We
are in
the transition region from low density QCD, described by the DGLAP
evolution, to high density QCD where we can use the quasi-classical gluon
field approximation \cite{LEKOVN}.   In the transition region the
Mueller-Glauber approach is a natural way to obtain reliable estimates of
the SC. However, we need to study   corrections to
\eq{RGEN} more carefully.  The most important among them, are
the diffractive production of
nucleon
 excitation ( see \eq{EXIT}) and  of the $ q \bar g G $ system, which
gives
the dominant contribution to the diffractive cross section \cite{GLMSM}.

\subsubsection{Diffractive production of nucleon excitations}

 \centerline{ \em A. General approach}
We start with a trivial remark, that the nucleon excitations even in
DIS are closely related to long distance processes and, therefore, to
the ``soft" interaction which cannot be determined in QCD.
An alternate way of saying this is, to attribute these diffractive
processes to
nonperturbative QCD, for which at present we only have a
phenomenological approach. 
  Theoretically  ``soft"
diffraction  can be viewed \cite{FGW} as a typical quantum
mechanical process which occurs, since the  hadron states are not diagonal
with respect to the strong interaction scattering matrix.  In other words
 diffractive dissociation occurs,  because even at high energy hadrons are
not the correct
degrees of freedom  for the strong nonperturbative
interaction. Unfortunately, we do not know the correct degrees of freedom
and
below we will discuss  some models for them.  We
denote by $n$ the correct degree of freedom or the set of quantum numbers
which characterizes the wave function $\Psi_n$. These function $\Psi_n$
are
diagonal with respect to the strong interaction
\beq \label{GD1}
A_{n,n'}\,\,=\,\,\big< \Psi_n \big| {\mathbf{T}}\big|\Psi_{n'} \big>
\,\,=\,\,A_n\,\,\delta_{n,n'}\,\,,
\eeq
where parentheses denote all needed integrations and $\mathbf{T}$ is the
scattering matrix.

Note, that only for amplitudes $A_n$ do we have the
unitarity constraints in the form of \eq{UNITARITY}, namely,
\beq \label{UNITARITYG}
Im\,A^{el}_n(s, b_t)\,\,\,=\,\,| A^{el}_n(s, b_t)
|^2\,\,\,+\,\,\,G^{in}_n(s,b_t)
\eeq
which has solutions of \eq{U1} and \eq{U2} for mainly imaginary $A_n$ at
high energies:

\begin{eqnarray}
& 
A^{el}_n\,\,\,=\,\,i\,\{\,\,1\,\,\,-\,\,\,e^{-
\frac{\Omega_n(s,b_t)}{2}}\,\,\}\,\,;& \label{UG1}\\
&
G^{in}_n\,\,\,=\,\,\,1\,\,\,-\,\,\,e^{-\Omega_n(s,b_t)}\,\,.& \label{UG2}
\end{eqnarray}

The wave function of a hadron is
\beq \label{GD2}
\Psi_{hadron}\,\,=\,\,\sum^{\infty}_{n =1} \,\,\alpha_n \,\,\Psi_n\,\,.
\eeq
For a dipole - hadron interaction the wave function is equal to
$\Psi_{dipole}(r_{\perp})\,\times\,\Psi_{hadron}$ before collision. After
collision the scattering  matrix $\mathbf{T}$ leads to a new wave
function,namely
\beq \label{FIWF}
\Psi_{final}\,\,=\,\,\Psi_{dipole}(r_{\perp})\,\times\,\sum^{\infty}_{n=1}\,\,
\alpha_n \,\,A_n\,\Psi_n\,\,.
\eeq
 
From \eq{FIWF} we obtain the elastic amplitude
\beq \label{ELA}
a^{el}_{dipole}\,\,=\,\,\big< \Psi_{final} \big|
\Psi_{dipole}(r_{\perp})\,\times\,\Psi_{hadron} \big>\,\,
=\,\,\sum^{\infty}_{n=1}
\,\,\alpha^2_n \,A_n (s,b_t)\,\,\,,
\eeq
while for the total cross section of the diffractive nucleon excitations
we have
\begin{eqnarray}   
\si^{DD}_{N^*}\,\,&=&\,\,\big< \Psi_{final} \big| \Psi_{final} \big>^2
\,\,\,-\,\,
\big< \Psi_{final} \big|
\Psi_{dipole}(r_{\perp})\,\times\,\Psi_{hadron}\big>^2\,\,; 
\label{DDGEN1}\\
 &=&\,\,\sum^{\infty}_{n=1}
\,\,\alpha^2_n \,|A_n (s,b_t)|^2 \,\,\,-\,\,\{\,\sum^{\infty}_{n=1}
\,\,\alpha^2_n \,A_n (s,b_t)\,\}^2\,\,. \label{DDGEN2}
\end{eqnarray}

Therefore, using \eq{PGSIMPLE} 
instead of \eq{RGEN},   we obtain a generalized
formula
\beq \label{KOLEGEN}
\Re\,\,=\,\,\frac{\si^{DD}}{\si_{tot}}\,\,\,=
\,\,\,\frac{\sum^{\infty}_{n = 1}\,\alpha^2_n\,\,\int\,d^2 \,b_t
\int\,\frac{dr^2_{\perp}}{r^4_{\perp}}\,\{\,1\,\,-\,\,e^{-
\frac{\Omega^P_n(r^2_{\perp},y;b_t)}{2}}\,\}^2}{2\,\,\sum^{\infty}_{n =
1}\,\alpha^2_n\,\,\int\,d^2 \,b_t\int\,\frac{dr^2_{\perp}}{r^4_{\perp}}
\,\{\,1\,\,-\,\,e^{-
\frac{\Omega^P_n(r^2_{\perp},y;b_t)}{2}}\,\}}\,\,,
\eeq
 One can see  that this generalized formula
has all the attractive features of \eq{RGEN}, and at small $x$ the ratio
tends
to $1/2$, since the normalization constraint
\beq \label{NORM}
\sum^{\infty}_{n=1} \,\,\alpha^2_n \,\,=\,\,1
\eeq
We can estimate $\Omega^P_n$ using the same \eq{OMEGA} here
\beq \label{OMEGAN}
\Omega_n\,\,=\,\,\kappa^n_{dipole}\,\,\,e^{ -
\frac{b^2_t}{R^2_n}}\,\,,
\eeq
where
\beq \label{KAPPAN}
\kappa^n_{dipole}\,\,\,=\,\,\,\frac{\pi^2
\as}{3\,\pi\,R^2_n }r^2_{\perp}\,\, x_PG_n(x_P,\frac{4}{r^2_{\perp}})\,\,.
\eeq
Therefore, the main difficulty with \eq{OMEGAN} and \eq{KAPPAN} is to 
determine $R_n$ and $xG_n(x,\frac{4}{r^2_{\perp}})$. We only have direct
experimental information  for the proton radius and the  gluon
distribution
in the proton. Unfortunately, we cannot evaluate \eq{KOLEGEN} without 
developing some model for the diffractive excitations.

~

\centerline{\em B. Two channel model for diffractive nucleon
excitations.}

~
The main idea \cite{GLMCOM} of this model is to replace the many
final states of diffractively produced hadrons by one state ( effective
hadron ). In this case  the general \eq{GD2} reduces to the simple form
\beq \label{TWO1}
\Psi_{hadron}
\,\,\,=\,\,\alpha_1\,\,\Psi_1\,\,\,+\,\,\,\alpha_2\,\,\Psi_2\,\,,
\eeq
with the condition $ \alpha^2_1\,\,+\,\,\alpha^2_2\,\,=\,\,1 $  from
\eq{NORM}.
 
The wave function of the produced effective hadron is equal to
\beq \label{TWO2}
\Psi_D\,\,\,=\,\,- \,\alpha_2\,\,\Psi_1\,\,\,+\,\,\,\alpha_1\,\,\Psi_2\,,
\eeq
which is orthogonal to $\Psi_{hadron}$.

   \eq{KOLEGEN} can be rewritten in the form
\begin{eqnarray}
\Re\,\,&=&\,\,\frac{\si^{DD}}{\si_{tot}}\,\,\,=\,\,\  \label{KOLETWO}\\
&= &\,\,\,\frac{
\,\,\,\int\,d^2 \,b_t
\int\,\frac{dr^2_{\perp}}{r^4_{\perp}}\,\left(\,
\alpha^2_1\{\,1\,\,-\,\,e^{-\frac{\Omega^P_1(r^2_{\perp},y;b_t)}{2}}\,\}^2\,\,
+\,\,\alpha^2_2\,
\,\{\,1\,\,-\,\,e^{-\frac{
\Omega^P_2(r^2_{\perp},y;b_t)}{2}}
\,\}^2\,\right)
 }{2\,\,\int\,d^2 \,b_t\int\,\frac{dr^2_{\perp}}{r^4_{\perp}} \,\left(  
\,\{\,1\,\,-\,\,e^{-
\frac{\Omega^P_1(r^2_{\perp},y;b_t)}{2}}\,\}\,\,+\,\,\alpha^2_2\,\,
e^{- \frac{\Omega^P_1(r^2_{\perp},y;b_t)}{2}}
\,\{\,1\,\,-\,\,e^{-\frac{\Delta  
\Omega^P(r^2_{\perp},y;b_t)}{2}}
\,\}\,\right)
}\,\,;\nonumber
\end{eqnarray}
with $\Delta \Omega \,\,=\,\, \Omega_2 \,-\,\Omega_1 $.

\begin{figure}[htbp]
\begin{tabular}{c c }
  \epsfig{file=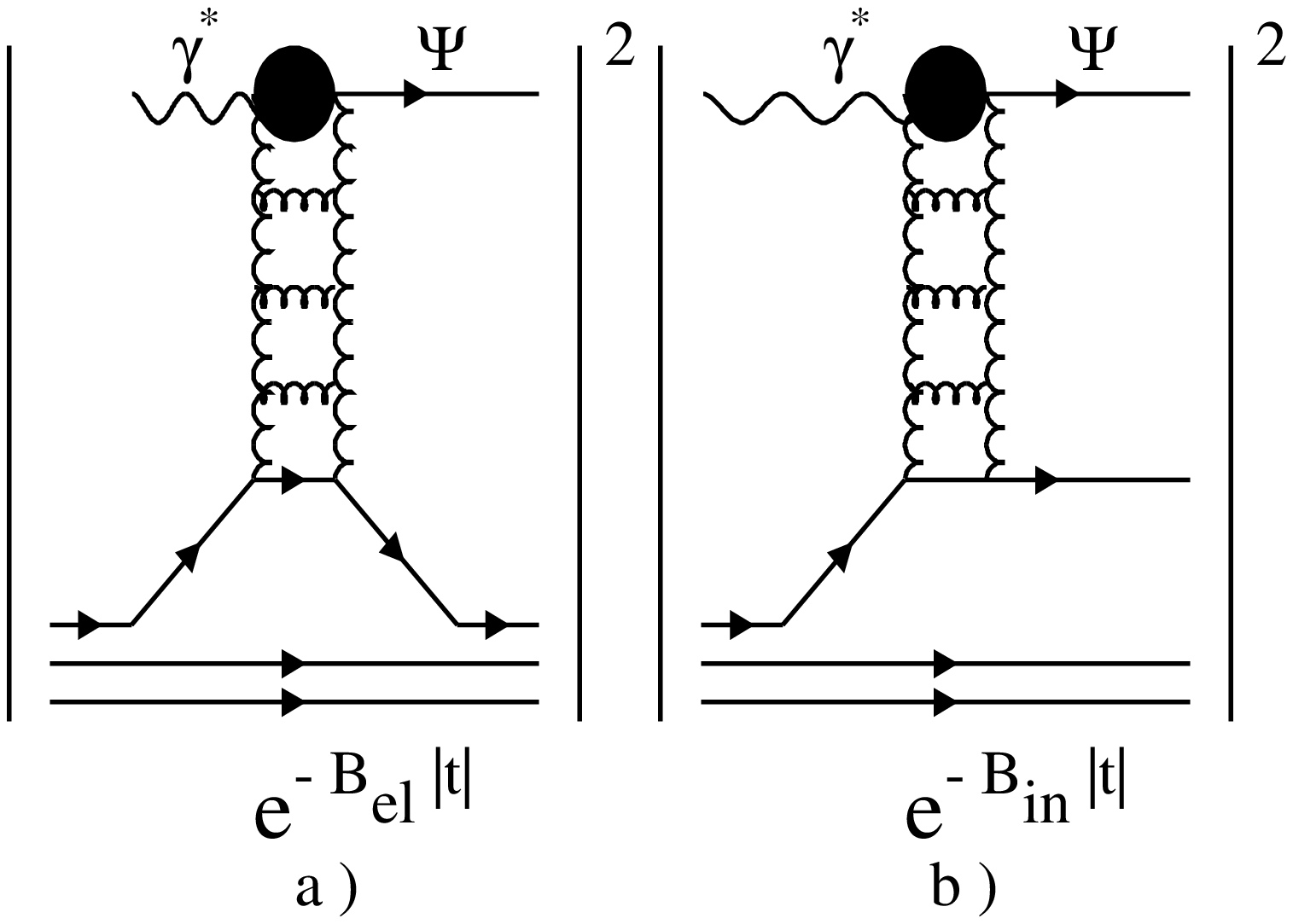,width=80mm,height=60mm} &
\epsfig{file=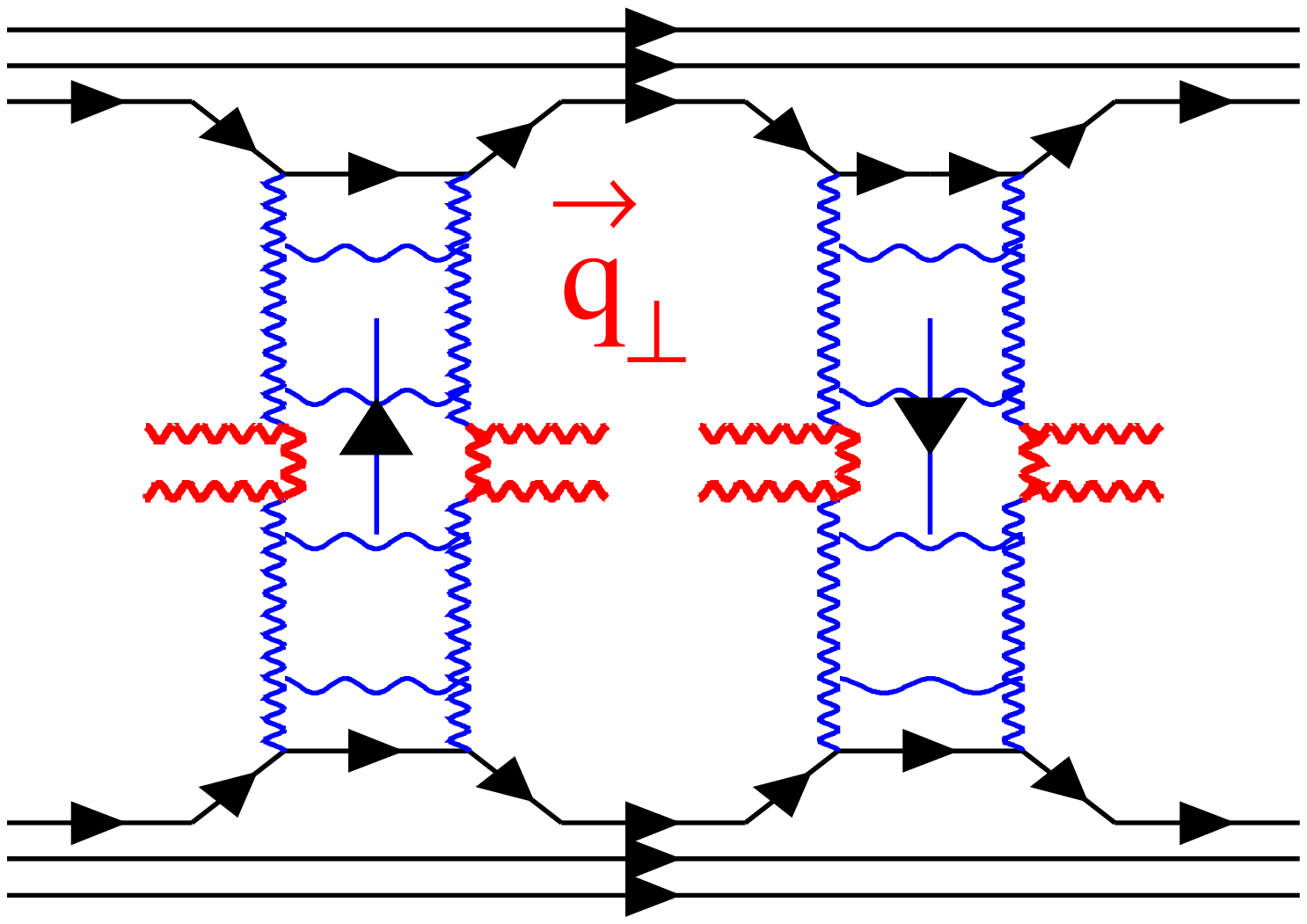,width=80mm,height=60mm}\\
 Fig. 6-a &
Fig. 6-b\\
\end{tabular}
  \caption[]{\it Mueller diagrams \protect\cite{MUDIA} for J/$\Psi$
photoproduction ( Fig.6-a )  
and for double parton interaction ( Fig.6-b ). }
\label{fig6}
\end{figure}
For $\Omega_1$ and $\Delta \Omega $ we use the  parameterization
given by \eq{OMEGAN} and \eq{KAPPAN},   and  the following experimental
data 
and phenomenological observations:
\begin{enumerate}
\item\,\,\, Data for single diffraction in proton - proton collisions
lead to $\alpha^2_2\,\,\approx\,\,0.2 $  Ref.
\cite{GLMCOM}\,\,;
\item\,\,\,Data for J/$\Psi$ photoproduction at HERA \cite{HERAPSI} 
( see Fig.6a ) show that $t$-dependance is quite different for elastic
and inelastic photoproduction. The measured values are $B_{el}
\,\,=\,\,4\,\,GeV^{-2}$ while $B_{in}\,\,=\,\,1.66\,\,GeV^{-2}$.
Therefore, we take $R^2_1 = 2\,B_{el} = 8 \,\,GeV^{-2}$ in $\Omega_1$ 
and $R^2_D  = 2 B_{in} = 3.32 \,\,GeV^{-2}$, where $R^2_D$ is the radius
in the exponential parameterization for $\Delta \Omega$\,;
\item\,\,\,The energy behaviour of  diffractive J/$\Psi$ photoproduction
shows that we can consider this process as a typical ``hard" process
which occurs at short distances \cite{AC}\,;
\item\,\,\,We use  experimental evidence \cite{HERAPSI} that 
the cross section for elastic and inelastic diffractive J/$\Psi$
photoproduction are equal. From this fact we can conclude that 
\beq \label{DELTAOM}
\alpha_2 \,( 1 - \alpha_2) (\Delta \Omega) (b_t = 0)^2\,\, R^2_D\,\,=\,\,
( 1 - \alpha_2 )^2 \, \Omega^2_1(b_t = 0)\,\, R^2_1\,\,,
\eeq
which gives 
\beq \label{DELTAOMEST}
\Delta \Omega(b_t = 0 ) \,\,=\,\,\sqrt{\frac{ ( 1 -
\alpha^2_2 )}{\alpha^2_2}}\,\frac{R_1}{R_D}\,\,\Omega_1(b_t = 0)\,\,.
 \eeq
Substituting in \eq{DELTAOMEST}  we find  $\Delta
\kappa_{dipole} \,\,\approx\,\,3.4\,\,\kappa^1_{dipole}$.
\end{enumerate}

\begin{figure}[htbp]
\begin{tabular}{c c }
  \epsfig{file=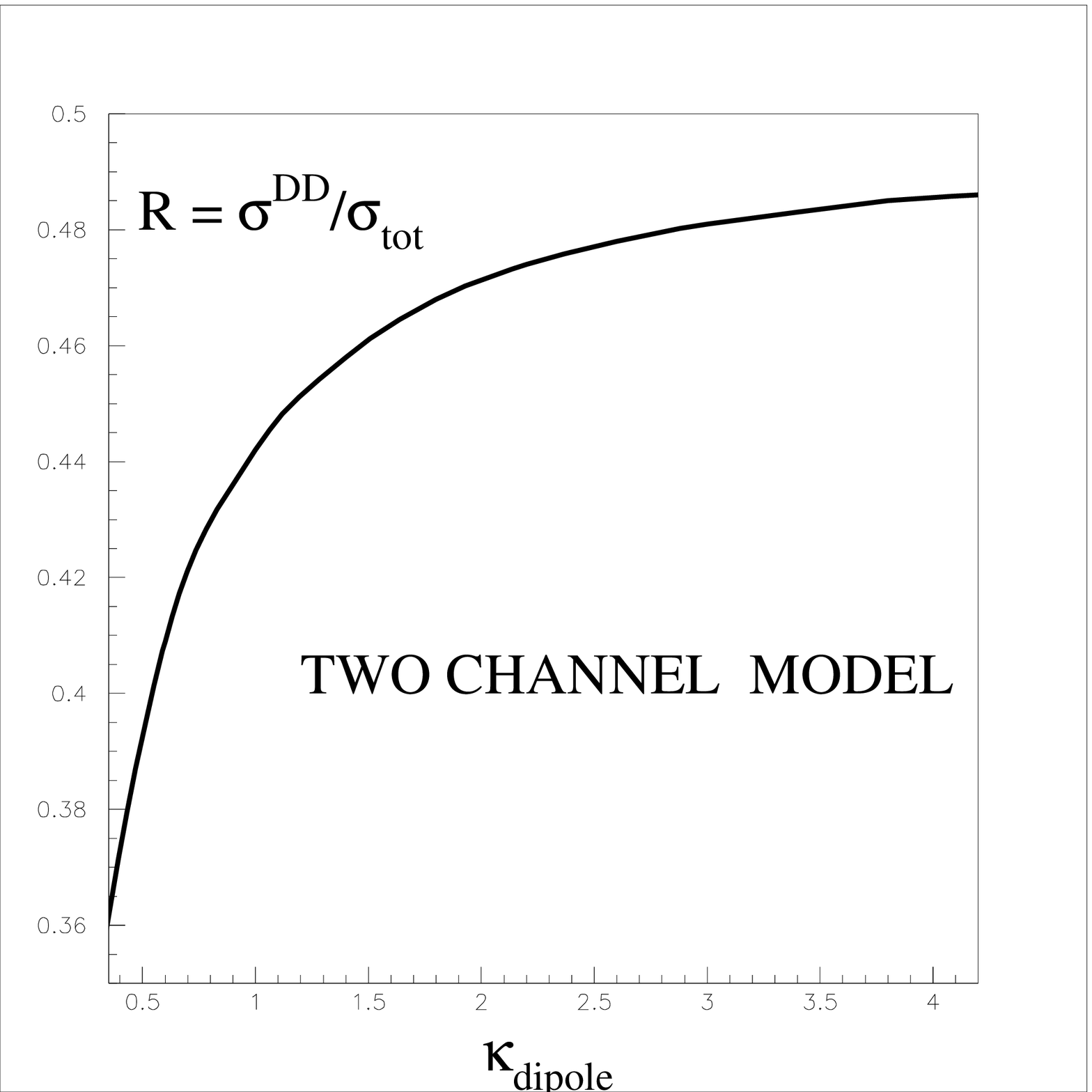,width=80mm} &
\epsfig{file=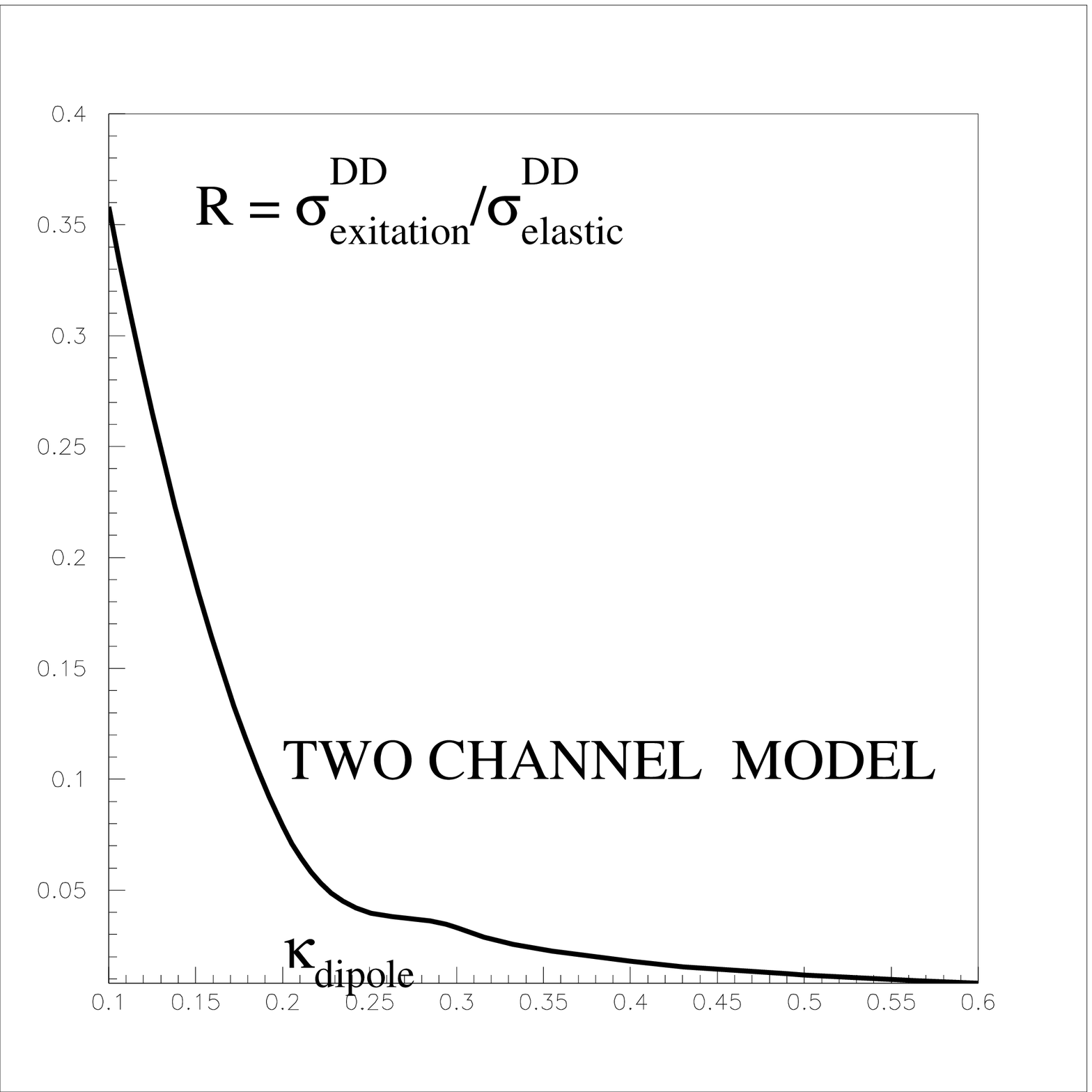,width=80mm}\\
 Fig. 7-a &
Fig. 7-b\\
\end{tabular}
  \caption{\it Ratio $\sigma^{DD}/ \sigma_{tot}$  (Fig.7a) and
ratio $ \sigma^{DD}_{excitation}/\sigma^{DD}_{elastic}$ (Fig. 7b)
versus $\kappa_{dipole} $
calculated in two channel model for nucleon excitations for $q \bar q $
diffractive production. }
\label{fig7}
\end{figure}

In Fig. 7 we display the ratio $R$ as a function of $\kappa_{dipole}$.
Comparing this figure with Fig. 4a we can conclude that in the two channel
model,  R increases more or less at the same rate as in the Eikonal
model, but the value of the ratio depends crucially on the model for the
diffractive excitation of the nucleon. Fig. 7b shows the contamination of
the total diffractive cross section by the nucleon excitations. One can
conclude that the two channel model gives a small fraction of the
excitation cross section at sufficiently large $\kappa_{dipole}$.
Experimentally\footnote{We thank Henry Kowalski for discussing 
this data and many  problems stimulated by these data with us},$
\sigma^{DD}_{excitation}/\sigma^{DD}_{elastic}$ = $ 35\, \pm\, 15 \%$.
Fig.7b supports a low value for this ratio, but this question should be
reconsidered after taking into account the diffractive production of 
 the $q\bar q G $ system.

~
\newpage
\centerline{\em C. Diffractive production in the Additive Quark Model}

~

As we have mentioned, the main problem of dealing with the ``soft" high
energy interaction,
is to find the correct degrees of freedom, and to incorporate them in the
general formalism for high energy scattering.  The two channel model gives
an estimate of the importance 
   of proton excitation processes, but it is too
phenomenological to be instructive and not totally reliable.  Here, we
consider
the nucleon excitation in the Additive Quark Model \cite{AQM}. In this
model the correct degrees of freedom at high energies are the constituent
quarks. In spite of a certain  naivity this model has not been abandoned 
and it is
included in
the standard Donnachie - Landshoff Pomeron approach\cite{DL}  for ``soft"
processes at high energies.     

 Inherent in this model is the assumption, that  the
 $\gamma^* $ - constituent quark interactions dominate,
 while other interactions e.g. the interaction of $\gamma^*$
with two constituent quarks, are suppressed by factor $r^2_Q/R^2_N$ ,
here
$r_Q$ is the size of the constituent quark and $R_N$ is the radius of the
proton. It is obvious that in the  AQM we have the same \eq{RGEN} ( or
\eq{RAPRED} ) where $\Omega = \Omega^P_Q$ describes the interaction of a
colour dipole
with the constituent quark.  In the AQM  the gluon distribution
 of the constituent quark is equal to $xG_Q(x,Q^2)
\,=\,\frac{1}{3}\,xG_N(x,Q^2)$ and, therefore
\beq \label{OMEGAQ}
\Omega^P_Q(x,r_{\perp};b_t)\,\,\,=\,\,\kappa^{Q}_{dipole}\,\,
e^{-\frac{b^2_t}{r^2_Q}}\,\,=\,\,\frac{\pi^2
\as}{9\,\pi\,r^2_Q } r^2_{\perp}\,\,
x G_N(x ,\frac{4}{r^2_{\perp}})\,\,e^{-\frac{b^2_t}{r^2_Q}}\,\,.
\eeq

Consequently, we find  $r_Q$  using the same AQM
to describe
the double parton cross section measured by the CDF \cite{CDFDP} (see Fig.
7b ). The CDF collaboration has measured the inclusive cross section for
the production of two ``hard"  pairs of jets,  with large and almost
compensating transverse momenta in each pair, and  with similar values of
rapidity. Such processes cannot occur in a
one parton shower, and  only originate from two parton shower
interactions as shown in Fig.7b.

The double parton cross section can be written in the form \cite{CDFDP}
\beq \label{DP}
\si_{DP}\,\,=\,\,m\,\frac{\si_{inel}(2 jets) \,\,\si_{inel}(2 jets)}{2
\si_{eff}}\,\,
\eeq
where factor $m$ is equal to 2 for different pairs of jets, and to 1 for
identical pairs. The experimental value\cite{CDFDP}  of
$\si_{eff}\,\,=\,\,14.5\,\,\pm\,\,1.7 \,\,\pm\,\,2.3\,\,mb$.

In the AQM (see Fig.7b ) $\si_{eff} $ can be easily calculated and it is
equal to
\beq \label{SIEFFQ}
\si_{eff}\,\,=\,\,9\,\times\,2 \pi \,r^2_Q \,\,,
\eeq
where factor 9 reflects the quark counting and $2 \pi r^2_Q$  comes from
the integration over $b_t$. Comparing \eq{SIEFFQ} with the experimental
value of $\si_{eff}$ we obtain $r^2_Q\,\,=\,\,0.66 \,\pm\,0.16\,GeV^{-2}$.

Substituting this result in \eq{OMEGAQ} we find  $\kappa^Q_{dipole}
\,\,\approx\,\,5\, \kappa^N_{dipole}$ with the same $x$ and $r_{\perp} $
dependance. In Fig. 8 one can see the prediction for the ratio
$\si^{DD}/\si_{tot}$.  This approach gives the ratio which depends 
smoothly
 on $\kappa_{dipole}$ for $\kappa_{dipole} \,>\, 0.75$. 

Comparing Fig.8 and Fig.4a we can conclude that the diffractive
production of the nucleon excitations gives about 30\% of the total
diffractive cross section at $\kappa_{dipole} \,\approx \,0.2 - 0.6$ and
  becomes very small at $\kappa_{dipole} \,\approx\,1.5 - 2 $.
\begin{figure}
\begin{tabular}{c c }
  \epsfig{file=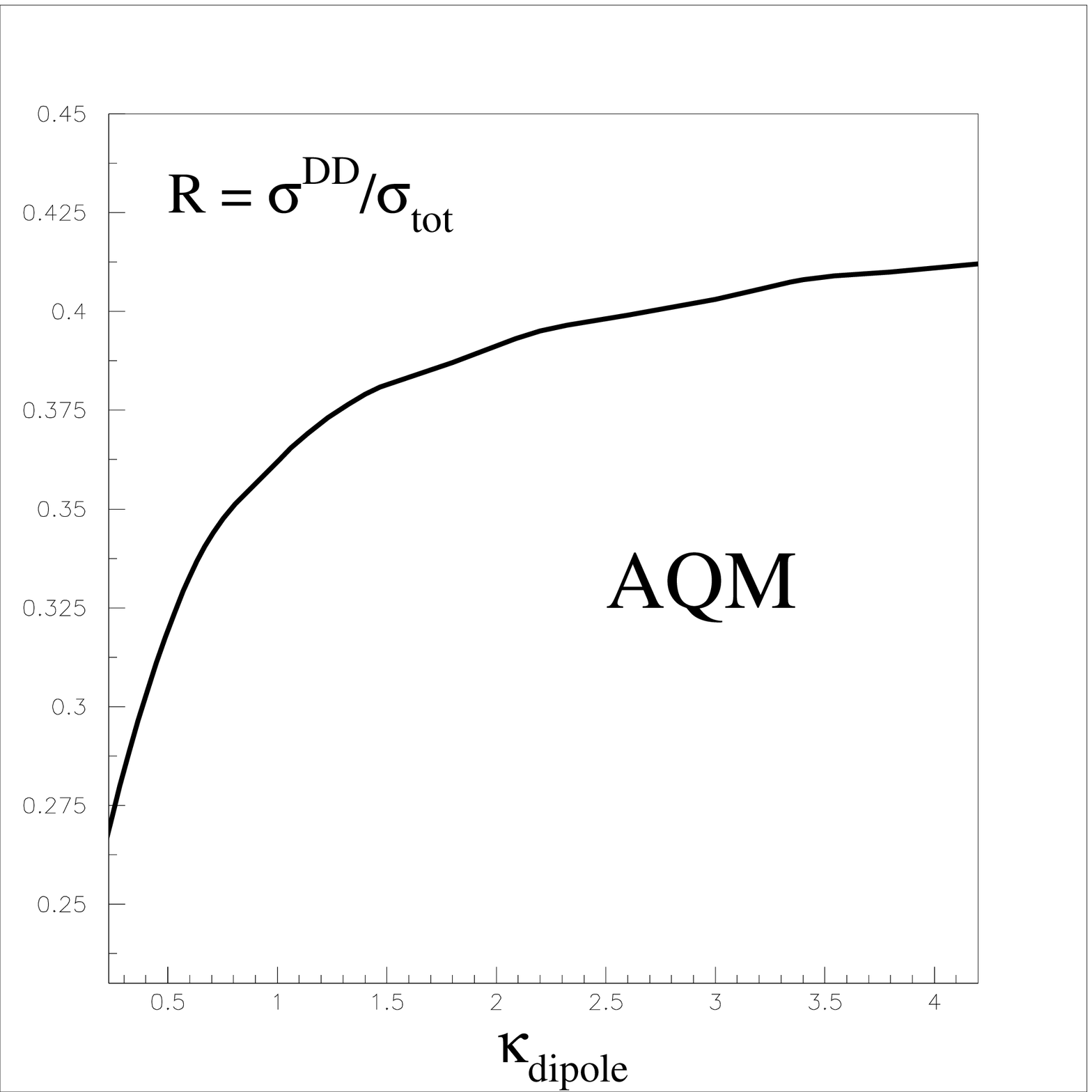,width=80mm} &
\epsfig{file=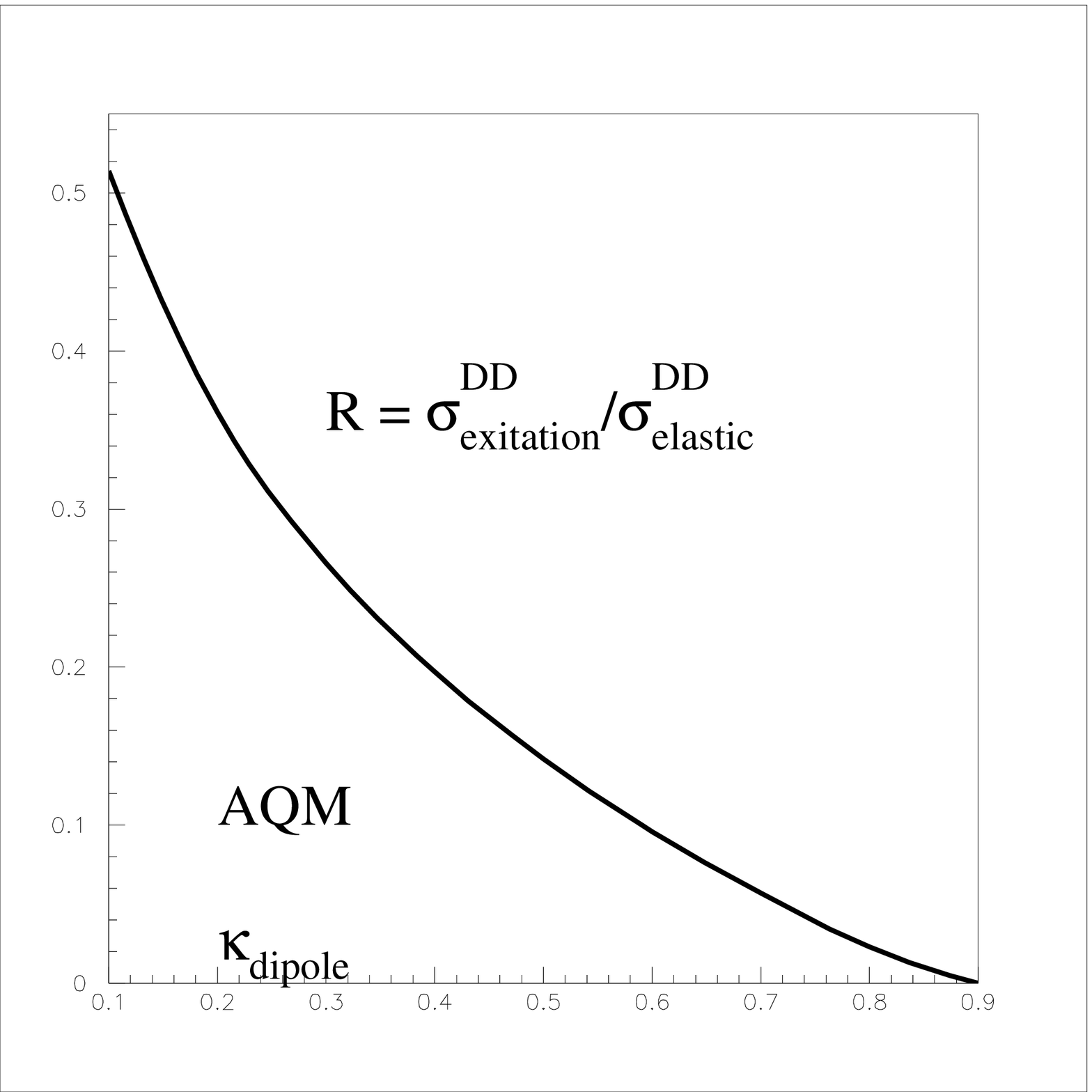,width=80mm}\\
 Fig. 8-a &
Fig. 8-b\\
\end{tabular}
  \caption{\it Ratio $\sigma^{DD}/\sigma_{tot}$  (Fig.8a) and 
ratio $ \sigma^{DD}_{excitation}/\sigma^{DD}_{elastic}$ (Fig. 8b)
versus $\kappa_{dipole} $
calculated in the Additive Quark Model for nucleon excitations for $q \bar
q $   
diffractive production. }
\label{fig8}
\end{figure}

\subsubsection{AGK cutting rules and diffractive production}

In this section we derive \eq{RGEN} in the Mueller-Glauber approach
exploiting
the AGK cutting rules \cite{AGK}. This derivation is  more
complicated than the previous derivation which was  based directly on
the $s$-channel unitarity constraints.  As  we intend using the
AGK cutting rules for calculating the 
diffraction production of the $q \bar q G $ system,
 we think it  instructive to start with a simple
example.

 The AGK cutting rules provide a prescription of how to
calculate the cross sections for the processes with different
multiplicities
of the produced particles, if one knows the structure of the Pomeron exchange,
and the expression for the  total cross section in terms of  multi
Pomeron exchanges \cite{AGK} \cite{BARY}. In the Mueller-Glauber approach
we have such an expression for the total colour dipole - proton cross
section, namely, (see Fig.9)
\beq \label{CRAGK}
\si_{dipole}\,\,\,=\,\,2\,\int\,d^2b_t\,\,\{\,1\,\,-\,\,e^{-
\frac{\Omega^P(x,r_{\perp};b_t)}{2}}\,\}\,\,,
\eeq
where $\Omega^P$ is given by \eq{OMEGA}.   We first need to 
 define the Pomeron structure i.e.  to specify the  kind
of inelastic processes that are described by $\Omega^P$. 
From \eq{OMEGA} we see that $\Omega^P$ describes the
inelastic processes with large average multiplicity ( $< n_P > $ )  of
produced partons, since it is   related to the DGLAP parton cascade.
For example, at large $Q^2$ and low $x$,  $< n_P >\,\,=\,\, 2\sqrt{ \bas
\ln
Q^2 \ln(1/x)}\,\,\gg\,\,1$.  The AGK cutting rules allow us to calculate
the cross sections for the processes with average multiplicities 2$< n_P >
$, 3 $< n_P > $ and so on, as well as the process for the diffractive
dissociation with multiplicity much smaller that $< n_P > $.

\eq{CRAGK} can be rewritten in the form
\beq \label{SERAGK}
\si_{dipole}\,\,\,=\,\,2\,\int\,d^2b_t\,\,\sum^{\infty}_{n =1}\,\,C_n\,( -
1 )^{n + 1}\,\,\left(\,\frac{\Omega^P}{2}\,\right)^n\,\,,
\eeq
where each term corresponds to the exchange of $n$ Pomerons.
The AGK rules are
\begin{eqnarray}
 \si^n_{dipole}(k < n_P > ) \,&=&\,\int\,d^2b_t\, C_n ( -1 )^{n 
 -k}\,\frac{n!}{(n - k)!\, k!}\,\,\left(\,\Omega^P\,\right)^n\,
\,\,;\label{AGK1}\\
 \si_{dipole}^n ( DD )\,&=&\, \int\,d^2b_t\, C_n ( - 1)^n\,
\{\,\left(\,\Omega^P\,\right)^n\,\,-\,\,
2\,\left(\,\frac{\Omega^P}{2}\,\right)^n\,\}\,\,;\label{AGK2}
\end{eqnarray}  
where $\si^n_{dipole}(k < n_P > )$ and $\si_{dipole}^n ( DD )$  are the 
contributions of $n$-Pomeron
exchange to the cross section for the process with average multiplicity
$k
< n_P > $, and to the cross section for the diffractive dissociation
processes with  small multiplicity. 

\begin{figure}[htbp]
\begin{center}
  \epsfig{file=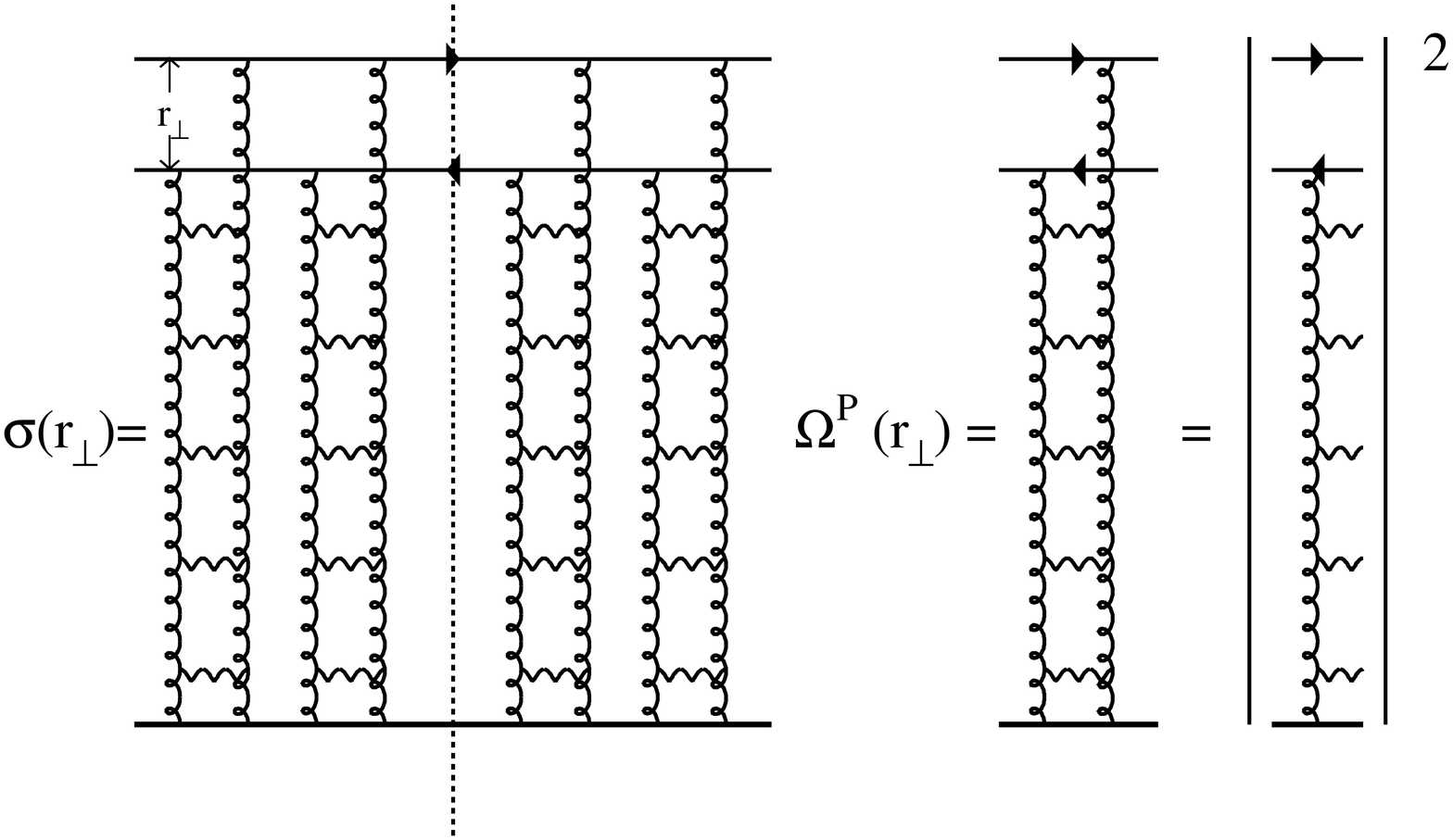,width=160mm,height=55mm}
  \caption[]{\it  Total cross section for the dipole - nucleon interaction
in 
the Mueller-Glauber approach. The dashed line shows the diffraction
dissociation cut. }
 \end{center}
\label{fig9}
\end{figure}

Summing over $n$ in \eq{AGK2} we obtain
\begin{eqnarray} 
\si^{DD}_{dipole}
(r_{\perp})\,\,&=&\,\,\int\,d^2b_t\,\sum^{\infty}_{n=1}\,\, C_n\,(-1 )^n\,
\{\,\left(\,\Omega^P\,\right)^n\,\,-\,\,
2\,\left(\,\frac{\Omega^P}{2}\,\right)^n\,\}\,\,;\label{DDAGK1}\\
&=&\,\,\int\,d^2b_t\,\left(\,2\,\{\,1 \,\,-\,\,e^{-
\frac{\Omega^P}{2}}\,\}
\,\,-\,\,\{\,1 \,\,-\,\,e^{-  \Omega^P }\,\}\,\right)\,\,;\label{DDAGK2}\\
&=&\,\,\int\,d^2b_t\,\left(\,1 \,\,-\,\,e^{-
\frac{\Omega^P}{2}}\,\right)^2\,\,.\label{DDAGK3}
\end{eqnarray}

One can see that \eq{DDAGK3} is just the same equation  for diffractive 
production that we have obtained from  unitarity ( see \eq{RGEN} ).
From \eq{AGK1} we can find the cross section for the production of $k$ -
parton showers
 which is equal to
\beq \label{PSHOW}
\si_{dipole}(k < n_P > )\,\,\,=\,\,\int \,\,d^2
\,b_t\,\,\,\frac{\left(\,\Omega^P\,\right)^k}{k!}\,\,e^{ - \Omega^P}\,\,.
\eeq
 \eq{PSHOW} will be very useful below, when we 
discuss the diffraction production of the  $q \bar q G $ system.
\subsection{Cross section for $\mathbf{q \bar q G }$ diffractive
production
 with SC}
\subsubsection{First correction to the Mueller -  Glauber formula  for the
total cross
section}
As shown  in Fig.9, the Eikonal approach takes into account only
rescatterings of the fastest colour dipole.  In this subsection we 
will extend the formalism so as also to include
 the rescatterings with the target of the two fastest color
dipoles: the initial colour dipole and the  fastest gluon in Fig.9.   
Our goal is  to include all diagrams of the type shown in Fig.10.
Dashed lines in Fig.10 indicate the diffraction dissociation cuts.
These diagrams include the diffractive production of $q
\bar q G $ system as well as a quark - antiquark pair.

Since \eq{U1} and \eq{U2} give the general solution to the unitarity
constraint, our problem is to find an expression for $\Omega$ which will
be more general that \eq{OMEGA} with $\kappa^{DGLAP}$ defined by
\eq{KAPPA}.

 \begin{figure}[htbp]
\begin{center}
  \epsfig{file=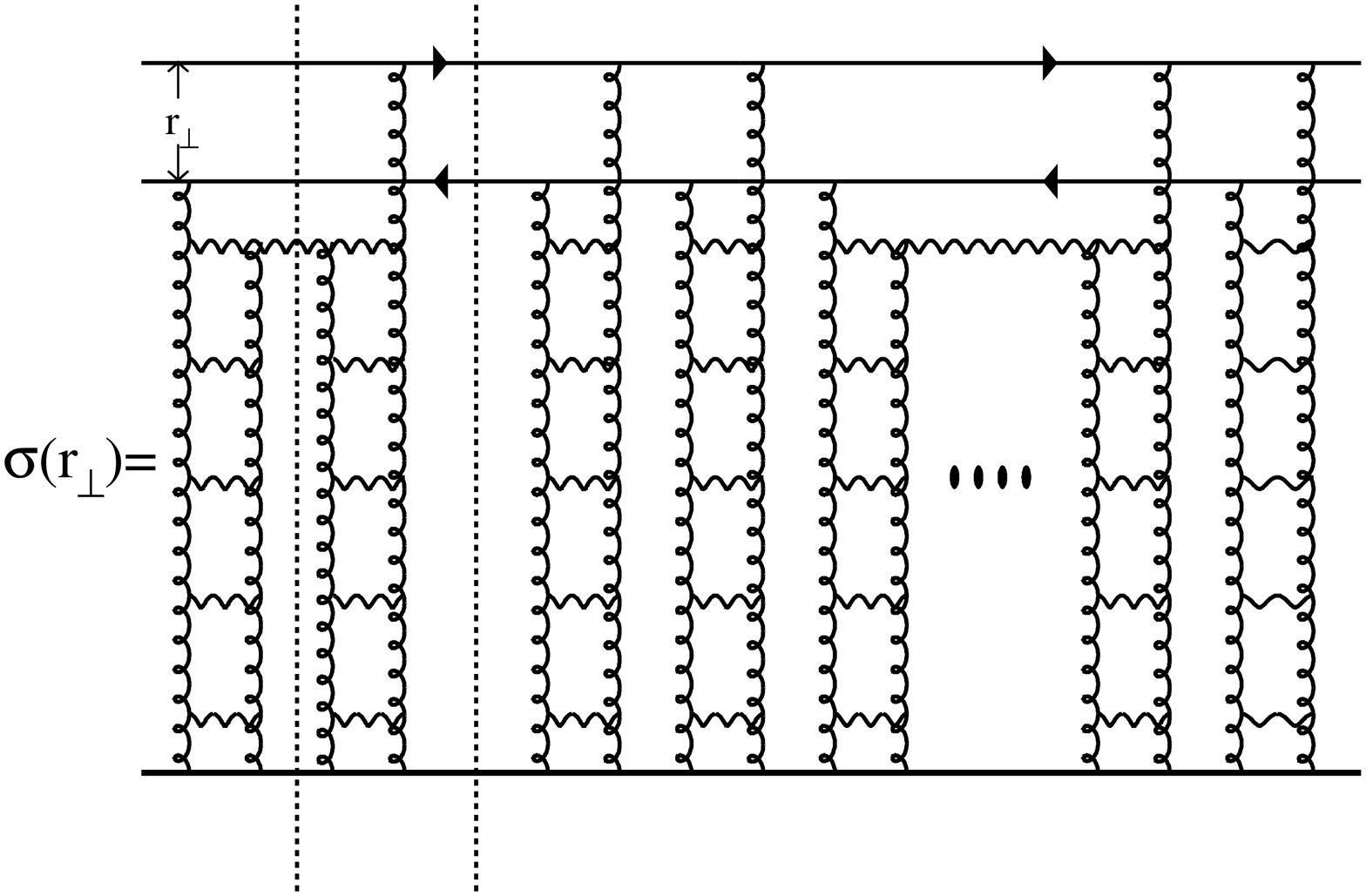,width=160mm,height=55mm}
  \caption[]{\it  Total cross section for dipole - nucleon interaction in
the first iteraction of the
Mueller-Glauber approach. The dashed lines show the diffraction
dissociation cut. }
 \end{center}
\label{fig10}
\end{figure}

The natural generalization of $\kappa^{DGLAP}$ is  to substitute in
\eq{KAPPA} the Mueller-Glauber formula for the gluon structure function
\cite{MU90} \cite{AGL}, namely\footnote{It should be stressed that $x
G(x,\frac{4}{r^2_t};b_t)$ is introduced in a such way that $\int \,d^2b_t
xG(x,\frac{4}{r^2_t};b_t) \,=\,xG(x,\frac{4}{r^2_t})$.}

 \beq \label{MGGL}
\frac{1}{\pi R^2}\,\,x
G^{DGLAP}(x,\frac{4}{r^2_{\perp}})\,\,e^{-\frac{b^2_t}{R^2}}\,
\,\longrightarrow\,\,x
G^{MG}(x,\frac{4}{r^2_{\perp}};b_t)\,\,\,=\,\,
\eeq
$$
\frac{4}{\pi^3}\,
\,\int^1_x\,\frac{d x'}{x'} \int^{\infty}_{r^2_{\perp}}\,\,\frac{ d
\,r'^2_{\perp}}{r'^4_{\perp}}\,2\,\{\,1\,\,\,-\,\,e^{-
\frac{\Omega^P_G(x',r'_{\perp};b_t)}{2}}\,\}\,\,,
$$
where 
\beq \label{OMEGAGL}
 \Omega^P_G\,\,=
\,\, \frac{3 \pi^2 \as}{4 \pi R^2}
\,\,r^2_{\perp}\,\,G^{DGLAP}(x,\frac{4}{r^2_{\perp}})\,\,e^{-
\frac{b^2_t}{R^2}}\,\,.
\eeq
\eq{MGGL} takes into account the rescattering of the gluon with the
target in the Eikonal approach ( see Fig.10),  however, the question
arises of 
why   
we need to only include  gluon rescattering for the $ q \bar q G$ system.
To understand  the physics of \eq{MGGL} it is advantageous to consider the
equation
that describes the rescattering of all partons. Kovchegov \cite{KOV}
 using two principle ideas suggested by A. Mueller
\cite{MUDIPOLE}
proved that the GLR nonlinear equation \cite{TEOID} is able to describe
such rescatterings. The principles are: 
\begin{enumerate}
\item\,\,\,The QCD interaction at high energy does not change the
transverse size of
interacting colour dipoles, and thus they can be considered
as the correct degrees of freedom at high energies;
\item\,\,\, The process of interaction of a dipole with the target has two
clear stages:
\begin{enumerate}
\item\,\,\, The transition of the dipole into two dipoles, the
probability for this is given by
\beq \label{PSIDIPOLE}
| \Psi(\,
{\mathbf{x_{01}}}\,\rightarrow\,{\mathbf{x_{02}}}\,\,+\,\,{\mathbf{x_{12}}}\,)
|^2\,\,\,=\,\,\,\frac{1}{z}\,\,\frac{{\mathbf{x^2_{01}}}}{{\mathbf{x^2_{02}}}\,
{\mathbf{x^2_{12}}}}\,\,,
\eeq
where $\mathbf{x_{ik}}$ denotes the size of the dipoles, and $z$  the
fraction
of energy of the initial dipole that the final dipole carries;
\item\,\,\,The interaction of each  dipole with the target has an
amplitude $a^{el}({\mathbf{x}},b_t,y=\ln(1/x))$.
\end{enumerate}

\end{enumerate}

The equation is illustrated in Fig.11 and it has the following analytic
form:
\beq \label{GLRINT}   
\frac{d a^{el}_{dipole}({\mathbf{x_{01}}},b_t,y)}{d y}\,\,\,=\,\,\,-
\,\frac{2\,C_F\,\as}{\pi} \,
\ln\left( \frac{{\mathbf{x^2_{01}}}}{\rho^2}\right)\,\,
a^{el}_{dipole}({\mathbf{x}},b_t,y)\,\,\,+  
\frac{C_F\,\as}{\pi^2}\,\,\int_{\rho} \,\,d^2 {\mathbf{x_{2}}}\,
\frac{{\mathbf{x^2_{01}}}}{{\mathbf{x^2_{02}}}\,
{\mathbf{x^2_{12}}}}\,\,
\eeq
$$
\{2\,a^{el}_{dipole}({\mathbf{x_{02}}},{ \mathbf{
b_t -
\frac{1}{2}
x_{12}}},y)\,\,\,-\,\,\,a^{el}_{dipole}({\mathbf{x_{02}}},{ \mathbf{ b_t -
\frac{1}{2}
x_{12}}},y)\,\,a^{el}_{dipole}({\mathbf{x_{12}}},{ \mathbf{ b_t -
\frac{1}{2}
x_{02}}},y)\,\}\,\,.
$$

The first term on the r.h.s. of the equation gives the contribution of
virtual corrections, which appear in the equation due to the  
normalization of the partonic wave function of the fast colour dipole (
see Ref. \cite{MUDIPOLE} ). The second term describes the decay of the
colour dipole of size ${\mathbf{x_{01}}}$ into two dipoles of
sizes
${\mathbf{x_{02}}}$ and ${\mathbf{x_{12}}}$, and their  interactions
 with the target in the impulse approximation ( notice factor 2 in
\eq{GLRINT} ). The third term corresponds to the simultaneous interaction
of
two produced colour dipoles with the target and describes the Glauber-type
corrections for scattering of these dipoles.

\begin{figure}[htbp]
\begin{center}
  \epsfig{file=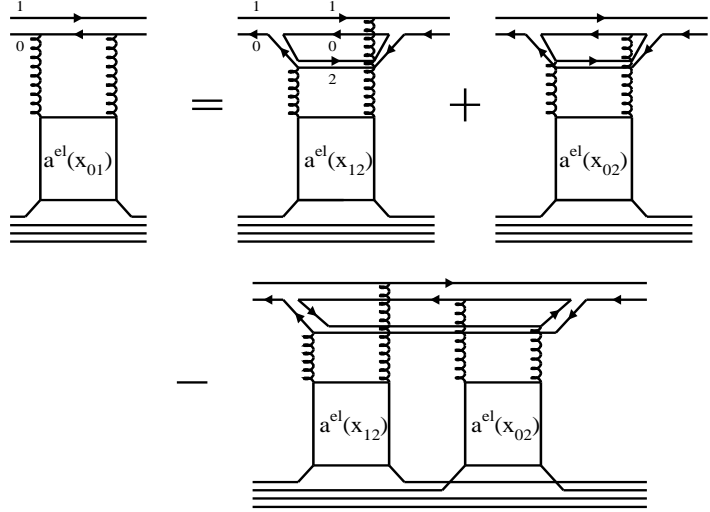,width=120mm, height=70mm}
  \caption[]{\it  Pictorial representation of the nonlinear evolution
equation that takes into account the rescattering of all partons with the
target. }
 \end{center}
\label{fig11}  
\end{figure}

The initial condition for this equation \cite{KOV} is given by \eq{U1}
with $\Omega = \Omega^P$ from \eq{OMEGA} at $x = x_0$.

In DIS the dominant contribution  comes from the decay of a
small dipole into two large dipoles. Therefore, we can reduce the kernel
of \eq{GLRINT}    
to \cite{KOV}
\beq \label{DLR}
\int_{\rho} \,\,d^2
{\mathbf{x_2}}\,\,\frac{{\mathbf{x^2_{01}}}}{{\mathbf{x^2_{02}}}\,
{\mathbf{x^2_{12}}}}\,\,\,\longrightarrow\,\,\,{\mathbf{x^2_{01}}}
\,\pi\,\,\int^{
\frac{1}{\Lambda^2_{QCD}}}_{{\mathbf{x^2_{01}}}}\,\,\frac{d
{\mathbf{x^2_{02}}}}{( \,{\mathbf{x^2_{02}}}\,)^2}\,\,.
\eeq

We  make the first iteration of \eq{GLRINT}, by substituting the 
Mueller - Glauber formula for color dipole rescattering ( see \eq{U1} 
with $\Omega = \Omega^P$ from \eq{OMEGA} ), and  obtain

\begin{eqnarray}
a^{el}_{dipole}( \,\,first\,\,\,\,\,iteration\,\,)\,\,&=&\,\,
C_F\,\as \,\mathbf{x^2_{01}}\,\,\{2\,( 1 - e^{- \frac{\Omega^P}{2}} ) -
( 1 - e^{- \frac{\Omega^P}{2}} )^2\,\}\,\,\label{FSTIT}\\
&=&\,\,C_F\,\as\,\, \mathbf{x^2_{01}}
\,\,\{\,1\,\,\,-\,\,\,e^{- \frac{2\,\Omega^P}{2}}\,\,\}\,\,.\nonumber
\end{eqnarray}
 \eq{FSTIT} gives the Mueller - Glauber formula for
the gluon structure function of \eq{MGGL} \footnote{ \eq{GLRINT} is
written for large number of colours $ N_c \,\gg\,1 $. For finite $N_c$,
$2\,\Omega^P$ in \eq{FSTIT} should be replaced by
$2\,\Omega^P\,\rightarrow\,\,\frac{9}{4} \Omega^P$.} . Note, that two
assumptions
have been made in deriving \eq{FSTIT}: (i) $b_t\,\,\gg\,\,\mathbf{
x_{12}}$
or
$\mathbf{ x_{02}}$ and (ii) we neglected the first term in \eq{GLRINT}.
Both approximations hold in the so called double log approximation of pQCD
\cite{KOV}. 

 \eq{MGGL} describes the passage of the $q \bar q G $
system through the target, as it corresponds to the interaction of two
colour dipoles, which is our $q \bar q G $ system, but not the gluon.
However,
in the parton cascade in the DIS kinematic region, the  gluon always
corresponds
to two colour dipoles of the same size.
By changing $\Omega^P \,\rightarrow \Omega^{MG}$
we take into account the fact that initial quark - antiquark pair can
fluctuate in the $q \bar q G $ system many times during the   passage
through
the target. This is illustrated in Fig.10.

 Finally, we obtain the following formula for the total cross
section:
\beq \label{TOTCRFRSIT}
\sigma^{(1)}_{dipole}\,\,\,=\,\, \int d^2 b_t
2\,\{\,1\,\,-\,\,e^{ -
\frac{\Omega^{MG}(x,r_{\perp};b_t)}{2}}\,\}\,\,,
\eeq
where
\beq \label{OMEGAMG}
\Omega^{MG}(x,r_{\perp};b_t)\,\,\,=\,\,\,\frac{\pi^2
\,\as}{3}\,\,r^2_{\perp}\,\,x G^{MG}(x,r^2_{\perp};b_t)\,\,
\eeq
with $x G^{MG}(x,r^2_{\perp};b_t)$ given in \eq{MGGL}.

\subsubsection{Cross section for diffractive production}
We would like to obtain the cross section for the  diffractive production
of
both $q \bar q $ and $q \bar q G$ final states using the AGK cutting
rules. In Fig.10 one can see which cuts in the total cross section are
related to the diffracrive processes. They are shown in Fig.10 by dashed
lines. First, we have to generalized the AGK cutting rules, since in
\eq{AGK1} and \eq{AGK2} we have used the property
 of one Pomeron exchange  i.e. one
gluon ``ladder"  exchange  , namely
\beq \label{POMPROP}
2 \, \Omega^P\,\,=\,\,G_{in}\,\,,
\eeq
where $G_{in}$ stands for the inelastic cross section with
large multiplicity $< n_P > $ ( see Fig.9 ). In \eq{TOTCRFRSIT}
$\Omega^{MG}$ itself has a more complicated structure , which can be
recovered using the AGK rules of \eq{AGK1} and \eq{AGK2}. 
 \eq{AGK2} for $\Omega^{MG}$ gives 
\beq \label{DDOMMG}
\si^{DD}_{MG} (b_t)\,\,\,=\,\,\frac{N_c
\as}{\pi}\,r^2_t\,\,\int^1_x\,\frac{d x'}{x'}
\,\int_{r^2_t}\,\frac{d
r^2_{\perp}}{r'^4_{\perp}}\,\,\left(\,1\,\,\,-\,\,\,e^{
- \frac{\Omega^P_G}{2}}\,\right)^2\,\,,
\eeq
with $\Omega^ P_G$ from \eq{OMEGAGL}. 
 
We  denote by $\Omega^{MG}_{cut}$ the contribution of all processes
given by the AGK rules for $\Omega^{MG}$. Using this notation 
 the simple generalization of the AGK cutting rules gives 
the contributions of
different processes to the total cross section of  \eq{TOTCRFRSIT};
\begin{eqnarray}
\si^{(1)}_{dipole}\,&=& \int \,d^2 b_t\,\sum^{\infty}_{n =1}
\,\,C_n\,(-1)^{n +
1}\,\left(\,\frac{\Omega^{MG}}{2}\,\right)^n\,\,;\label{GAGK1}\\
\si^{(1)}_{dipole}(n;k)\,&=& \,\int\,d^2 b_t \,C_n \,( - 1)^{n
-k}\,\frac{n!}{(n - k)!\, k!}
\,\,(\,\Omega^{MG}\,)^{ n - k}
(\,\Omega^{MG}_{cut}\,)^k\,\,;\label{GAGK2}\\
\si^{(1)}_{dipole}(n;0)
\,&=& \int\,d^2b_t\, C_n ( - 1)^n\,
\{\,\left(\,\Omega^{MG}\,\right)^n\,\,-\,\,
2\,\left(\,\frac{\Omega^{MG}}{2}\,\right)^n\,\}\,\,;\label{GAGK3}
\end{eqnarray}
where we denote by $\si^{(1)}_{dipole}(n;k)$ the contribution of $k$ - cut
$\Omega^{MG}$ to the $n$-th term of \eq{GAGK1}, $ \si^{(1)}_{dipole}(n;0)$
is the cross section for the diffraction production of a quark - antiquark
pair
for the $n$-th term of \eq{GAGK1} ( see Fig.10). The structure of the
inelastic processes for  each term $\si^{(1)}_{dipole}(n;k)$ is rather
complicated but is well defined by the AGK rules for $\Omega^{MG}$. 
However, we need  only take  the cross section of the diffractive
process
from each $\Omega^{MG}_{cut}$ or, in other words, we should replace
$\Omega^{MG}_{cut}$ by $ \si^{DD}_{MG} (b_t)$ in \eq{GAGK2}. Performing
the 
summation over $n$ and $k$ we obtain:
\beq \label{QQGDD}
\si^{DD}_{dipole}( q \bar q \,\rightarrow\, q \bar q\,G )\,\,\,=\,\,\,
\int\,d^2b_t \,\,e^{ -
\Omega^{MG}(x,r_{\perp};b_t)}\,\,\left(\,e^{\si^{DD}_{MG}
(b_t)}\,\,-\,\,1\,\right)\,\,,
\eeq
where $\Omega^{MG}$ defined in \eq{OMEGAMG}. \eq{QQGDD} gives the
contribution to the diffractive dissociation cross section of emission of
gluons and $k$ is the number of gluons that we summed over.
 \eq{GAGK3} leads to
\beq \label{QQDD}
\si^{DD}_{dipole}( q \bar q \,\rightarrow\, q \bar q  )
\,\,\,=\,\,\,
\int\,d^2b_t \,\,\left(\,1\,\,\,-\,\,e^{ - \frac{ 
\Omega^{MG}(x,r_{\perp};b_t)}{2}}\,\,\right)^2\,\,.
\eeq

In this paper we are only interested in production of one extra gluon
which corresponds to \eq{GAGK2} with $k$ = 1. Therefore 
\beq \label{QQGDDFIN}
\si^{DD}_{dipole}( q \bar q \,\rightarrow\, q \bar q\,G )\,\,\,=\,\,\,
\int\,d^2b_t \,\,e^{ -
\Omega^{MG}(x,r_{\perp};b_t)}\,\si^{DD}_{MG}(b_t)\,\,,
\eeq
which we will use for our numerical calculation.

Finally, collecting \eq{TOTCRFRSIT},  \eq{QQDD}, and \eq{QQGDDFIN}  we
obtain the generalization of \eq{RGEN}, which takes into account the
diffractive production of both a quark - antiquark pair, and a quark -
antiquark
pair plus an extra gluon final states:
\beq \label{FIN}
\Re\,\,=\,\,\frac{\si^{DD}}{\si_{tot}};
\eeq
\begin{eqnarray}
& &\si^{DD}\,\,=\,\,
 \int\,d^2 b_t \,\int\,d^2 r_{\perp}\, P^{\gamma^*}(z, r_{\perp};Q^2) \times
\nonumber \\ \nonumber \\ 
& &\left( \{\,1\,-\,e^{ - \frac{
\Omega^{MG}(x,r_{\perp};b_t)}{2}}\,\,\}^2\,+ \,
\,e^{ -
\Omega^{MG}(x,r_{\perp};b_t)}\,r^2_{\perp}\frac{2\as(r_{\perp})}{3\pi}\,\int^1_x
\frac{d
x'}{x'}\,\int^{\infty}_{r^2_{\perp}}\frac{dr'^2_{\perp}}{r'^4_{\perp}}\,\{
\,1\,-\,e^{-
\frac{\Omega^P_G(x',r'_{\perp};b_t)}{2}}\,\}^2\,\right); \nonumber
\end{eqnarray}
$$
\si_{tot}\,\,=\,\,
2\,\,\,  \int\,d^2 b_t \,\int\,
d^2r_{\perp}\, P^{\gamma^*}(z, r_{\perp};Q^2)\,\,\left(\,\,1\,\,\,-\,\,\,e^{ - \frac{ 
\Omega^{MG}(x,r_{\perp};b_t)}{2}}\,\,\right).
$$

We note  that \eq{FIN} was derived in the double log
approximation to the DGLAP evolution equations, in which the colour
dipoles in the produced $q \bar q G$ system are much larger than the
initial
quark - antiquark dipole.

The factor 
$e^{ - \Omega^{MG}(x,r_{\perp};b_t)}$ in the second term in \eq{FIN}, 
 leads to the suppression of diffractive production of the
 the $q \bar q G $ system.
 Therefore, in the asymptotic limit at large values of
$\kappa_{dipole}$,
only elastic rescattering of the quark - antiquark pair survives, leading
to
the value of the ratio
$\sigma^{DD}/\sigma_{tot}\,\,\longrightarrow\,\,1/2$.  
 It turns out that this factor  is important for numrerical
calculations in the HERA kinematic region  ( see the next section).
 It was omitted in Ref.\cite{GOWU2}

\section{Numerical calculation for $\mathbf{ \sigma^{DD}/\sigma_{tot}}$}
\setcounter{equation}{0}
In this section we present the numerical result for the ratio of the
total diffractive dissociation cross section to the total cross section in
DIS.   We  postpone to the next section the  consideration of the
influence of the 
experimental mass cutoff on this ratio.

The main parameters that determine the value of $\kappa_{dipole}$, are
 the value of $R^2$ and the value of the gluon distribution.
We choose $R^2\,=\,10\,GeV^{-2}$ since it is the value which is obtained
from ``soft" high energy phenomenology \cite{DL} \cite{GLMSOFT} and 
  is in agreement with HERA data on J/$\Psi$ photoproduction
\cite{HERAPSI}. 
For $xG(x,Q^2)$ we use the GRV'94 parameterization and 
the leading order solution of the DGLAP evolution equation \cite{GRV}.
We have two reasons for this choice:
\begin{enumerate}
\item\,\,\, Our goal in this paper is to understand the influence of the
SC on the energy behaviour of the ratio  $ \sigma^{DD}/\sigma_{tot}$.
However,  experience shows that we are able to describe almost all HERA
data by changing the initial conditions of the DGLAP evolution equations.
Unfortunately, we have no theoretical restrictions on this input for any
of the 
parameterizations on the market. On the other hand, we know  theoretically
\cite{TEOID} that the SC corrections work in a such manner that they alter 
the initial conditions for the DGLAP evolution, making it impossible to
apply
them  at fixed $Q^2 = Q^2_0$. With SC we have to solve the DGLAP equations
starting with $Q^2 = Q^2_0(x)$ where $Q^2_0$ is the solution of the
equation $\kappa^{DGLAP}_{dipole}(x,Q^2_0(x))\,=\,1$. Therefore, we are in
controversial situation: we require a GLAP input but, if we take it from 
the so called global fits, there is a  danger that we will incorporate
all
the effects of the SC
in the initial condition of these parameterizations. Only data taken after
the
 1995 runs are at energies sufficiently high to effect the low $x$
behaviour of
the initial inputs. 
 GRV'94 is based on the experimental data at rather large $x$, so we
hope  that  SC are minimal in this
parameterization;
\item\,\,\, The GRV parameterization starts with rather low virtualities (
as low as $Q^2 \,\approx\,0.5 \,GeV^{-2}$ ). This is a major weakness 
 of this approach, since one cannot guarantee that only the leading
twist contribution is dominant in the DGLAP evolution
equations.  We agree with this criticism, but the low value of $Q^2_0$
leads to the solution of the DGLAP equation which is closer  to the
leading log approximation, in which we can guarantee the accuracy of our
master equation ( see \eq{FIN} ).

\end{enumerate}
Before we present our numerical results we have to make an important
comment. It concerns the substitution in \eq{MGGL}, where the iterarted
 gluon density
$xG^{MG}(x,4/r^2;b)$ is defined.
 Due to technical problems, two modifications of the formula
are made in the actual numerical calculations. 
First of all we assume the exponential $b$ factorization of  $xG^{MG}(x,4/r^2;b)$:
$$
xG^{MG}(x,4/r^2;b)\,\,=\,\, \frac{1}{\pi\,R^2} \,xG^{MG}(x,4/r^2)\, e^{-\frac{b^2}{R^2}},
$$
where
\beq
\label{mal1}
xG^{MG}(x,4/r^2)=\int d^2b\, xG^{MG}(x,4/r^2;b)
\eeq
We  justify the factorization by
the following two arguments. The first one is numerical.
We actually have  checked that  factorization  holds numerically with
  satisfactory
accuracy. The second  argument is that  having assumed  factorization
for $xG^{DGLAP}$ on the  same basis 
 we can assume it for $xG^{MG}$.
The second modification, made with \eq{MGGL}
is due to the  fact that the GRV parameterization we are working with does not
satisfy the DGLAP equation in DLA (see Ref. \cite{AGL} for
 the discussion of the problem).
The authors of   \cite{AGL} suggested a modification of \eq{mal1},
 which we implement
in all our numerical calculations. The final formula for $xG^{MG}$ has the form:
\begin{eqnarray}
xG^{MG}(x,Q^2)\,=\, \frac{4}{\pi^2}\int_x^1\frac{dx^\prime}{x^\prime}\
\int_{4/Q^2}^\infty \frac{dr^{\prime 2}}{r^{\prime 4}} \int\, db^2\,
2\,(1-e^{-\Omega^{GRV}(x^\prime,r^\prime,b)})\,+ \nonumber \\ \\
xG^{ GRV}(x,Q^2) - \frac{\as N_c}{\pi}\int_x^1\int_{Q_0^2}^{Q^2}\frac{dx^\prime}{x^\prime}
\frac{dQ^{\prime 2}}{Q^{\prime 2}}\, x^\prime G^{ GRV}(x^\prime,Q^{\prime 2}). \nonumber
\end{eqnarray}

In all the figures representing our numerical results
the upper solid line corresponds to the full answer, the widely spaced 
dashed line to
 diffractive production of $q \bar q $ pair,
while the narrowly spaced dashed line
 to the  diffractive production of $q \bar q G $ .

\begin{figure}[htbp]
\begin{tabular}{c c}
 $\mathbf{Q^2 = 8 \,GeV^2 }$ & $\mathbf{Q^2 = 14 \,GeV^2 }$ \\
  \epsfig{file=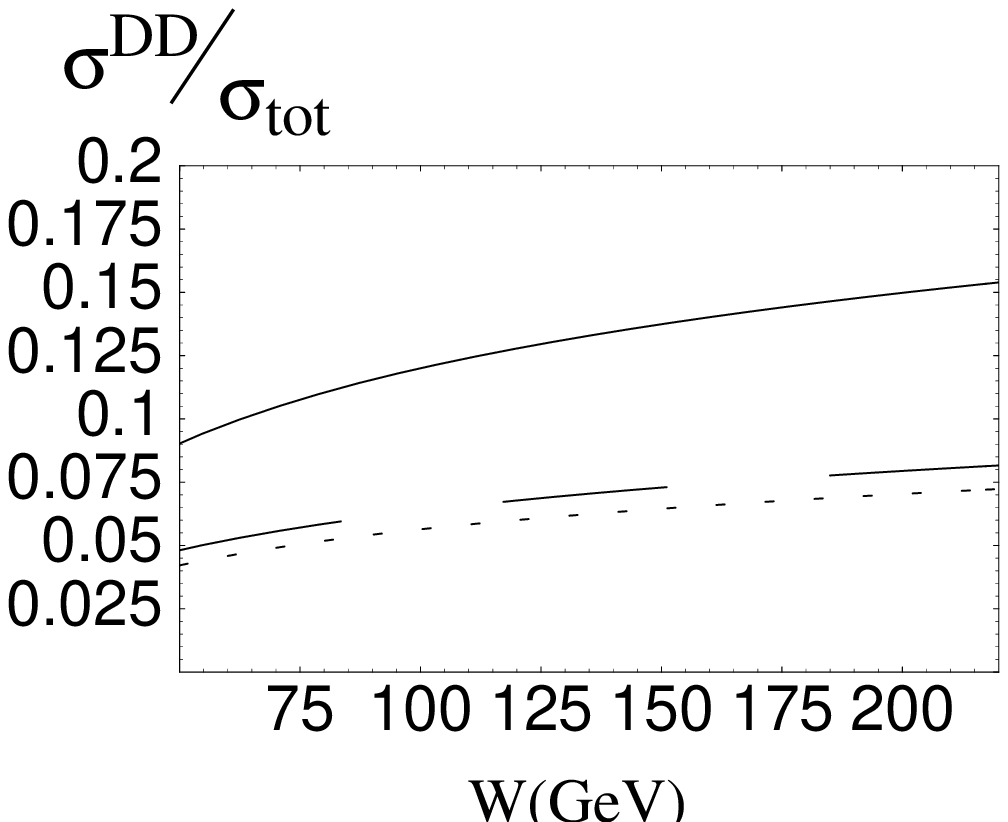,width=80mm, height=60mm}&
\epsfig{file=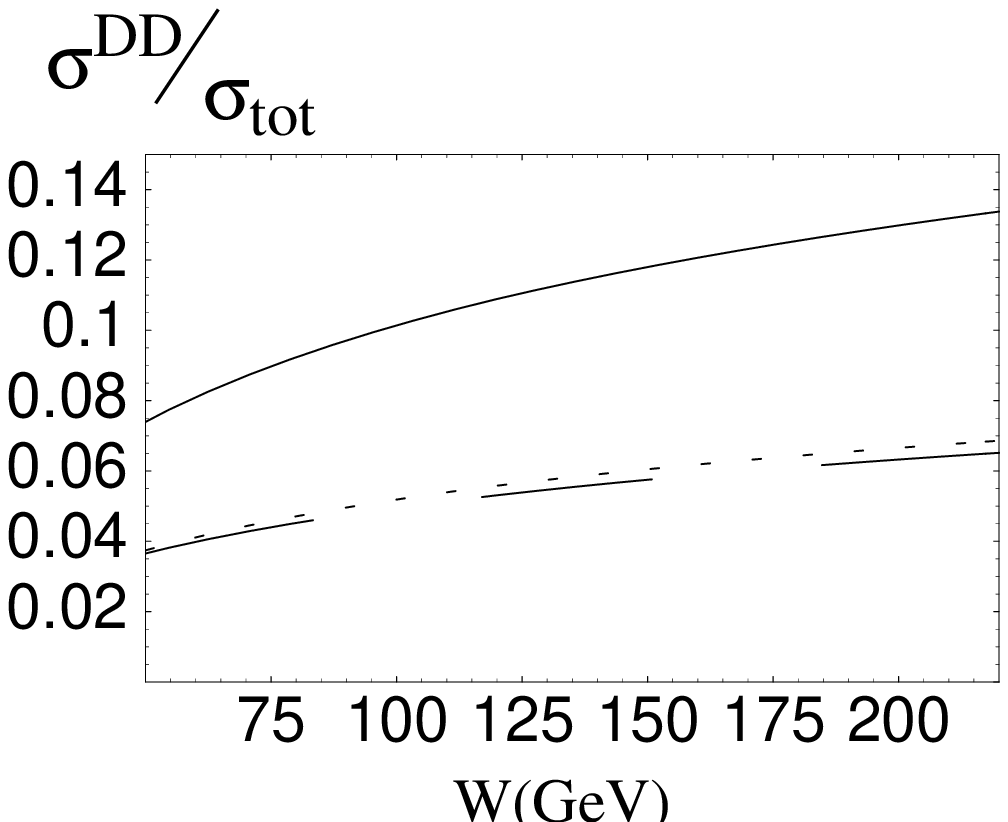,width=80mm, height=60mm}\\
& \\ & \\
$\mathbf{Q^2 = 27 \,GeV^2 }$ & $\mathbf{Q^2 = 60 \,GeV^2 }$ \\
 \epsfig{file=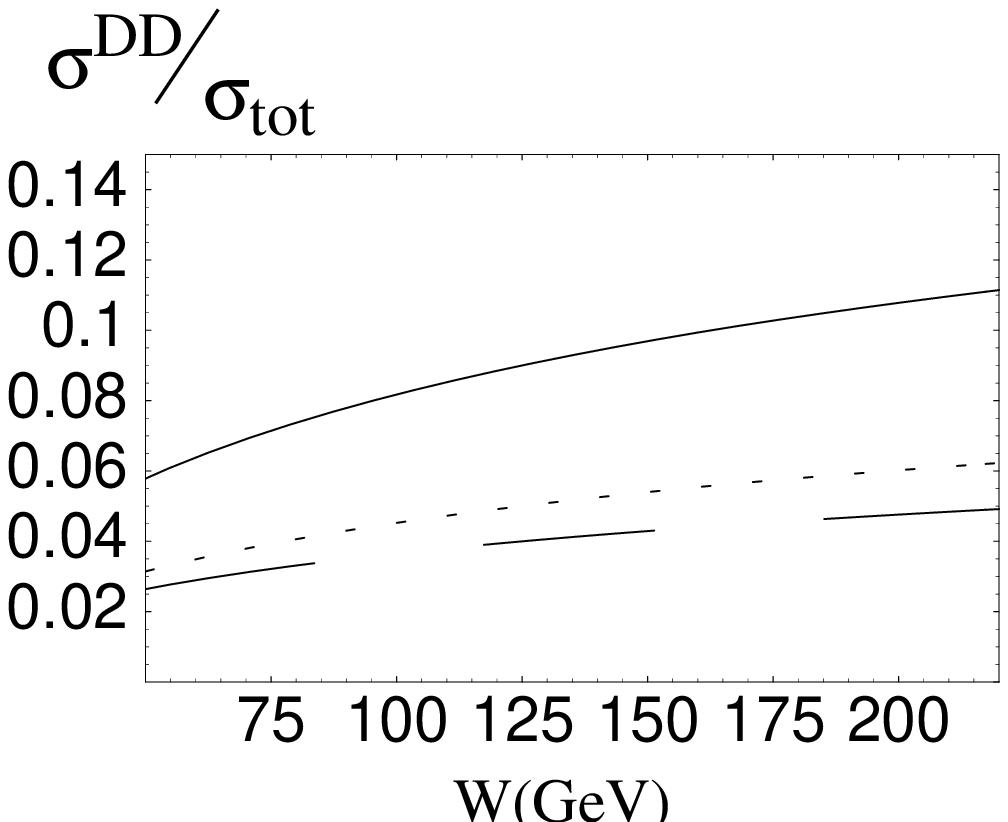,width=80mm, height=60mm}&
\epsfig{file=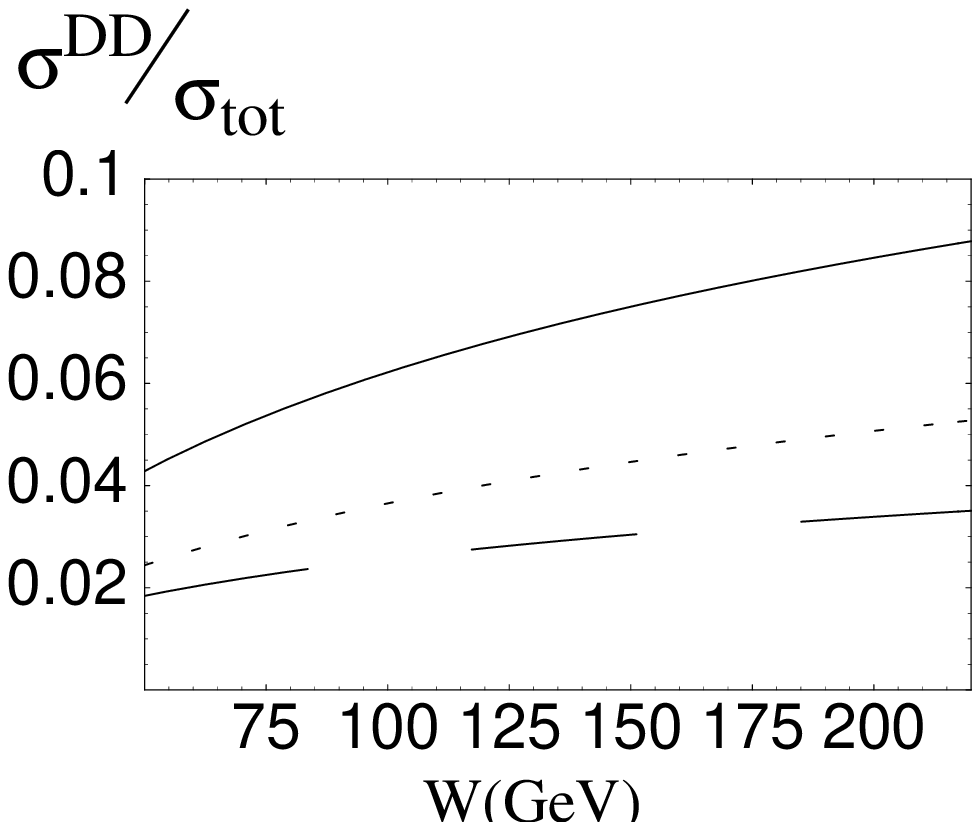,width=80mm, height=60mm}\\
\end{tabular}
  \caption[]{\it  The ratio $\sigma^{DD}/\sigma_{tot}$ versus $W$ for
different values of  $Q^2$ ( $R^2 \,=\, 10\,{\rm GeV}^{-2}$ ).
The upper solid line corresponds to the full answer, the widely spaced
 dashed line to
 diffractive production of $q \bar
q $ pair,  while the narrowly dashed
line  to the  diffractive production of $q
\bar q G $ . }
\label{fig12}
\end{figure}

\begin{figure}[htbp]
\begin{tabular}{c c}
$\mathbf{Q^2 = 1 \,GeV^2 }$ & $\mathbf{Q^2 = 2.5 \,GeV^2 }$ \\
 \epsfig{file=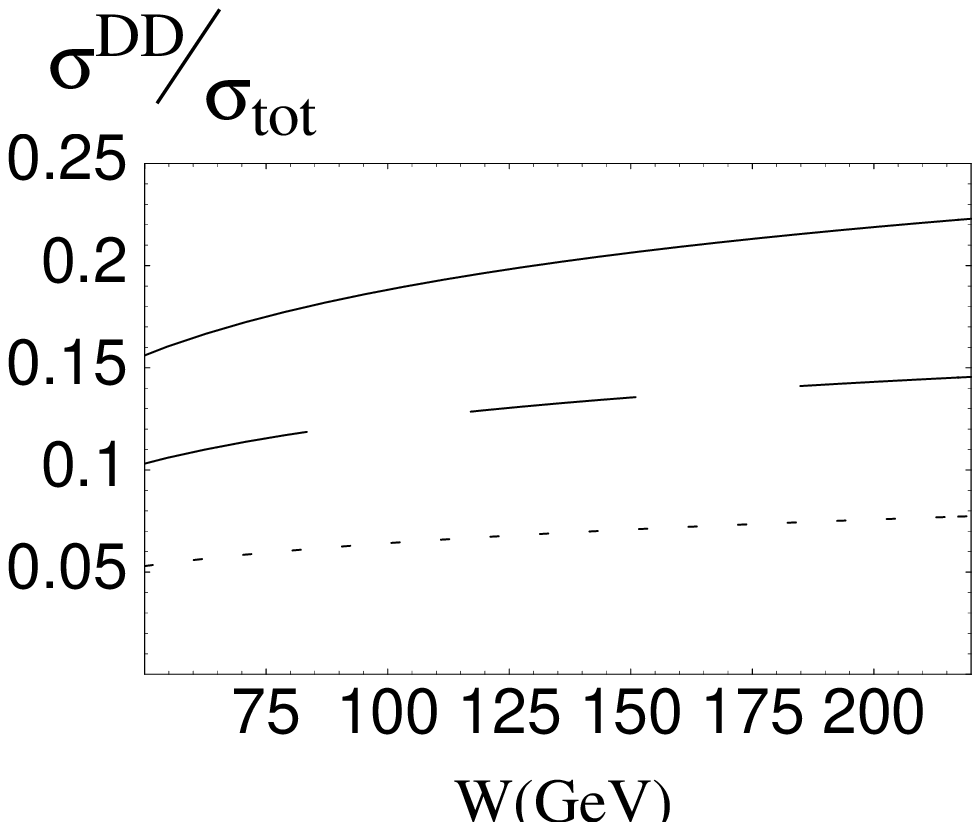,width=80mm, height=60mm}&
\epsfig{file=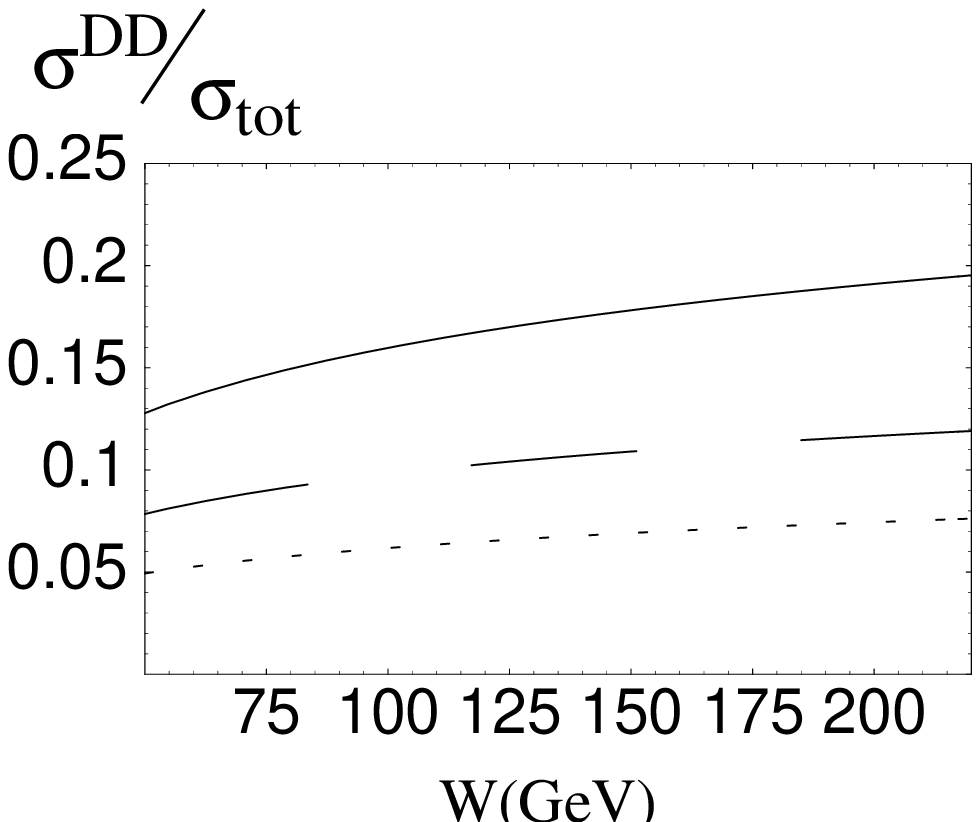,width=80mm, height=60mm}\\ 
\end{tabular}
  \caption[]{\it  The ratio $\sigma^{DD}/\sigma_{tot}$ versus $W$ for
small  values of  $Q^2$  ( $R^2 = 10\,GeV^{-2}$  )
The upper solid line corresponds to the full answer, the widely
spaced dashed
 line to  diffractive production of $q \bar
q $ pair,  while the narrowly spaced  dashed line to
  the  diffractive production of
$q \bar q G $ . }
\label{fig13}
\end{figure}

In Fig. \ref{fig12} we plot the results of our calculations using  \eq{FIN}.
We have not added any contribution from the nucleon excitations, based on
our
estimates given in Fig.7b.
In Fig.\ref{fig13} we show our predictions for small values of $Q^2$. The first
observation is the fact that the ratio increases considerably,  reaching
the value of about 25 \%. It should be stressed that even at rather small values of
$Q^2$, we do not see any sign of saturation of the energy behaviour of
this  ratio, which is continously
increasing   in the HERA kinematic region.

It is interesting to consider separately    the quark - antiquark
diffractive production and
the production of the $q\bar q G$ final state.
Both cross sections increase as a function of the energy $W$. At small $Q^2$ (Fig. \ref{fig13})
the total diffractive cross section is dominated (about 70 \%) by the diffractive production
of the quark - antiquark pair. The situation changes when higher values of $Q^2$
are considered. Then the $q\bar q G$ final state contibutes even more than the $q \bar q $ pair.
Indeed, at high energy $W = 200 \, GeV$,
$q \bar q G $ is produced in  approximately $ \,25 \%$ of all diffractive
events
at $Q^2\, =\, 8 \, GeV^2 $. For smaller values of $Q^2$ this fraction 
decreases,
being only 10\% at $Q^2 = 1\, GeV^2$.  This is a expected behaviour
if the SC play an important role.
The  results presented show that the contribution of the
extra gluon emission is crucial for the predictions, and may
lead to a hundred percent enhancement of the ratio.

At small $Q^2$, the fraction of $q\bar q G$
  diffractive production to the total cross section
increases sufficiently slow due to the suppression factor
$e^{-\,\Omega^{MG}(x,r_\perp;b)}$
in \eq{FIN}. For larger $Q^2$, $\Omega^{MG}$ 
 becomes small and the fraction of $q \bar q G$ diffractive
production increases faster with the energy.

\begin{figure}[htbp]
\begin{tabular}{c c}
 $\mathbf{Q^2 = 1 \,GeV^2 }$\,\, $\mathbf{ R^2 = 5\,GeV^{-2}}$ &
$\mathbf{Q^2 = 2.5\,GeV^2 }$\,\,$ \mathbf{R^2 = 5\,GeV^{-2}}$  \\
  \epsfig{file=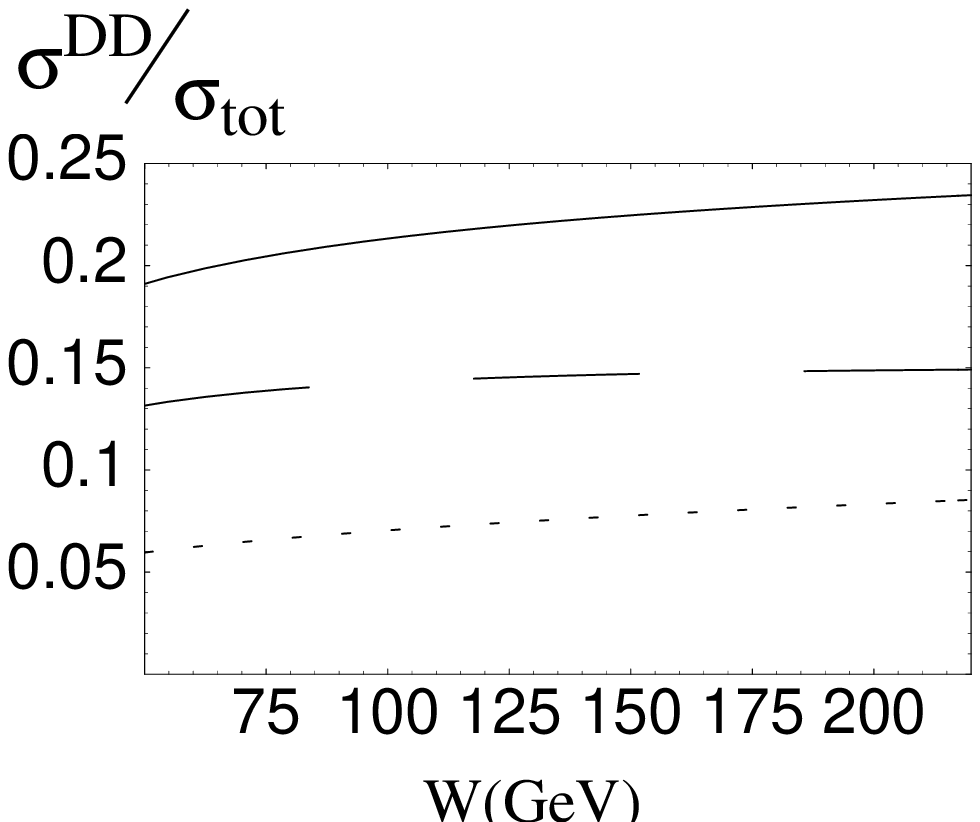,width=80mm, height=60mm}&
 \epsfig{file=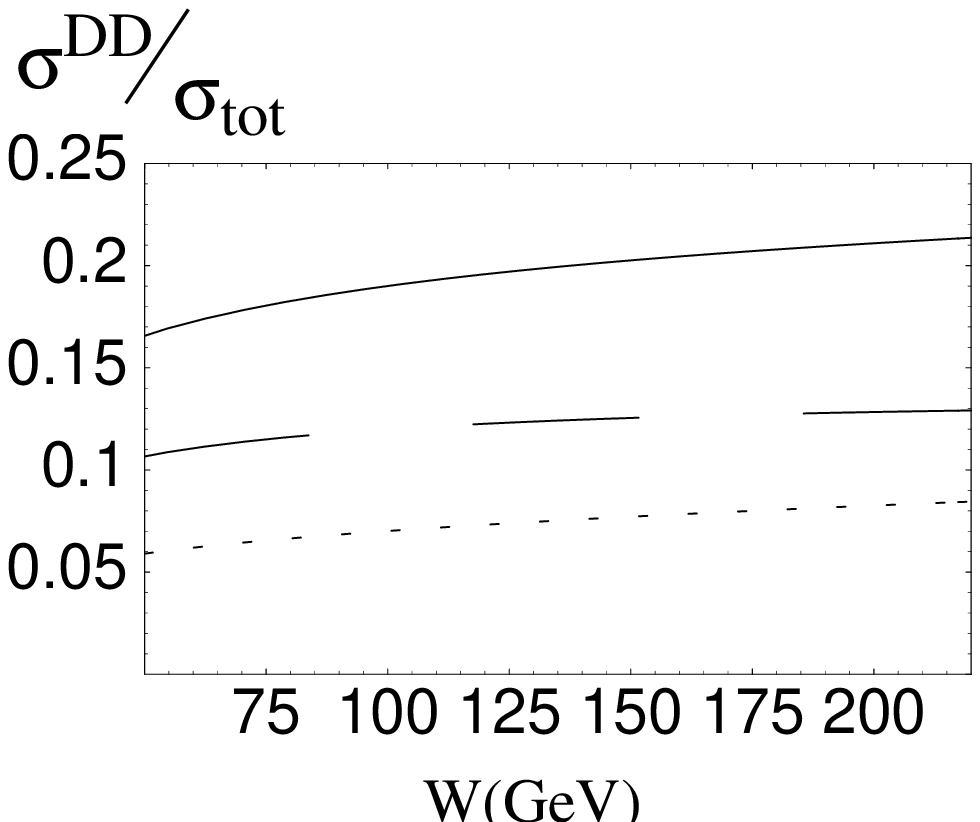,width=80mm, height=60mm}\\
& \\
$\mathbf{Q^2 = 8 \,GeV^2 }$\,\,$\mathbf{ R^2 = 5\,GeV^{-2}}$ &
$\mathbf{Q^2 =
14\,GeV^2 }$\,\,$\mathbf{ R^2 = 5\,GeV^{-2}}$ \\
 \epsfig{file=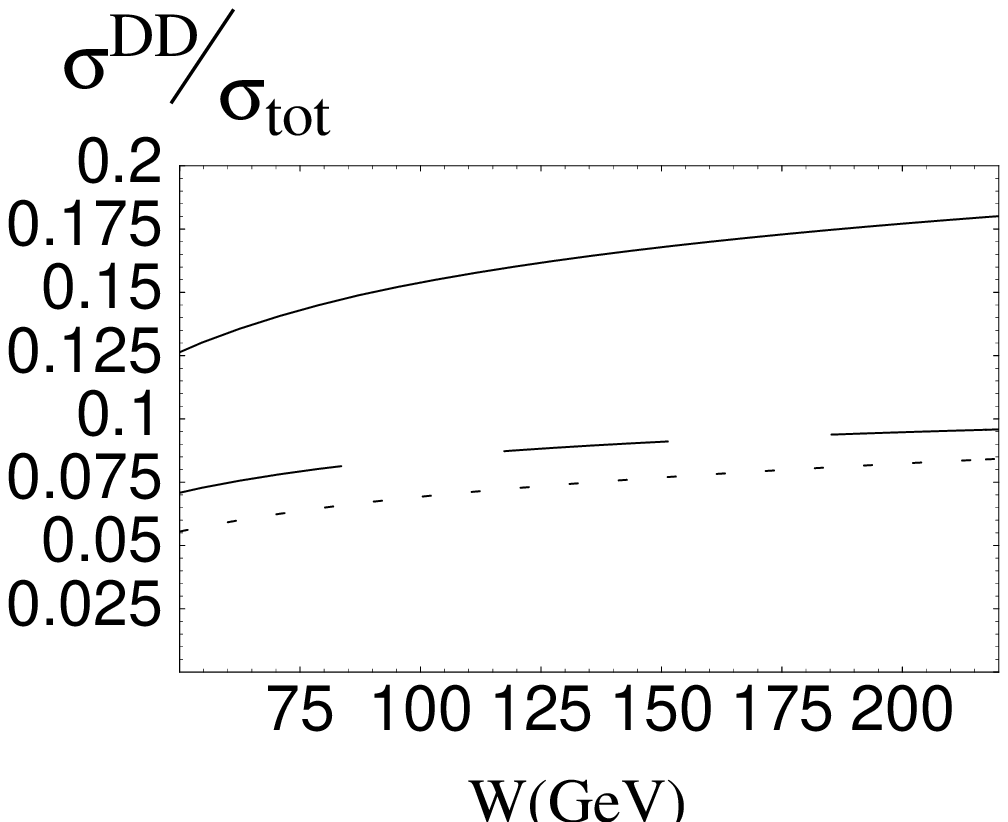,width=80mm, height=60mm}&
\epsfig{file=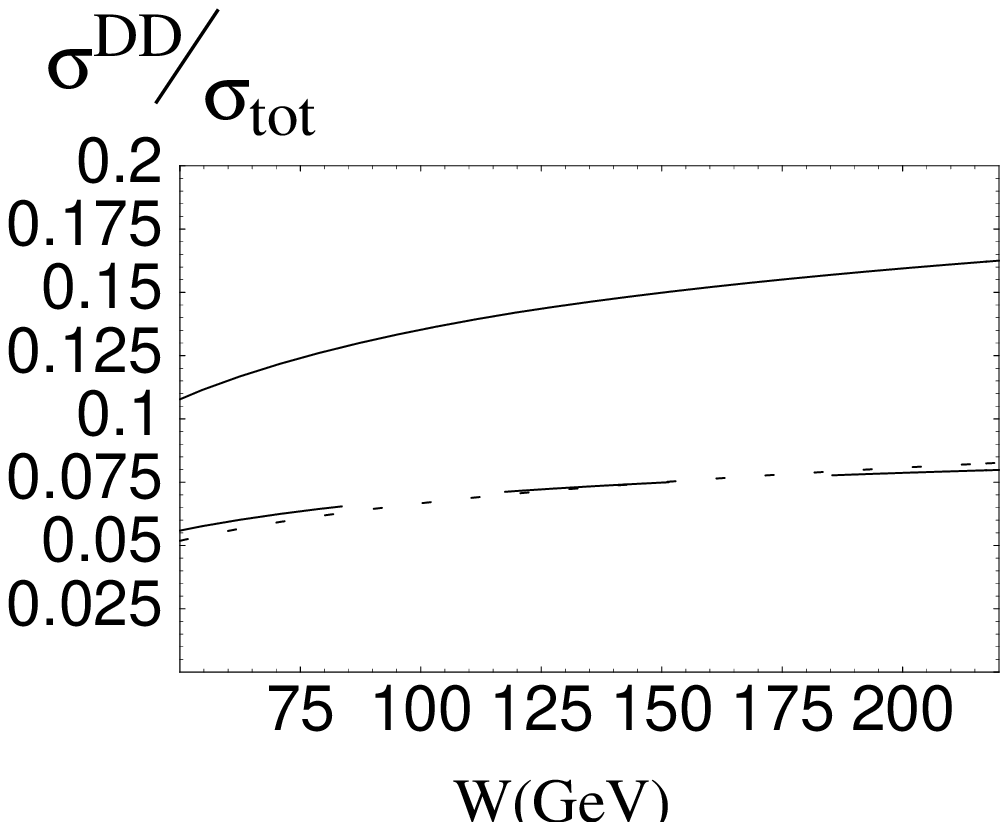,width=80mm, height=60mm}\\ & \\
 $\mathbf{Q^2 = 27 \,GeV^2 }$\,\, $\mathbf{ R^2 = 5\,GeV^{-2}}$ &
$\mathbf{Q^2 = 60\,GeV^2 }$\,\,$\mathbf{ R^2 = 5\,GeV^{-2}}$  \\
\epsfig{file=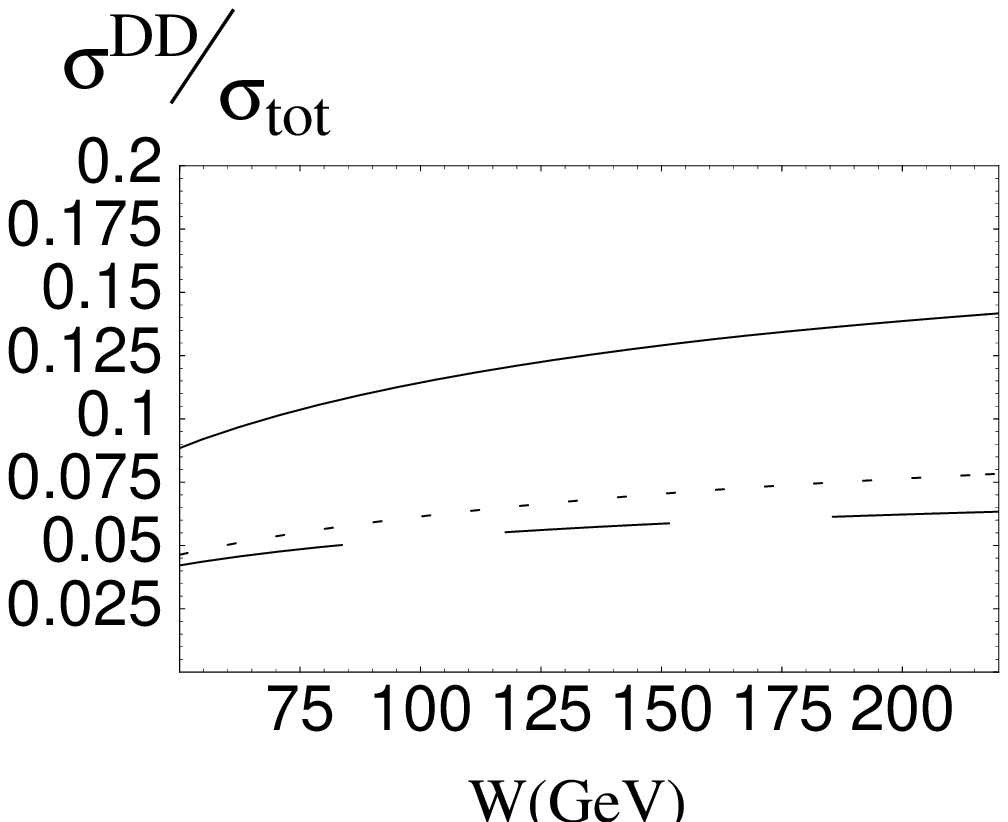,width=80mm, height=60mm}&
\epsfig{file=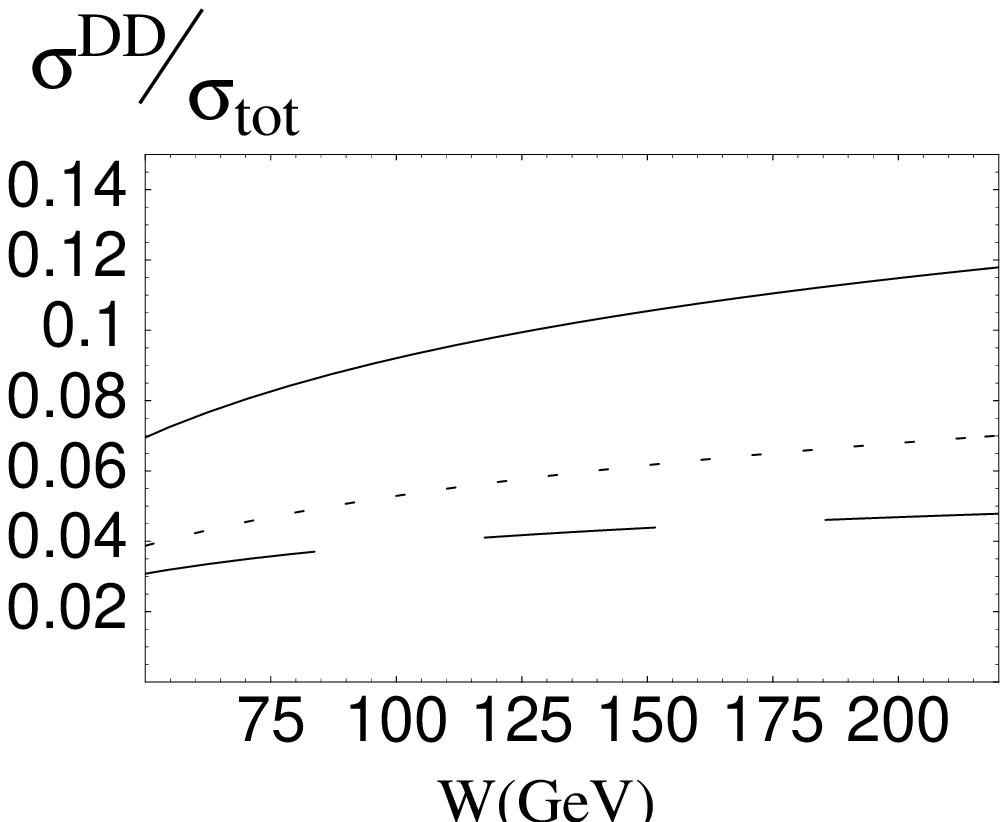,width=80mm, height=60mm}\\
\end{tabular}
  \caption[]{\it  Ratio $\sigma^{DD}/\sigma_{tot}$ versus $W$ for $R^2\,=\,5 \,{\rm GeV}^{-2}$. }
\label{fig14}
\end{figure}

In Fig. \ref{fig14} we illustrate the dependence of our calculation
 on the value of
$R^2$. We would like to recall that  two values of the radius $R$, which 
we  use, have the following physics behind them:
\begin{enumerate}
\item\,\,\,$R^2 \,=\,10\,GeV^{-2}$ is related to the Mueller-Glauber approach,
which corresponds to the Eikonal model for the nucleon target\,;
\item\,\,\,$R^2 \,=\,5\,GeV^{-2}$ is the average radius for the two
channel model which has been discussed in subsection 2.2.3(B)\,;
\end{enumerate}

One can see from Fig. \ref{fig14} that the value of the ratio
$\sigma^{DD}/\sigma_{tot}$ depends on $R^2$. However, the energy
dependence is still  very pronounced.

Therefore, the general conclusion of this section is that the SC fail to
reproduce  the constant ratio of $\si^{DD}/\si_{tot}$ for DIS  seen in
the HERA kinematic region.  

\section{Ratio  $\mathbf{ \sigma^{DD}/\sigma_{tot}}$ in the mass windows}
\setcounter{equation}{0}

In this section we examine a possibility that the mass interval will
induce an energy independent ratio $\si^{DD}/\si_{tot}$. As one can see in
Fig.1, the experimental measurements were made within some windows in
mass.

The full derivation of the mass dependent formulae is presented in the Appendix.
 The cross sections for the diffractive dissociation
production of   $q\bar{q}$ pair  and $q\bar{q}G$ parton system with a definite final state mass
are given by \eq{4} and \eq{10} respectively. Any mass window can be selected for the
mass integrals.

If summation over the whole infinite
mass intervals is performed, our formulae for the diffractive dissotiation as well as for the total
cross sections should in principle reproduce \eq{FIN}. However, the two sets
of formulas are different but consistent with each other  in the
leading $ \log(1/x) $ approximation where $ \log(1/x_P) \simeq \log(1/x_B)$. 
 When $x_P$
is replaced by $x_B$, \eq{4} and \eq{14} reproduce analytically
 (and numerically)
the corresponding expressions in \eq{FIN}. In the numerical computations
 we use $x_P$  instead of $x_B$ since this energy variable reflects the
real kinematics of the diffraction production process.
Since $x_P = x_B/\beta$ and the typical values of $\beta$ is not very
small for $q \bar q $ system, we do not expect large corrections due to
this substitution. The $ q \bar q G $
channel is more sensitive to the kinematic restriction (see \eq{10}).
 The changes, introduced in \eq{10}, 
concern both the energy variables as well as the integration limits.
These  cannot be justified in log(1/x) limit which we used in our
general formulae, but we have to introduced them to make our calculation
 reasonable for the diffractive production in the mass window.

\begin{figure}[htbp]
\begin{tabular}{c c}
$\mathbf{Q^2 = 8 \,GeV^2 }$ & $\mathbf{Q^2 = 14\,GeV^2 }$ \\
 \epsfig{file=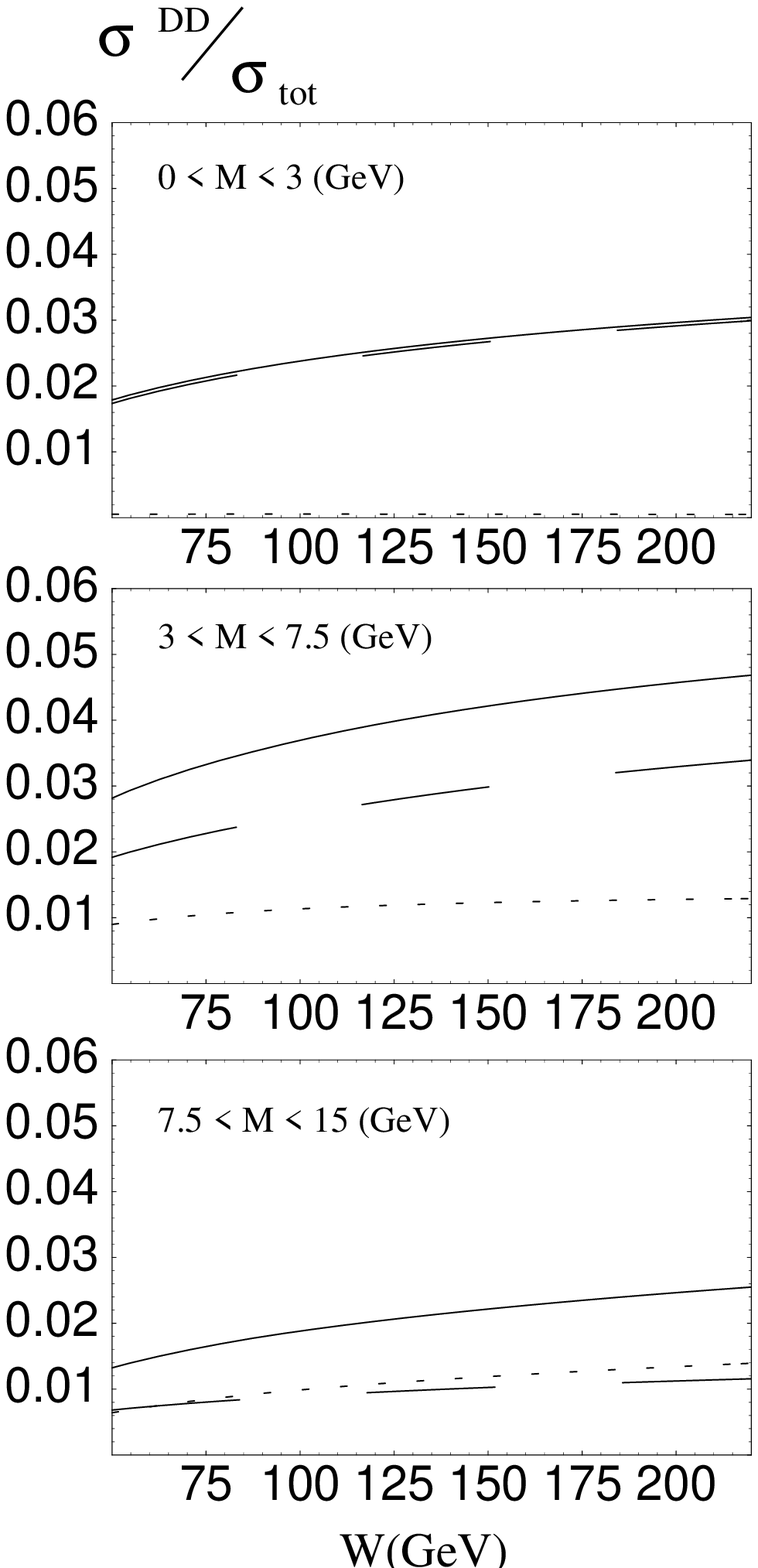,width=70mm, height=90mm}&
\epsfig{file=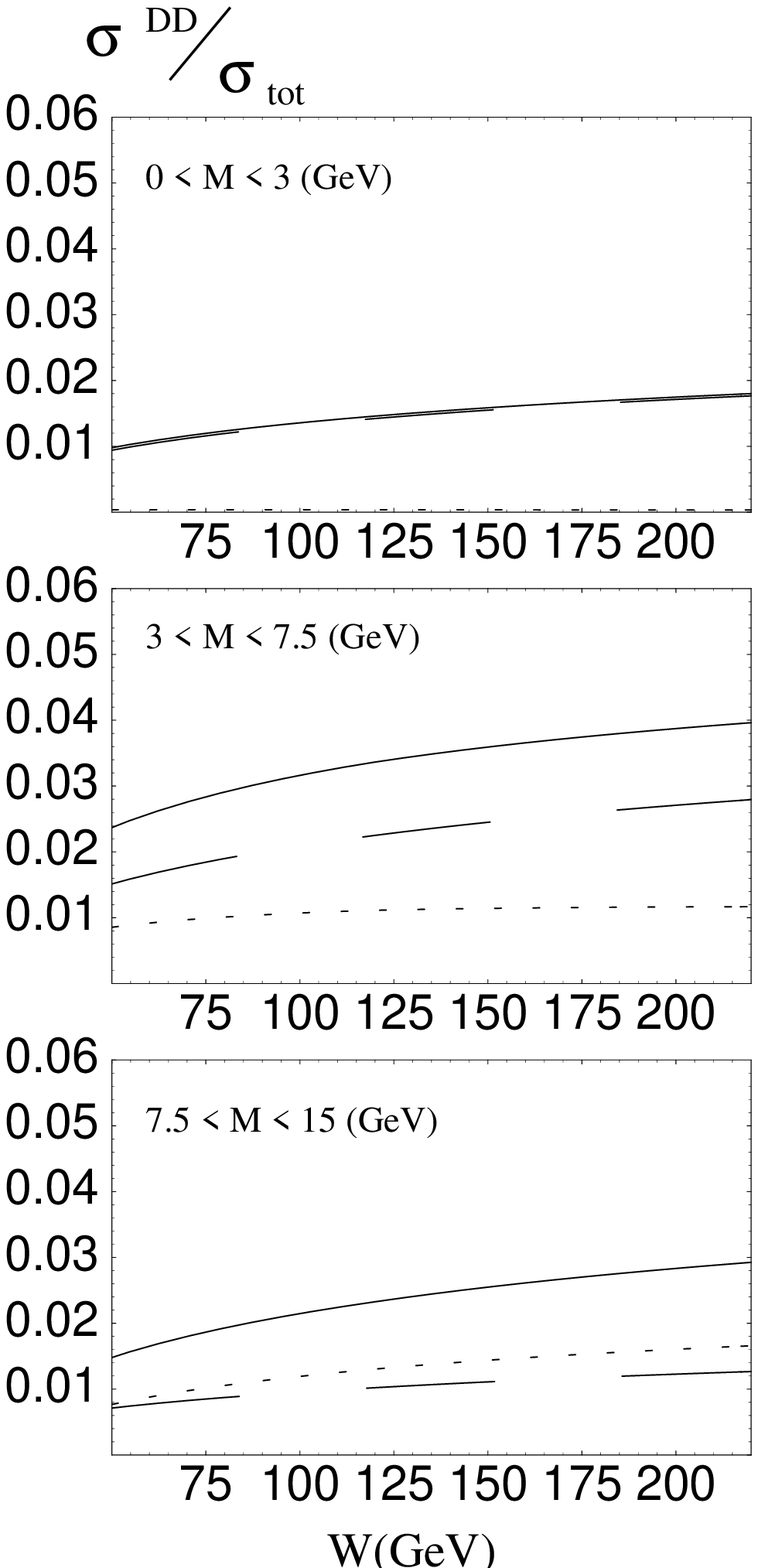,width=70mm, height=90mm}\\ 
$\mathbf{Q^2 = 27 \,GeV^2 }$ & $\mathbf{Q^2 = 60 \,GeV^2 }$ \\
 \epsfig{file=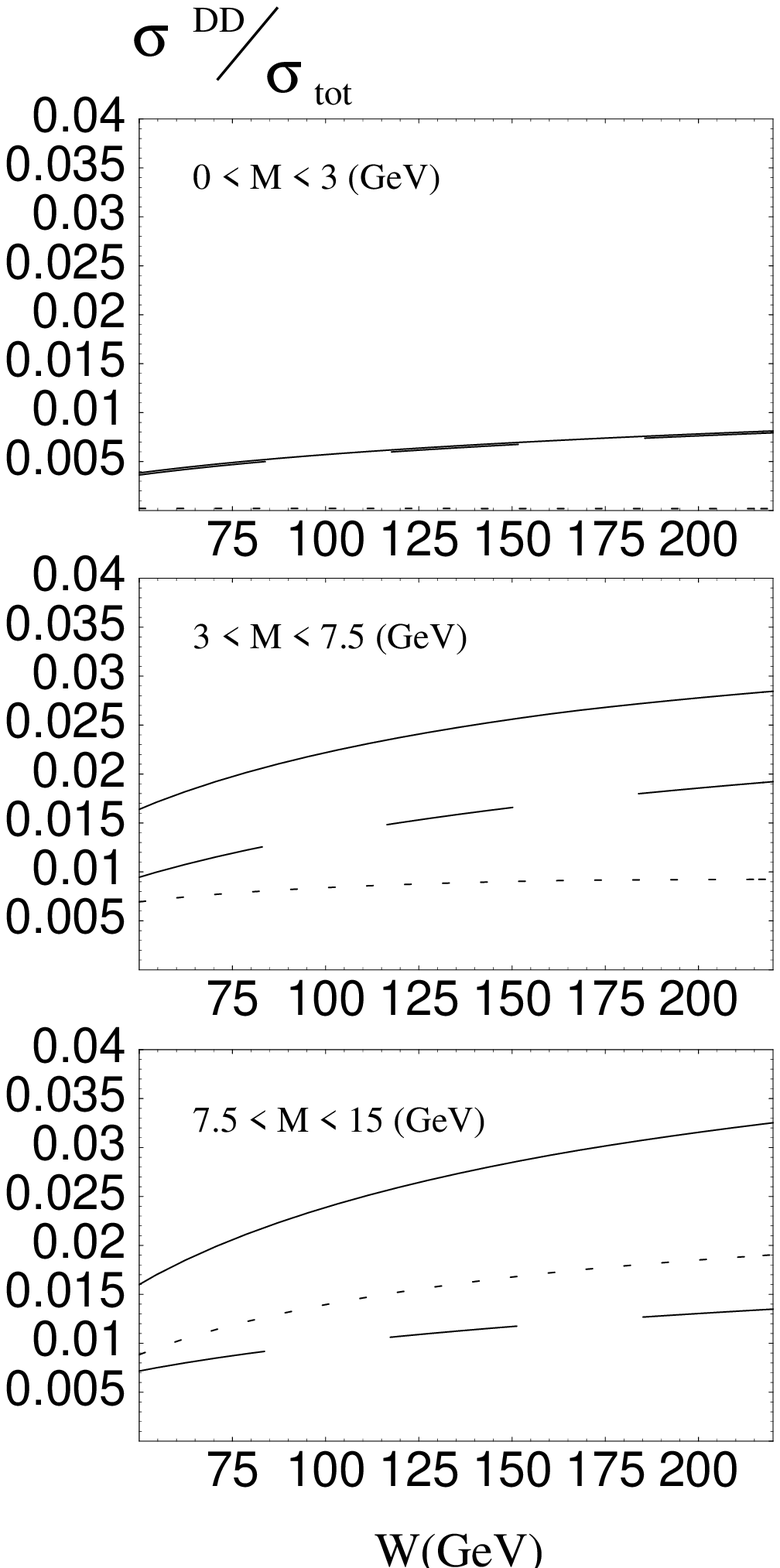,width=70mm, height=90mm}&
\epsfig{file=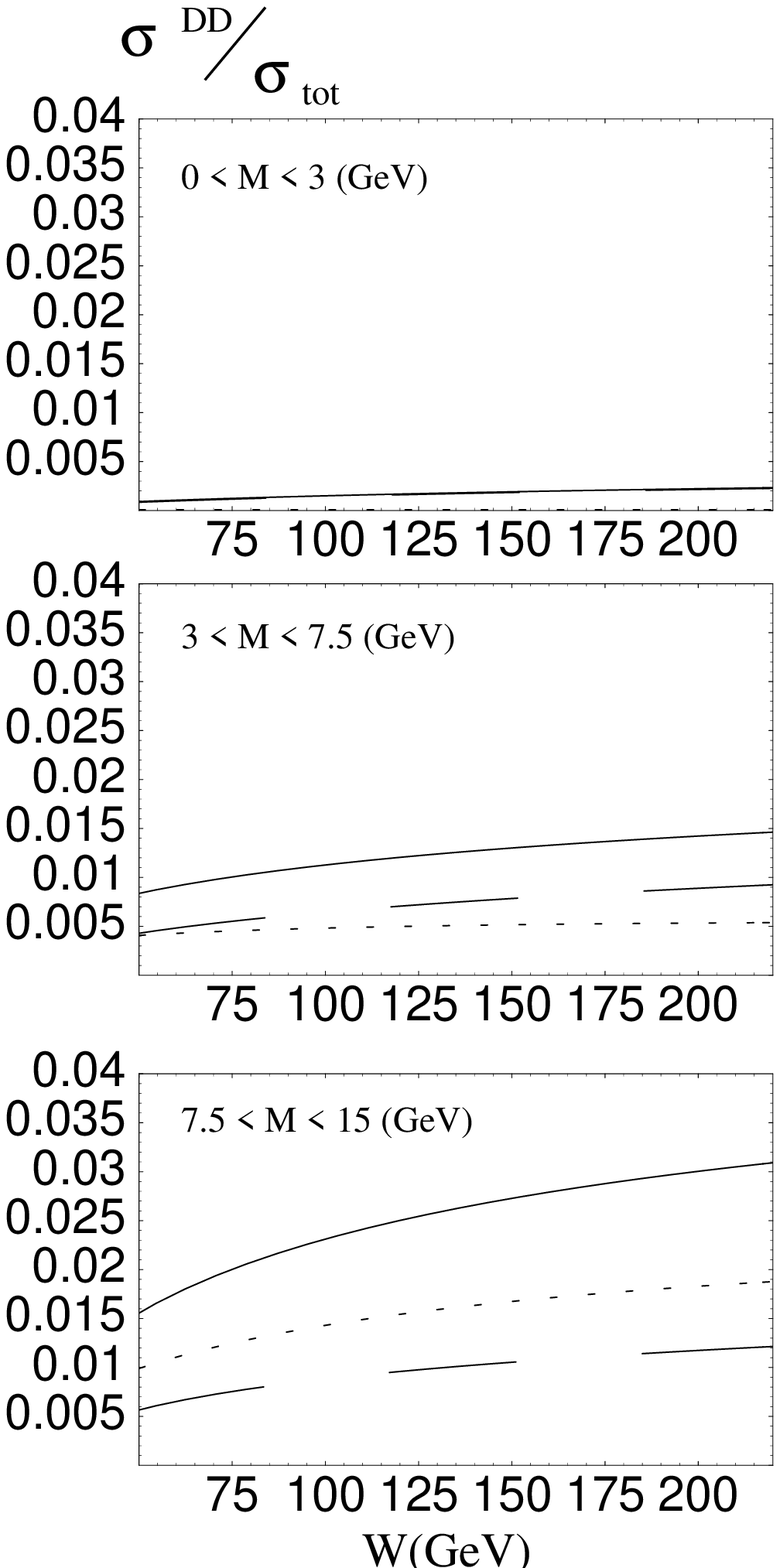,width=70mm, height=90mm}\\ 
\end{tabular}
  \caption[]{\it  The ratio $\sigma^{DD}/\sigma_{tot}$ as a function of $W$ for
differen mass bins ( $R^2 = 10\,GeV^{-2}$  ).
The upper solid line corresponds to the full answer, the widely dashed
 line to diffractive production of a $q \bar
q $ pair,  while the narrowly dashed line to the  diffractive production
of $q \bar q G $ . }
\label{figM}
\end{figure}

Figure \ref{figM}  presents our results for the  mass bins,
for which   experimental data exists (Fig. 1).   The ratio of the
diffractive dissociation cross section to the inclusive cross section is
plotted as a function
of the center of mass energy. The $q\bar q$ pair (tranverse plus longitudinal)
and $ q\bar q G$  contributions are shown separately. 

The  ratios obtained do not reproduce the  experimental data (Fig. 1). The
significant energy
dependence persists due to the growth of both  the  $q\bar q$ and $ q\bar
q G$  contributions.
In the wide range of the energies  our results are smaller
 than the experimental curves.
This observation is quite consistent, since our model excluded the target
excitations estimated
in the Chapter 2 by about 30\%.

As it should be, at small masses the main contributions come from the  $q\bar q$ pair with
more than 50\% given  by the longitudinal part. The $ q\bar q G$ production is suppressed
at small masses. Its contibution grows with the mass and dominates at large masses.

Summing  all results up to the mass $M=15$(GeV) we do not reproduce
the  inclusive  mass
results of the previous section. This means that even for $Q^2=8 {\rm GeV^2}$ we expect
contributions from higher masses.  At  $Q^2=8 ({\rm GeV^2})$ these
contributions  originated from
 $ q\bar q G$ production. At $Q^2=60 ({\rm GeV^2})$
both $q\bar q$  and $ q\bar q G$ will be significant above $M=15$(GeV). It is seen that only
about 50\% of the inclusive DD production is contributed by small masses up to $M=15$(GeV).  

We wish to point  that we have completely
disregarded  possible final state corrections. As an example of such corrections, the 
master formula \eq{FIN} does not
take into account the second diagram in  Fig. 2. This diagram is
believed to give a relatively small contribution compared to the first and
the third diagrams. 

\begin{figure}[htbp]
\begin{tabular}{c c}
 \epsfig{file=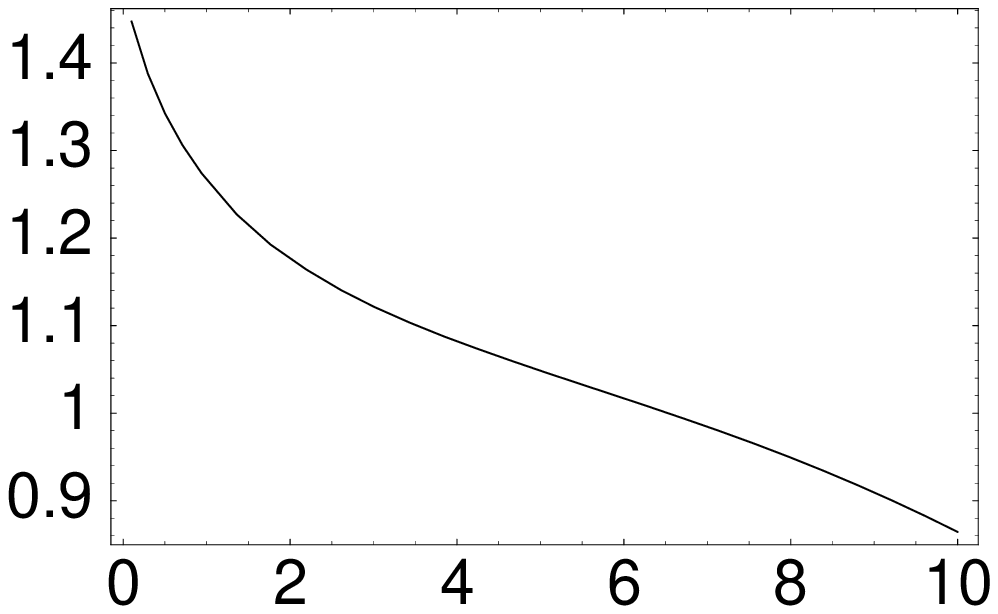,width=80mm, height=50mm}&
\epsfig{file=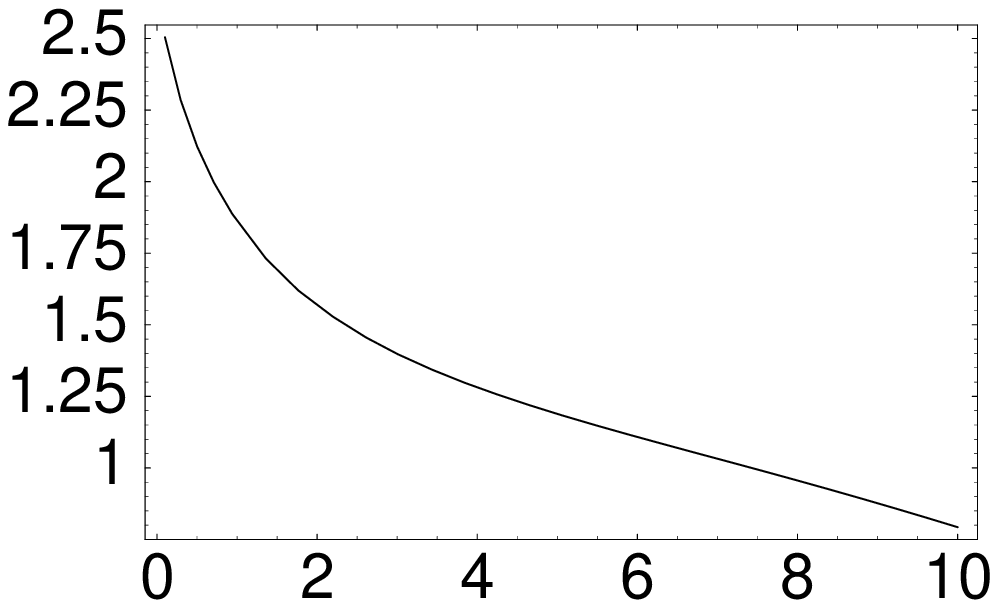,width=80mm, height=50mm}\\
 a) & b)\\
\epsfig{file=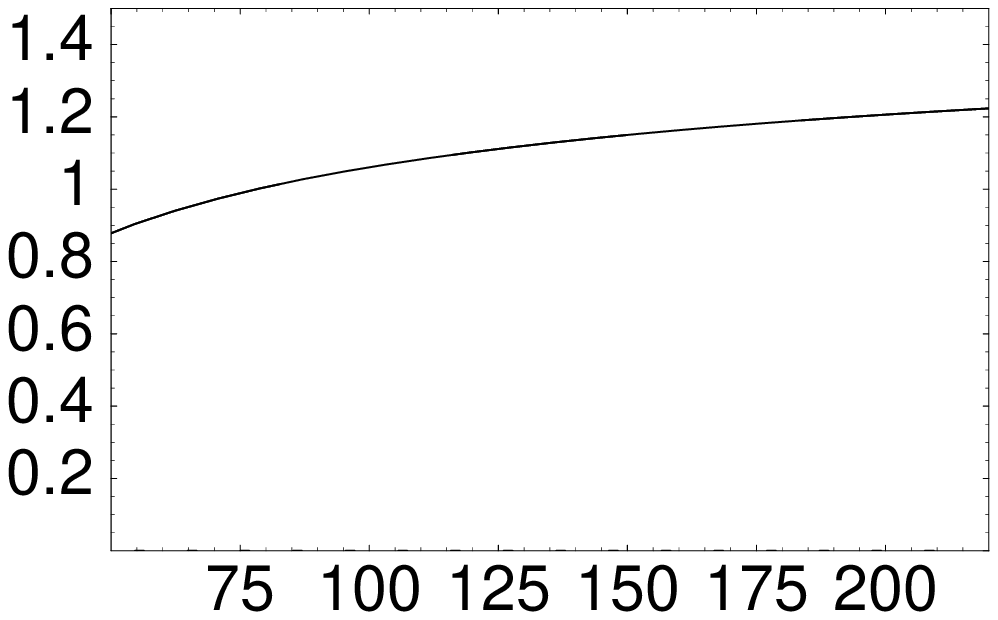,width=80mm, height=50mm}&
\epsfig{file=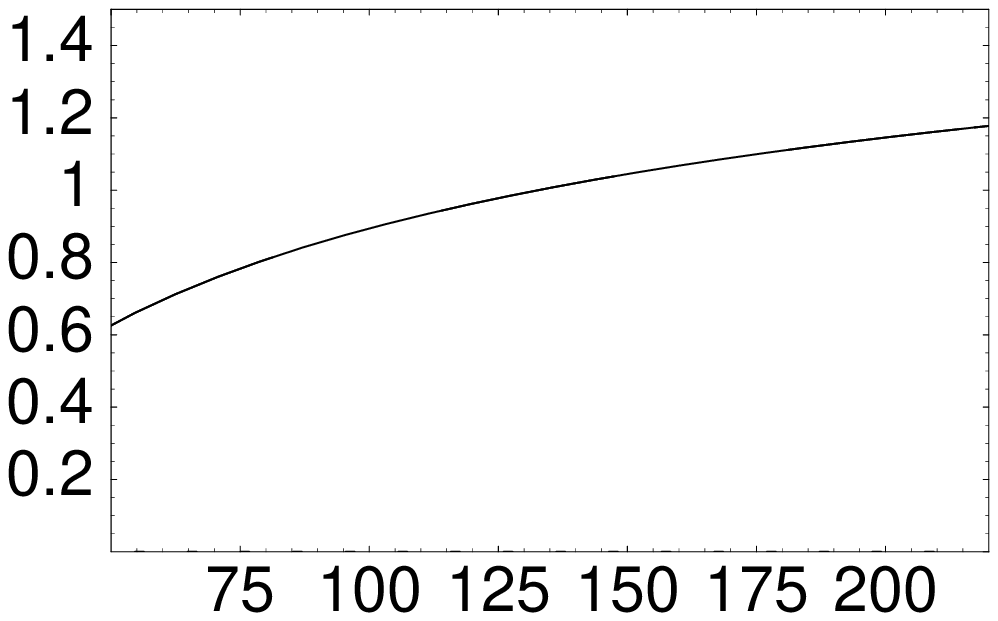,width=80mm, height=50mm}\\
c) & d) \\
\epsfig{file=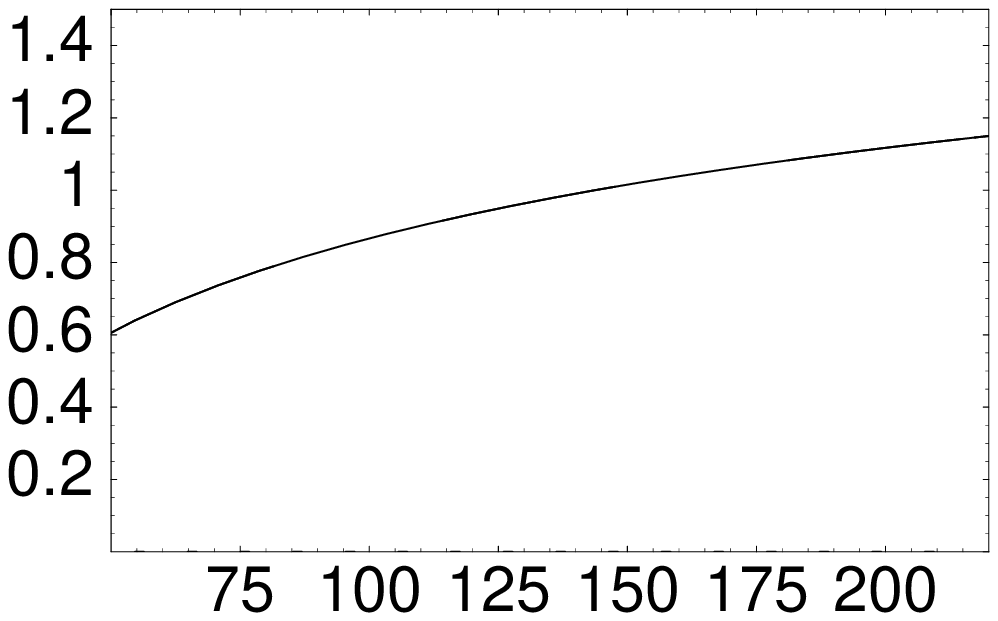,width=80mm, height=50mm}&
\epsfig{file=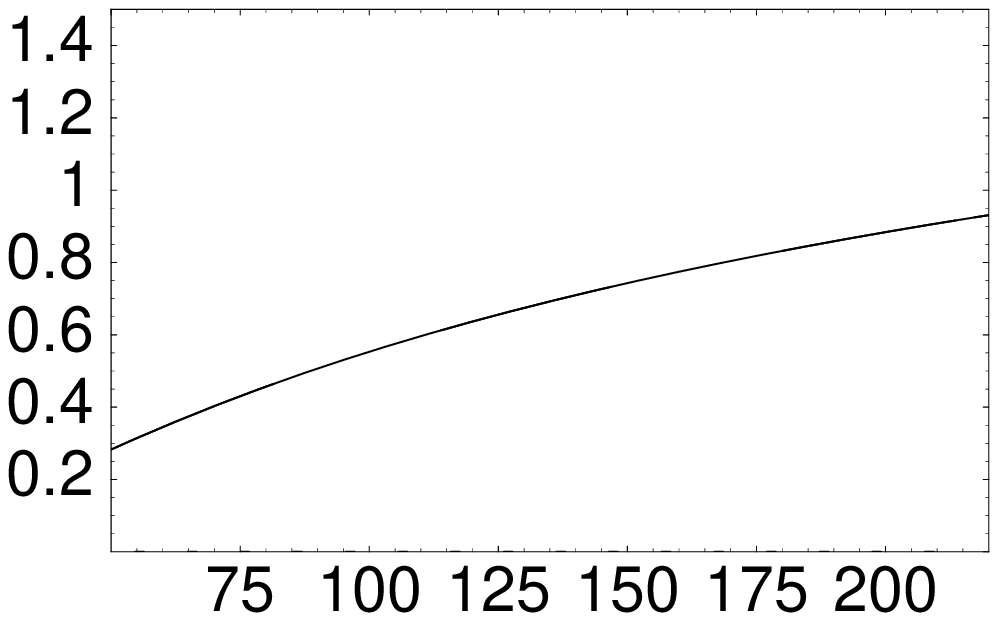,width=80mm, height=50mm}\\
 e) & f)\\
\end{tabular}
  \caption{\em A comparison between G-W model and our approach:
  a) The ratio of our calculation of the dipole cross sections to G-W
model  versus $r^2_{\perp}$. b) The same as Fig.16a but for the
diffractive
dissociation cross sections. All the graphs are plotted for $x=10^{-3}$.
c) The ratio of the total cross sections for $Q^2 = 8\,GeV^2$ as function
of energy $W$. d) The ratio of the diffractive cross sections for $Q^2 =
8\,GeV^2$ as a function of energy $W$. e) and f) the same as c) and d) 
for $Q^2 = 60\,GeV^2$. }
 \label{figWus}
\end{figure}

It is worthwhile comparing  our model with 
the Golec-Biernat  Wusthoff model \cite{GOWU1},
which successefully reproduces the experimental
data (Fig. 1). In the Golec-Biernat  Wusthoff model  
the effective dipole cross section
$\sigma_{GW}(x,r)$,
describing the interaction of the $q\bar q$ pair with a nucleon has the form:
\begin{eqnarray}
\sigma_{GW}(x,r_{\perp})\,=\,\si_0\,[1\,-\,\exp
(-r^2_{\perp}/(4\,R_0^2(x)))];
\,\,\,\,\,\,\,R_0(x)\,=
\, (x/x_0)^{\lambda/2}
\,({\rm GeV^{-1}}); \label{Wus} \\
 \si_0\,=\,23.03\, ({\rm mb});\,\,\,\, \,\,\lambda\,=\,0.288;\,\,\,\,\,\,
\, x_0\,=\,3.04\,\cdot \,10^{-4}.
\nonumber
\end{eqnarray}
In this model, the diffractive dissociation cross section is given by the
squaring of
$\sigma_{CW}$ in \eq{Wus}:
\begin{equation}
\sigma_{GW}^{DD}(x,r_{\perp})\,=\,\sigma_{CW}^2/(16\,\pi\,
B_D);\,\,\,\,\,\,\,
B_D\,=\,6\,{\rm GeV^{-2}}.
\label{WusDD}
\end{equation}
A comparison between CW model and the present work model is presented 
in Fig. \ref{figWus}.

We  found a significant difference between the two models .
The advantage of G-W model is that this model takes into account in the
simplest way a new scale: saturation momentum $Q^2_s(x) = 4/R^2_0(x)$, but
in doing so, this model loses its correspondence
 with the  DGLAP evolution
equation. Our approach has a correct matching with the DGLAP evolution and
we expected that we would be able to describe experimental data better
than
the G-W model. It turns out ( see Fig.16-a and Fig. 16-b ) that we do not
reproduce the $\sigma^{DD}/\sigma_{tot}$ ratio in contrast to the G-W
model,  mostly due to our `improvement' in the region of small
$r^2_{\perp}$.

The second remark is the 
 substantial difference in the way  we describe the $q
\bar q G  $ state. We  failed to find a correspondence between
 our formula for $q \bar q G$ production which follows from the AGK
cutting rules, and the G-W description of this process. However, 
 our failure to fit the experimental data is mostly due to a
large difference in the dipole  cross section, rather than in the 
different   
treatment of the   $q \bar q G$ diffractive production.

\begin{figure}[htbp]
\begin{tabular}{c c}
 \epsfig{file= 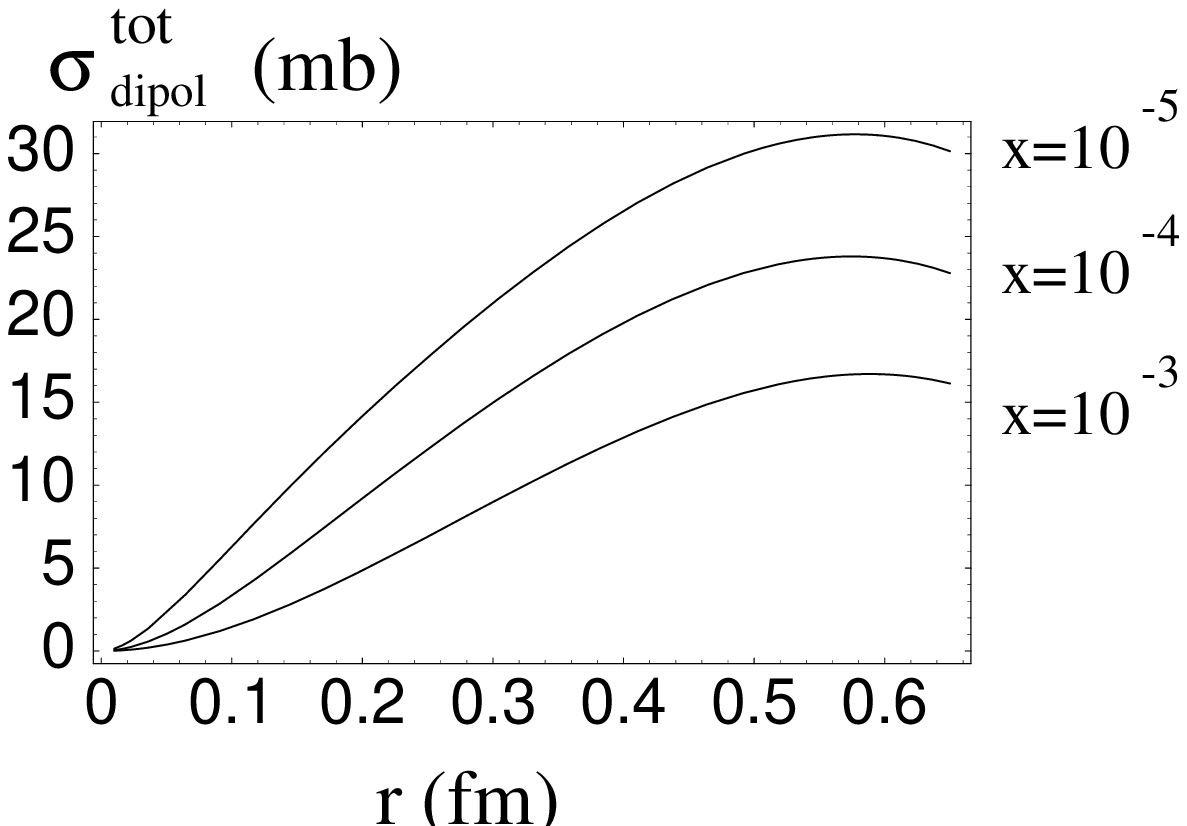,width=80mm, height=45mm}&
\epsfig{file=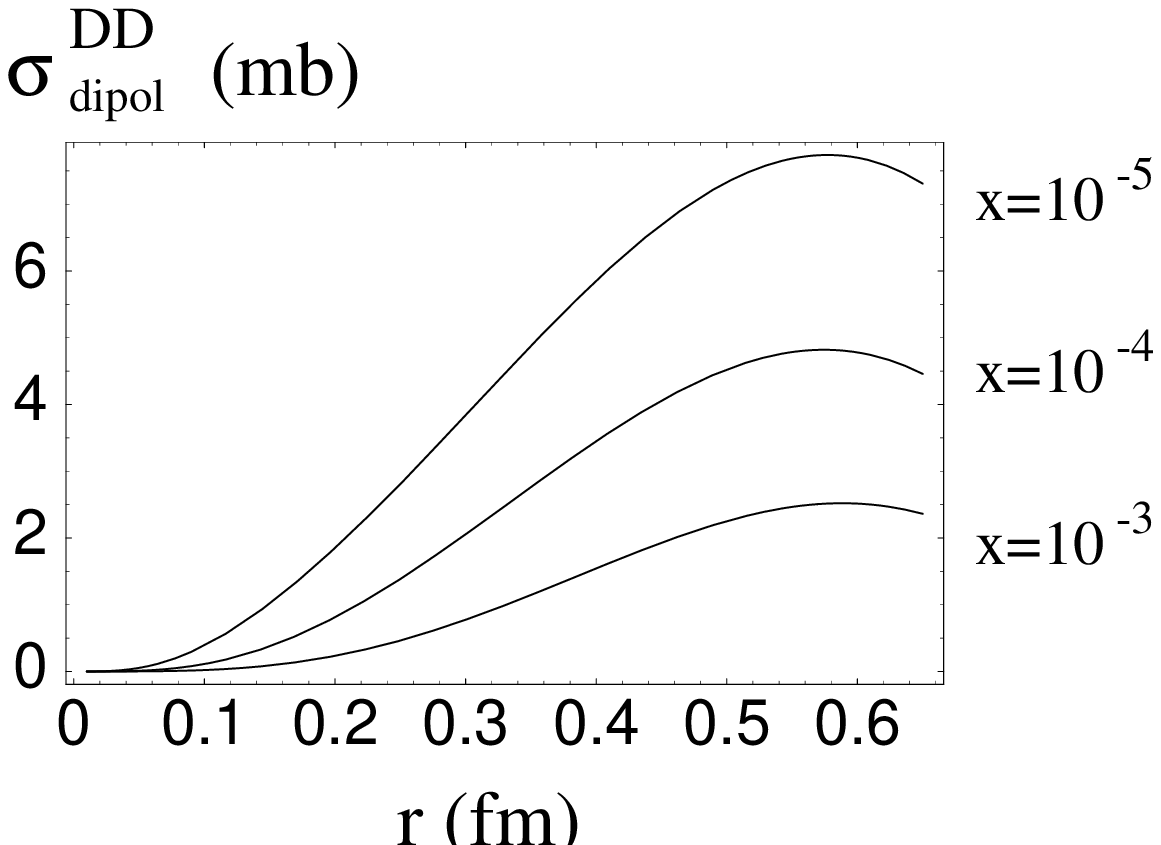,width=80mm, height=45mm}\\
 Fig.17-1 & Fig.17-2 \\
\epsfig{file=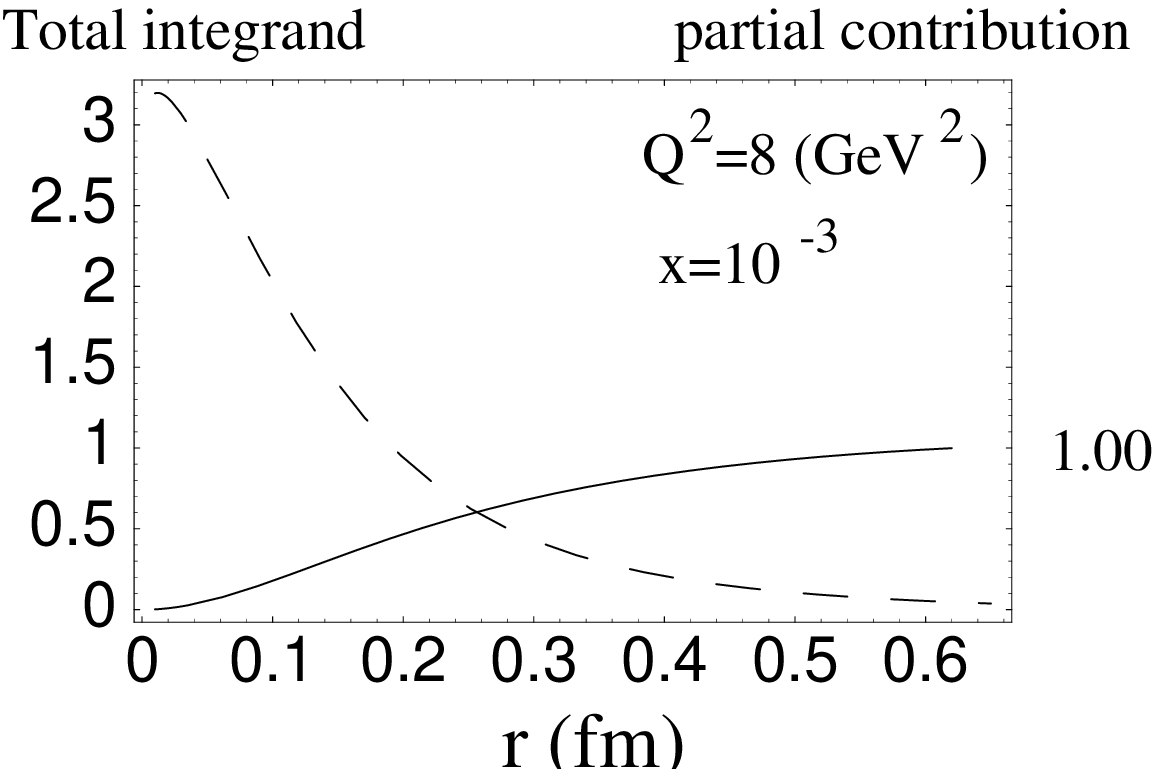,width=70mm, height=45mm}&
\epsfig{file=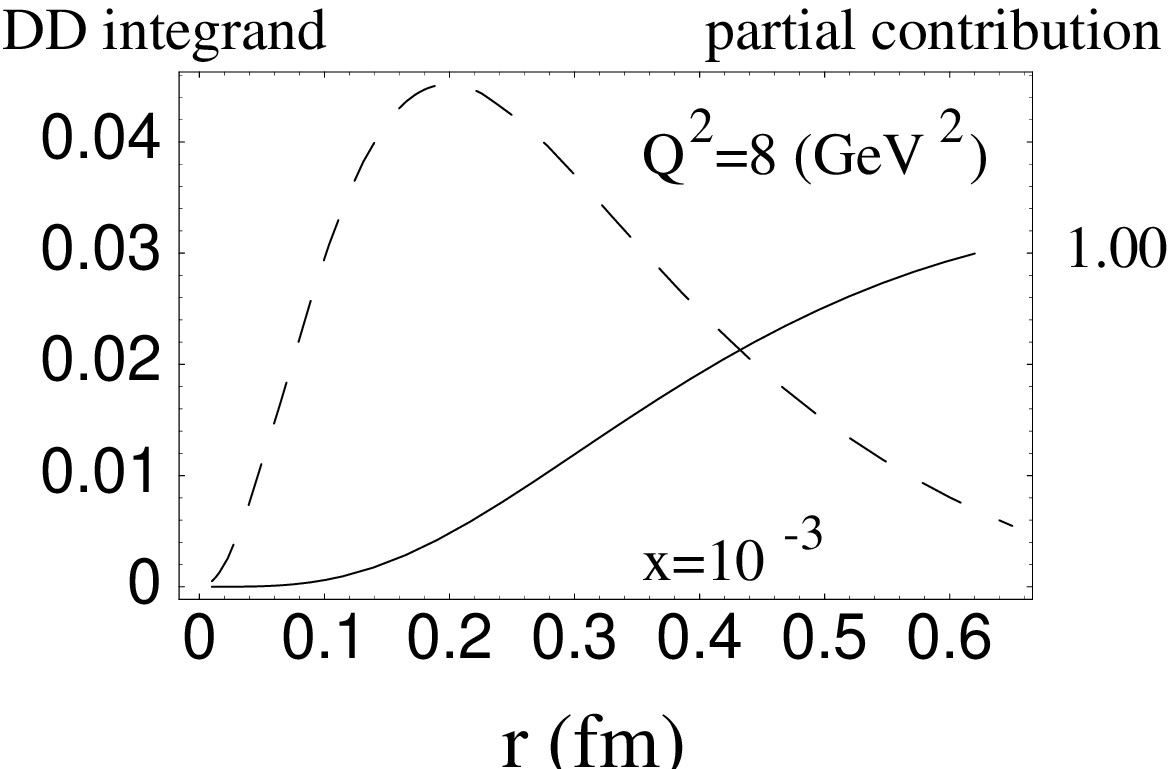,width=70mm, height=45mm}\\
Fig.17-3  & Fig.17-4 \\
\epsfig{file=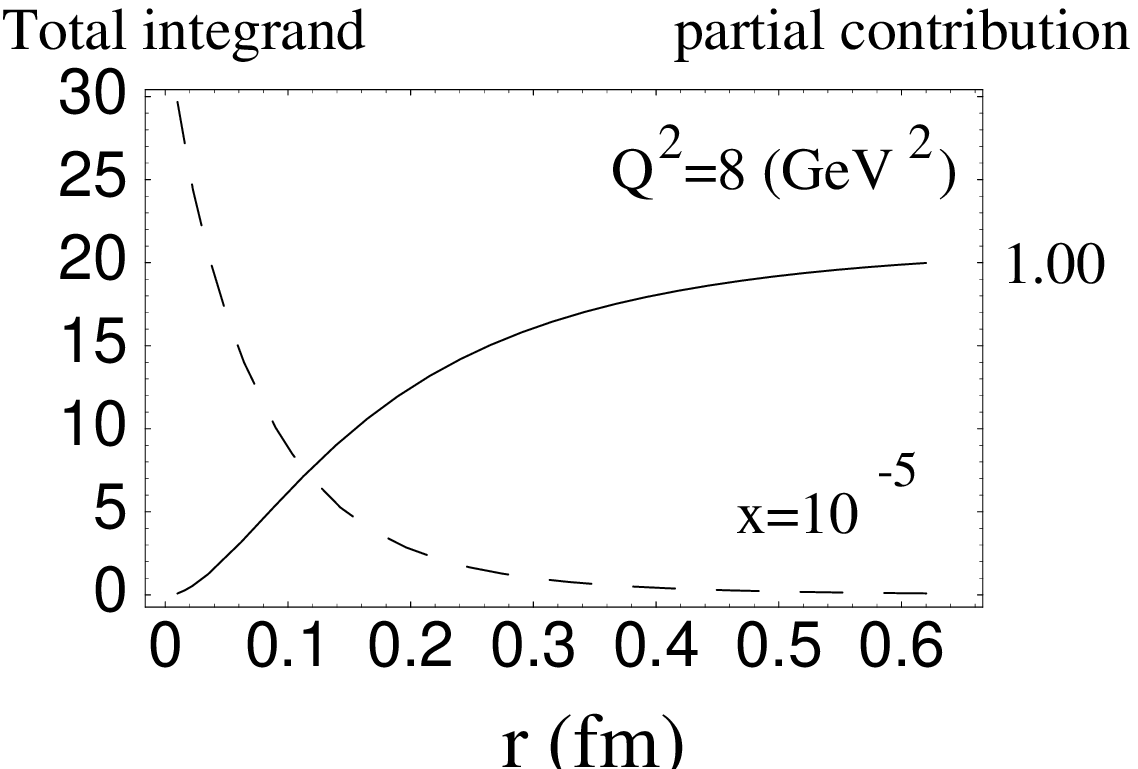,width=70mm, height=45mm}&
\epsfig{file=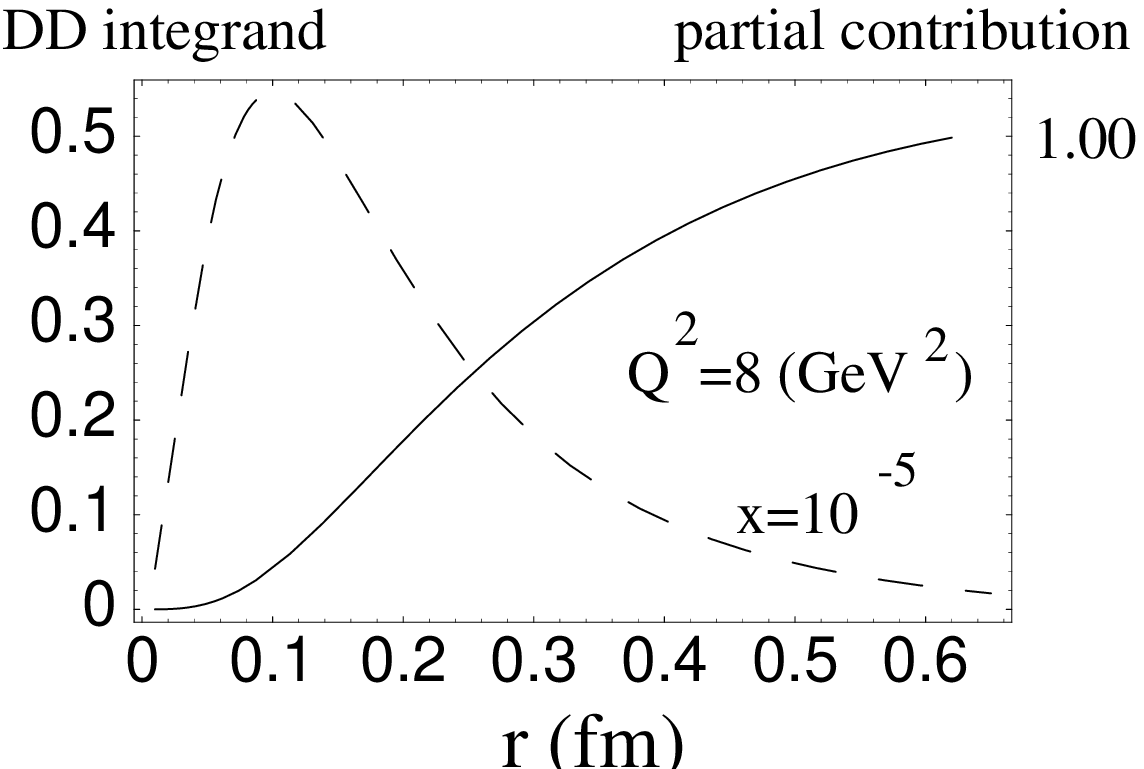,width=70mm, height=45mm}\\
Fig.17-5 & Fig.17-6 \\
\epsfig{file=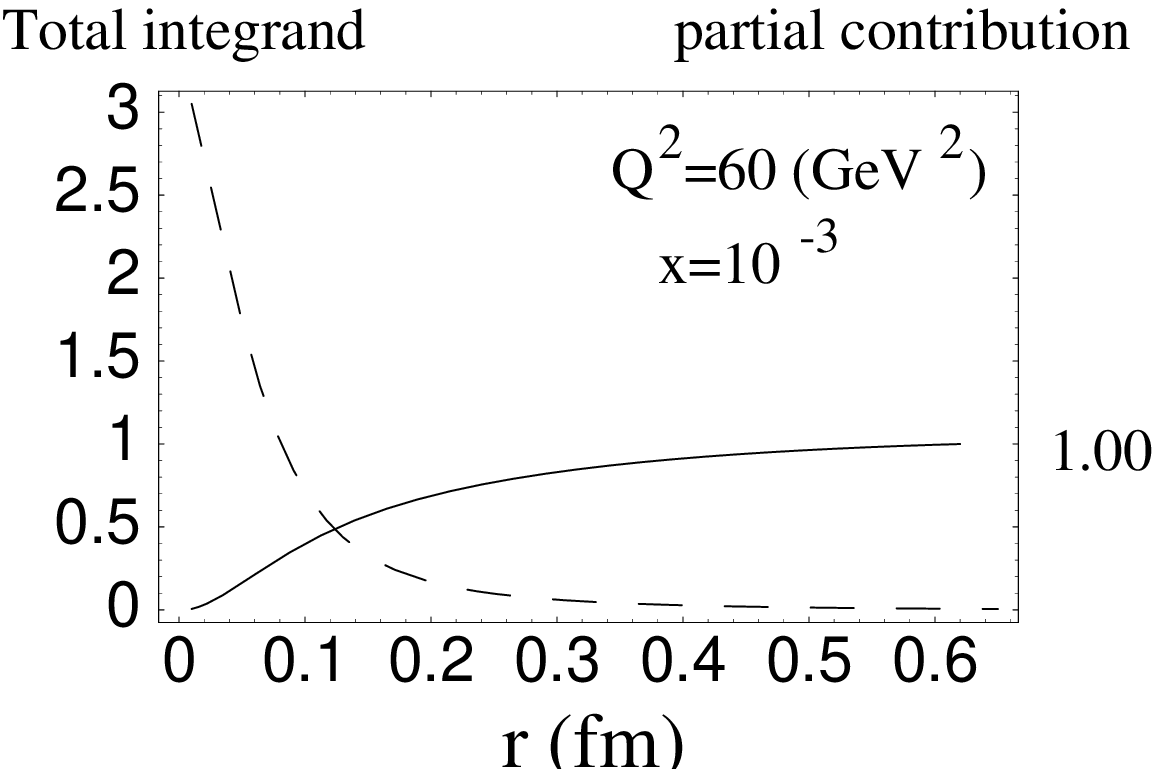,width=70mm, height=45mm}&
\epsfig{file=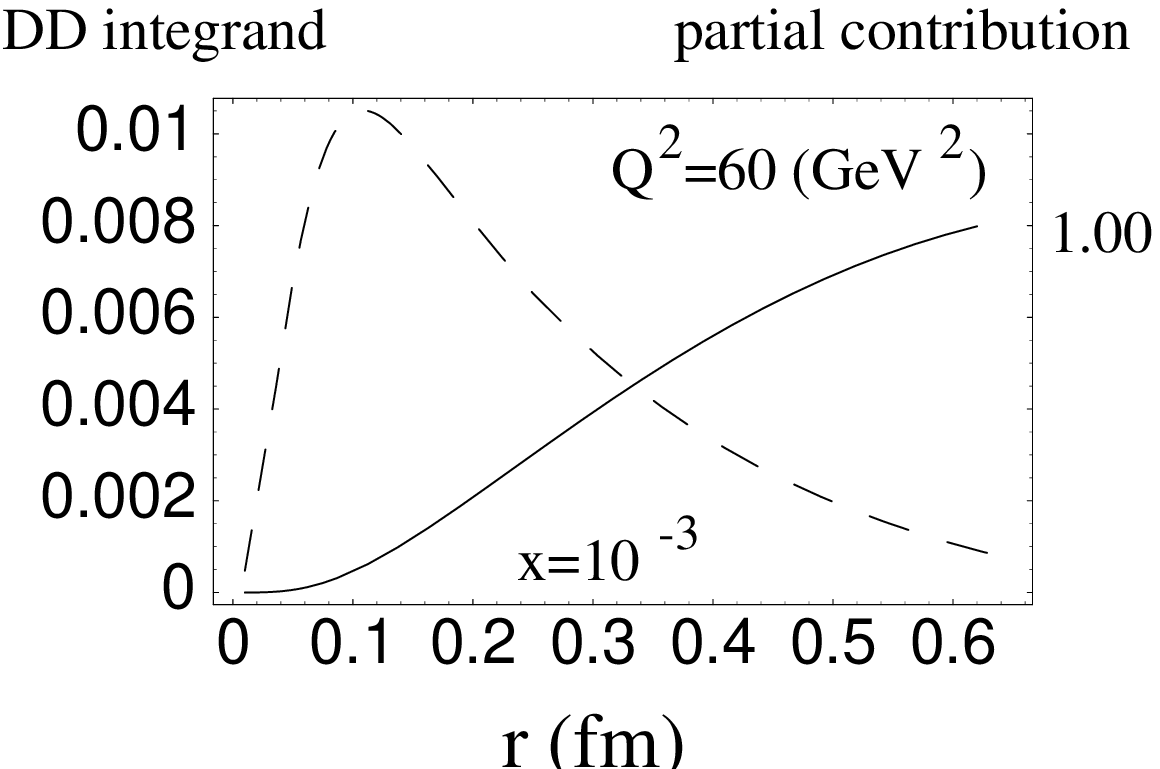,width=70mm, height=45mm}\\
 Fig.17-7 & Fig.17-8\\
\end{tabular}  
  \caption{\em $\sigma_{dipole}^{tot}(r_t;x)$ ( Fig.17-1 ) and
$\sigma^{DD}_{dipole}(r_t;x)$ (Fig.17-2) versus $r_t$ at different values
of $x$. All others figures are the integrants and partial contributions to
the cross sections versus distances $r_t$ at diffeent values of $x$ and
$Q^2$.}
 \label{figparr}
\end{figure}

Figs.17-1 and 17-2 show our dipole cross sections at different energies.
The main difference with the Golec-Biernat and Wusthoff model is the
energy rise of the cross sections which follows from  $b_t$ dependance 
included in the Eikonal formulae but neglected in the Golec-Biernat and
Wusthoff model. Fig.17-3 - Fig.17-10 show whast distances are essential in
our calculations. One can see that  the main contributions in diffractive
cross sections stem from longer distances than in the total cross sections
as have been predicted theoretically \cite{TEOID}. As has been known for
long time the typical distances in diffractive cross section is of the
order of the saturation scale $Q_s(x)$  ($\kappa_{dipole}( r_t =
2/Q_s(x))=1$ ) in contrast with total cross section were typical
distances are much shorter , about $1/Q$.. Therefore, one of the reason
why we failed to describe the
ratio of interest could be that our model cannot describe the dipole cross
section in the vicinity of the saturation scale. However, it was
demonstrated in AGL papers ( see Ref. \cite{TEOID} ) that our Eikonal
model gives a good approximation to the correct non-linear evolution
equations at that particular distances which are  essential accordingly to
Fig.17.

\section{Summary and discussions}

In this paper we developed an approach to  the diffractive dissociation in
DIS , based on three main ideas:
\begin{enumerate}
\item\,\,\, The dominant contribution to diffractive dissociation
processes in DIS stems from rather short distances \cite{GLMSM} and,
therefore, we can use pQCD to describe them;
\item\,\,\, The final state in  diffractive dissociation can be simplified
in the HERA kinematic region by only considering
 the production of $q \bar q$ and
$q \bar q G$ (\cite{GLMSM} );
\item\,\,\, The shadowing corrections which are essential for the
description of the diffractive processes, can be taken into account, using 
Mueller-Glauber approach \cite{MU90}.
\end{enumerate}
Using pQCD  and Mueller-Glauber approach we derived  a generalization of
Kovchegov-McLerran formula \cite{KOLE} for the ratio
$\sigma^{DD}/\sigma_{tot}$ ( see \eq{FIN} ) which is applicable to the
HERA experimental data on diffractive production at HERA.
However, we found that \eq{FIN} cannot describe the approximate energy
independence of the ratio $\sigma^{DD}/\sigma_{tot}$ , observed
experimentally. Our attempts to introduce the experimental cuts for
diffractive production does not change this pessimistic conclusion.

Therefore, we believe, this paper is a strong argument that the
nonperturbative QCD contribution is essential for diffractive
production  and our approach, based on pQCD, should be reconsidered.
However, we showed that the main source of the observed energy dependence
arises from
rather short distances (  see Figs. 16 and 17  ) where we did not see
any nonperturbative correction to the total DIS cross section.
In principle, it was pointed out in Ref. \cite{KL} that the scale anomaly
of QCD generates a nonperturbative contribution at high energy at
sufficiently short distances $r^2_{\perp} \approx 1/M^2_0 \sim 0.25
GeV^{-1}$ . We intend studying
 the influence of such  nonperturbative corrections 
in further publications.

We studied in detail the contribution of the excited hadronic states in
diffractive production. We found that the contribution of nucleon
excitations  should depend on $Q^2$ and $x$  leading to large cross
section
at bigger values of
$Q^2$ and higher $x$.

We hope that our paper will draw the attention of the high energy
community to the
beautiful experimental data on the energy dependance of the ratio 
$\sigma^{DD}/\sigma_{tot}$ which have  still not recieved an adequate
theoretical explanation, and which can provide a new insight to the 
importance of nonperturbative  corrections at sufficiently short
distances.

{\bf Acknowledgements:}  E.L.  would like to acknowledge the hospitality
extended to him at DESY Theory Group where this work was started. 

The research   was supported in part by  BSF \# 9800276
and by  the Israel Science
Foundation, founded by the Israeli Academy of Science and Humanities.

\newpage

 \begin{appendix}
\section{Appendix}
\setcounter{equation}{0}
In this Appendix we present a derivation of formulae for the cross
sections for the diffractive dissociation
production of  a  $q\bar{q}$ pair  and the $q\bar{q}G$ parton system when
a final state mass window
is selected. In order to preserve the unitarity relation between the DD
 cross section and the total
cross section, we modify the latter. All the formulae are written in the
leading $\log(1/x) $
aproximation of pQCD.  Below we only present  results derived for the
transversely
polarized photon. The results for the longitudinal part can be obtained
 by similar treatment.

\subsection{ $q\bar{q}$ contribution}
The $q\bar{q}$ DD cross section has the form \cite{MU90}:
\begin{eqnarray}
\label{1}
\si^{DD}_{qq}\,\,=\,16\,\pi\, \int \,\frac{dM^2}{M^2+Q^2}\,
\frac{\alpha_{em} \,N_c}{8\,(2 \pi)^2}\,\sum_f \,Z^2_f\,
\int\,\frac{d^2k_{\perp}}{(2 \pi)^2} \int^1_0 \,d z \,\,\,4\pi\, N_\lambda \,(M^2\,+\,Q^2) \times
\nonumber \\
\delta(M^2\,\,-\,\,\frac{k^2}{z\,(1-z)})\,\, |f_{qq}|^2,
\end{eqnarray}
where within our model for the dipole interaction the square of the amplitude is written:
\begin{eqnarray}
\label{2}
|f_{qq}|^2\,=\,\int d^2b \int
\frac{\,d^2 r}{2\pi}  \frac{d^2 r^\prime}{2\pi}\,\,\Psi^{\gamma^*}(r))^*
\Psi^{\gamma^*}(r^\prime) \{ 1 - e^{ - \frac{\Omega^{MG}(x^\prime,r;b)}{2}} \} 
\{ 1 - e^{ - \frac{\Omega^{MG}(x^\prime,r^\prime;b)}{2}} \} 
e^{ i\vec{k}(\vec{r}-\vec{r}^\prime)}
\end{eqnarray}
The notation  has been introduced in  section 2 (see also Fig. 3),
while 
the parameters are defined as follows.
$$x^\prime\,\,=\,\,(k^2\,/\,z(1-z)\,\,+\,\,Q^2)\,/\,W^2; \ \ \ \ \ \ 
 N_\lambda\,\, =\,\,z^2\,\,+\,\,(1-z)^2; \ \ \ \ \\ \ a^2\,\,=\,\,Q^2\, z\,(1-z) $$
It is important to note that in \eq{2} we introduce a correct energy argument $x^\prime$
in $\Omega$ since at fixed $M^2$ the energy of the
 dipole-proton interaction is $\beta\, W^2$.
The kinematic constraint forced by the delta function sets
$x^\prime\,=\,x_P$.
Performing the angle integrations we obtain
\begin{eqnarray}
\label{3}
|f_{qq}|^2\,=\, \,\int d^2b \,\left[ \int_0^\infty dr\, r\, a\, K_1(ar)\,J_1(kr)\,
\{ 1 - e^{ - \frac{\Omega^{MG}(x^\prime,r;b)}{2}} \}\right]^2,
\end{eqnarray}
Finally for the cross-section we  get the result
\begin{eqnarray}
\label{4}
\si^{DD}_{qq}\,= \,8\,\alpha_{em} 
\int \frac{dM}{M^3} 
\int_0^{M/2} \frac{dk\,k^3\,  N_\lambda}{\sqrt{1-4k^2/M^2}}
\int db^2 \left[ \int_0^\infty dr\, r\, a\, K_1(ar)\,J_1(kr)
\{ 1 - e^{ - \frac{\Omega^{MG}(x_p,r;b)}{2}} \}\right]^2
\end{eqnarray}
with
$$
N_\lambda\,\,\,=\,\,1\,\,-\,\,2\,k^2\,/\,M^2\,; \ \ 
\ \ \ \ \ \ \ \ \ \ \ \  a\,\,\,=\,\,\,Q\,k\,/\,M\,.
$$

\subsection{ $q\bar{q}G$ contribution}
Consider the diffractively produced $q\bar{q}G$  system with $z$ and
$z^\prime$ being
the fractions of the energy carried by quark and gluon respectively. We assume that the
transverse gluon momentum $l$ is much smaller than the quark transverse momentum $k$.
In the leading $\log(1/x)$ approximation of pQCD $z^\prime  \ll z$.
The kinematic constraint is  dictated by the final state mass:
$$M^2\,\,\,=\,\,\,k^2\,/z\,(1-z)\,\,+\,\,l^2\,/z^\prime . $$
The cross section for the $q\bar{q}G$ production is
\begin{eqnarray}
\label{5}
\si^{DD}_{qqG}\,=\,(16\pi)
\frac{\alpha_{em} N_c}{8(2 \pi)^2}\sum_f \,Z^2_f \int dM^2
\int\,\frac{d^2k}{(2 \pi)^2} \int^1_0 \,d z \,\,\,4\pi N_\lambda \,|f_{qqG}|^2,
\end{eqnarray}
with the square of the amplitude
\begin{eqnarray}
\label{6}
|f_{qqG}|^2\,=\, \int d^2b \int
\frac{\,d^2 r}{2\pi} \, \frac{d^2 r^\prime}{2\pi}\,\,\Psi^{\gamma^*}(r))^*
\Psi^{\gamma^*}(r^\prime)  e^{ - \frac{\Omega^{MG}(x^\prime,r;b)}{2}}  \,
e^{ - \frac{\Omega^{MG}(x^\prime,r^\prime;b)}{2}} \,
e^{ i\vec{k}(\vec{r}-\vec{r}^\prime)}\, \tilde{\kappa}/2.
\end{eqnarray}
$\tilde{\kappa}$ is defined as follows.
\begin{eqnarray}
\label{7}
\tilde{\kappa}\,=\,\frac{2}{\pi^2}\int_0^zdz^\prime\int_r^\infty \frac{d^2R}{\pi}
\frac{d^2R^\prime}{\pi}
\int_0^{k^2}\frac{d^2l}{(2\pi)^2}\delta(M^2-k^2/z(1-z)-l^2/z^\prime)\,
e^{i\vec{l}(\vec{R}-\vec{R}^\prime)} 
\times \nonumber \\ \nonumber \\
\Psi_g(R)\,
\Psi_g^*(R^\prime) \,(1- e^{ - \frac{\Omega^P_G(x_P,R;b)}{2}}  )\,
(1-e^{ - \frac{\Omega^P_G(x_P,R^\prime;b)}{2}})\, \sqrt{\as(r^2)}\,
\sqrt{\as(r^{\prime 2})}\,\frac{\pi^2\,\vec{r}\,\cdot\, \vec{r^\prime}}{3}
\end{eqnarray}
We use the small $z^\prime$ approximation of the gluon wave function
$$
\Psi_g^{m\,n}(R,z^\prime)\,\,\simeq\,\,\frac{\sqrt{2}}{\sqrt{z^\prime}\, R^2}\,
\left(\delta^{m\,n}\,-\,2\,\frac{R^m\,R^n}{R^2}\right)
$$
Performing the angle integration and
 removing the delta function by doing the $z^\prime$ integration we obtain
\begin{eqnarray}
\label{8}
|f_{qqG}|^2\,=\, \frac{2}{3} \frac{1}{M^2-k^2/z(1-z)}\int d^2b \int_0^{zM^2-k^2/(1-z)} 
\frac{d^2l}{(2\pi)^2}
\times \nonumber \\ \nonumber \\
 \left[\int \frac{\,d^2 r}{2\pi} \, a
K_1(ar) \sqrt{\as(r^2)} r J_1(kr) e^{ - \frac{\Omega^{MG}(x^\prime,r;b)}{2}} 
\int_{r^2}^\infty \frac{dR^2}{R^2} J_2(lR) (1-e^{ - \frac{\Omega^P_G(x_P,R;b)}{2}})\right]^2
\end{eqnarray}
As a result of the delta function integration we also find that
$k^2\,\leq\, M^2\, z\,(1-z)$. It should be stressed that $r$ is the size of the initial
quark -antiquark pair, while $R$ is the size of the produced two colour dipoles.
In our approximation, both dipoles have the same size with $R\,\gg\,r$.
Introducing   dimensionless variable $\tilde{l}$
\beq
\label{9}
l^2\,=\,(z\,M^2\,-\,k^2\,/\,(1-z))\, \tilde{l}^2
\eeq
we  finally arrive at the expression for the cross section:
\begin{eqnarray}
\label{10}
&&\si^{DD}_{qqG}\,\,=\,\,
\frac{\alpha_{em} }{6} \int dM^2
\int^1_0 \,d z \,z\, N_\lambda \int_0^{M^2z(1-z)}\,dk^2
\int db^2 \int_0^1d\tilde{l}^2 \times \\ \nonumber  \\ 
&&\left[\int \,dr\, r^2 \,a\,
K_1(ar) \sqrt{\as(r^2)} \, J_1(k\,r)\, e^{ - \frac{\Omega^{MG}(x^\prime,r;b)}{2}} \,
\int_{r^2}^\infty \frac{dR^2}{R^2}\, J_2(l\,R) \,
(1-e^{ - \frac{\Omega^P_G(x_P,R;b)}{2}})\right]^2. \nonumber
\end{eqnarray}
Contrary to the $q\bar{q}$ case, in the present expression  $x^\prime$,
 which is 
$ (k^2\,/\,z(1-z)\,+\,Q^2)\,/\,W^2$  is  not equal to $x_P$. 
The energy variables are different for
$\Omega^{GM}$ and $\Omega^P_G$, because they describe different physics.
Indeed, factor $e^{- \Omega^{MG}}$ is a probability that $q \bar q $ 
pair does not interact inelastically before emission of the extra gluon,
while $( 1 - e^{- \frac{\Omega^P_G}{2}} )^2$ stands for diffractive
production of $q \bar q G$ system ( two colour dipoles of size
$R$ ). For the rescattering of the $q \bar q$ colour dipole the 
energy is
$s =W^2 \,x_B$, while the rescattering of the colour dipoles of size
$R$ occurs at energy $s' = \beta s$.

\subsection{ Total cross section}

In order to preserve the unitarity relation between $q\bar{q}$ DD cross section and
the total cross section we modify the latter.
Similarly to the $q\bar{q}$ case we write for the total cross section
 \begin{eqnarray}
\label{11}
\si_{tot}\,\,=\,\,16\,\pi\, \int \,\frac{dM^2}{M^2+Q^2}\,
\frac{\alpha_{em}\, N_c}{8\,(2 \pi)^2}\,\sum_f \,Z^2_f \,
\int\,\frac{d^2k_{\perp}}{(2 \pi)^2}\, \int^1_0 \,d z \,\,4\pi\, 
N_\lambda\, (M^2+Q^2) \times 
\nonumber \\ 
\delta(M^2\,-\,\frac{k^2}{z(1-z)})\,\,|f_{tot}|^2
\end{eqnarray}
with the square of the amplitude 
\begin{eqnarray}
\label{12}
 |f_{tot}|^2&=& \int d^2b \int
\frac{\,d^2 r}{2\pi} \, \frac{d^2 r^\prime}{2\pi}(\Psi^{\gamma^*}(r))^*
\Psi^{\gamma^*}(r^\prime) 
\{ 2 \,-\, e^{ - \frac{\Omega^{MG}(x^\prime,r;b)}{2} }\,-\,
e^{ - \frac{\Omega^{MG}(x^\prime,r^\prime;b)}{2}} \} 
e^{ i\vec{k}(\vec{r}-\vec{r}^\prime)}\,\,\,= \nonumber \\  \,\,\,\,\,\ &&\,\,\   \\ 
  &&\int d^2b \int_0^\infty dr\, r\, dr^\prime \, r^\prime 
\,a^2\, K_1(ar)\,J_1(kr)\, K_1(ar^\prime)\,J_1(kr^\prime)
\{ 2\, -\, e^{ - \frac{\Omega^{MG}(x^\prime,r;b)}{2}}\,-\,e^{ - \frac{\Omega^{MG}
(x^\prime,r^\prime;b)}{2}} \}
\nonumber
\end{eqnarray}
One of the space integrations can be performed analytically noting that 
\begin{eqnarray}
\label{13}
\int_0^\infty dr\, r\, K_1(a\,r)\, J_1(k\,r)\,\,=\,\,\frac{k}{a\,(k^2\,+\,a^2)}
\end{eqnarray}
Substituting $a\,=\,Q\,k\,/\,M$ we finally obtain the total cross section
\begin{eqnarray}
\label{14}
\si_{tot}\,\,=\,\, 16\,\alpha_{em} 
\int \frac{dM}{M^2}\,  \frac{Q}{Q^2\,+\,M^2}
\int_0^{M/2} k^3\,dk \,\frac{ 1\,-\,2\, k^2\,/\,M^2}{\sqrt{1-4\,k^2\,/\,M^2}}
 \times \nonumber \\  \nonumber \\ 
 \int db^2  \int dr\, r\, K_1(a\,r)\,\,J_1(k\,r)\,\,
\{ 1 \,-\, e^{ - \frac{\Omega^{MG}(x_P,r;b)}{2}} \}
\end{eqnarray}
In the above expression for the total cross section the mass integration
  should be carried out over the
 whole infinite mass interval.
The  result obtained is consistent with the previous expression for the
total cross section.
If we change $x_P$ to $x_B$ and perform the  $M$ integration we reproduce 
the old result written in \eq{FIN}. A useful equality we use is
\begin{eqnarray}
\int dM  \,\frac{M^2}{Q^2\,+\,M^2}\,\, J_1(M\, r) \,\,=\,\, Q\, K_1(Q\,r)
\end{eqnarray}

\end{appendix}

\end{document}